\documentclass[usenatbib,psfig]{mn2e}

\usepackage{epsfig}
\usepackage{subfigure}
\usepackage{amssymb}

\begin{document}


\title[Environments of Interacting Transients]{Environments of interacting transients: Impostors and type IIn supernovae}
\author[S. M. Habergham et al.]{S. M. Habergham$^1$\thanks{E-mail:
s.m.habergham@ljmu.ac.uk (SMH)},  J. P. Anderson$^{2,3}$, P. A. James$^1$
\&  J. D. Lyman$^{1,4}$\\
$^1$Astrophysics Research Institute, Liverpool John Moores University, Liverpool, L3 5RF, UK\\
$^2$Departamento de Astronom\'ia, Universidad de Chile, Casilla 36-D, Santiago, Chile\\
$^3$European Southern Observatory, Alonso de C\'{o}rdova 3107, Casilla 19001, Santiago, Chile\\
$^4$Department of Physics, University of Warwick, Coventry, CV4~7AL, UK\\
}

\date{Accepted . Received ; in original form }

\pagerange{\pageref{firstpage}--\pageref{lastpage}} \pubyear{2013}

\maketitle

\label{firstpage}

\begin{abstract}
This paper presents one of the first environmental analyses of the locations of the class of `interacting transients', namely type IIn supernovae and supernova Impostors. We discuss the association of these transients with star formation, host galaxy type, metallicity, and the locations of each event within the respective host. Given the frequent assumption of very high mass progenitors for these explosions from various studies, most notably a direct progenitor detection, it is interesting to note the weak association of these subtypes with star formation as traced by H$\alpha$ emission, particularly in comparison with type Ic supernovae, which trace the H$\alpha$ emission and are thought to arise from high mass progenitors. The radial distributions of these transients compared to type Ic supernovae are also very different. This provides evidence for the growing hypothesis that these `interacting transients' are in fact comprised of a variety of progenitor systems. The events contained within this sample are discussed in detail, where information in the literature exists, and compared to the environmental data provided. Impostors are found to split into two main classes, in terms of environment: SN2008S-like Impostors fall on regions of zero H$\alpha$ emission, whereas $\eta$ Carina-like Impostors all fall on regions with positive H$\alpha$ emission. We also find indications that the Impostor class originate from lower metallicity environments than type IIn, Ic and IIP SNe.
\end{abstract}


\section{Introduction}

Studies of the host galaxies and local environments of supernova (SN) explosions have been integral in our understanding of the progenitors of these explosions. Initially host galaxy studies found the almost exclusive presence of Core-Collapse Supernovae (CCSNe) in star-forming galaxies \citep[e.g.][]{vanden05}, implying young stellar progenitors. The more central locations of type Ibc SNe compared to type II SNe also led to the suggestion that these events arise from higher metallicity environments \citep[e.g.][]{bois09,ande09}. These examples demonstrate the power of using host galaxy analysis, and studying the local environments of SNe, in constraining the progenitors of these explosions. These type of analyses will be utilised throughout this paper to concentrate on a sample of transients generally accepted to be interacting with a dense circumstellar medium (CSM), namely type IIn supernovae (SNIIn) and SN Impostors. The progenitor systems of these `interacting transients', and the cause of the dense CSM, may arise through a variety of routes. In this introduction we will discuss both types of transients in this sample, SNIIn and Impostors, the features which appear common to both groups, and the possible subgroups within the classes.

\subsection{SNIIn}

The IIn class is rare within the CCSN group ($\sim$7~per cent according to \citealt{li11}), and the nature of the explosions remains very uncertain, with the class being characterised principally by narrow lines in the spectrum \citep[e.g.][]{kiew12}. The class was originally defined by \citet{schl90}, though he also stressed that a fundamental progenitor characteristic was required for the objects to be a truly distinct class, something which remains elusive. This initial paper not only characterised the objects as having narrow spectral features, but also having a blue continuum and slow evolution (at least spectroscopically), characteristics of the group which remain prominent \citep{kiew12}. 

The source of the narrow emission lines is generally accepted to be the result of interaction between the supernova ejecta and a dense circumstellar medium immediately surrounding the explosion. This interaction creates hard emission which photo-ionises the surrounding unshocked material, resulting in an H$\alpha$ excess \citep[e.g.][]{chug91}. As SNIIn have been explored in more detail, the narrow spectral lines have been found to have multiple components; a `narrow' line (typically a few hundred kms$^{-1}$) caused by the photoionisation of the unshocked wind surrounding the progenitor, and an `intermediate-width' line (a few thousand kms$^{-1}$) from dense post-shock gas (e.g. \citealt{smit08}; \citealt{kiew12}). These compare to typical broad lines in SN ejecta with widths of 10,000 - 20,000 kms$^{-1}$ \citep{smit08}. 

\subsubsection{SN2005gl}
The only robust direct detection of a SNIIn progenitor was made for SN2005gl \citep{galy09}. The detection of a star at its location, and the subsequent disappearance, indicated that the progenitor was a luminous blue variable (LBV) star with a mass likely to be in excess of 50 M$_{\odot}$ \citep{galy09,smar09}. The calculated mass loss rates and wind speeds of well-studied SNIIn seem to reflect this progenitor channel with only LBV stars predicted to be able to reach high enough mass loss rates, yet with smaller wind speeds in general than Wolf-Rayet (WR) stars \citep[e.g.][]{kiew12,tadd13}. The lack of information available about the variability of the star pre-explosion does lead to some debate as to whether the pre-explosion observations of SN2005gl, which give rise to the high mass estimates of the progenitor, were taken during a quiescent period. Little is known about the long term variability of LBVs, though recent work by \citet{ofek14} has studied type IIn SNe detected by the Palomar Transient Survey \citep{law09,rau09} and found that half of IIn show at least one outburst up to $\sim$120 days prior to explosion, with multiple outbursts common over the preceding year. An example of this is the type IIn SN2010mc which experienced a huge mass loss event 40 days prior to explosion \citep{ofek13a}. Studies have also shown LBVs to have major outbursts on timescales of years \citep[e.g.][]{past10,szcz10} which suggest that there is a chance that the progenitor of SN2005gl was serendipitously in outburst on the single pre-explosion image.  Should the LBV have been in outburst when the one pre-explosion image was taken, it is possible that the quiescent star had an actual intrinsic brightness 3 magnitudes lower \citep{hump94} and consequently the mass would lie in the range of 20-25 M$_{\odot}$ according to the models of \citet{groh13b}.

There is more controversy over the assumption of LBV progenitors for SNe, not least because stellar evolution models have until recently been unable to end a star's lifetime in the LBV phase (e.g. \citealt{maed08}, though see \citealt{hirs10} for some possible scenarios). Stellar evolution models instead require the LBV star to lose the hydrogen envelope before becoming a WR star and then exploding \citep{dwar11}. However, \citet{groh13} found from their rotational stellar evolution models coupled with atmospheric models, that stars in the mass range 20-25 M$_{\odot}$ exploded in the LBV phase. Although no post-explosion spectra were computed, \citet{groh13} speculated from the surface abundance measurements pre-explosion and position of the star on the HR diagram, that explosions of stars of 20 M$_{\odot}$ are likely to be classified as type IIb rather than IIn SNe. However, the increased mass loss of the 25 M$_{\odot}$ stars is likely to produce a H-rich, dense CSM, and the interaction of the SN with this CSM could produce a type IIn SN. 

Aside from the case of SN2005gl, \citet{dwar11} argues that the assumptions of LBV progenitors may be flawed. The mass loss equations assume wind mass-loss rates and velocities to be constant with time \citep{dwar11}. An LBV phase, with numerous eruptions prior to explosion (required in order to have the dense medium for the ejecta to interact with), does not have a constant wind mass-loss rate. The one certainty regarding the progenitors is the need for a dense CSM (see \citealt{ofek14a} for a discussion). The assumption has traditionally been that this must be due to LBV-type eruptions, but \citet{dwar11} argue that it is possible that the dense CSM was produced over a longer timescale prior to the eventual eruption of the progenitor star, meaning an LBV could have progressed into its WR phase before explosion. Alternatively a clumpy CSM could produce a similar result, where interaction with a dense clump could cause narrow lines in the spectrum, but not represent an overall high-density CSM (see \citealt{dwar11} for a detailed discussion). The general consensus in the literature, however, appears to be that the most likely candidate progenitors of SNIIn are LBV stars, though there are likely to be several different progenitor types even within this small but diverse class.

The SNIIn class has therefore proved mysterious. Aside from the defining narrow emission lines, the objects within the class show diverse features, particularly light curves (LCs), and over time it has been discovered that many of the objects are not true CCSNe at all and have hence been reclassified. 

\subsubsection{Type Ibn SNe}
A rare ($<$1 per cent of all CCSNe; \citealt{past08,smit12}) subclass of transients have been termed type Ibn \citep{past07}; these appear to be type Ib/c CC-explosions embedded within a He-rich envelope \citep[e.g.][]{past08}. SN2006jc is often referred to as the prototype for this group \citep[e.g.][]{matt08}, but several others have been observed (e.g. SN1999cq, \citealt{modj99}; SN2000er, \citealt{chas00}; SN2002ao, \citealt{mart02}). These events have been suggested to be a distinct class of CCSNe \citep{fole07}, but more recent analysis has suggested that they are more likely SNIb/c explosions occurring in a high density CSM, formed either by pre-explosion mass loss of a very massive ($<$100M$_{\odot}$) single star, or an LBV and WR binary system \citep[e.g.][]{past07,past08}. They have distinctive spectroscopic characteristics showing strong signs of CSM interaction in the form of narrow lines, but these tend to be dominated by He rather than H \citep[e.g.][]{past08,smit12}. More recent events found to exhibit these characteristics (e.g. SN2011hw, \citealt{smit12}) indicate that the class may be a span a large range, with some events showing more H emission, which could occur if the progenitors exploded at different points along the transition from an LBV to a WR \citep{smit12}. The recent work of \citet{gorb13} obtained the earliest ever observations of a type Ibn SN, iPTF13beo, whose light curve showed a double peak structure. \citet{gorb13} interpret this as a massive star exploding in a dense CSM with the initial peak powered by SN shock breakout in the CSM, followed by the second peak representing the SN radioactive decay. However, this massive star origin has been complicated by recent observations of the type Ibn SN Pan-STARRS1-12sk which was found in an elliptical brightest cluster galaxy containing no star formation \citep{sand13}, suggesting that some of these events may have older stellar progenitors.

\subsubsection{Thermonuclear IIn}
A group within the IIn class has been identified as thermonuclear explosions within a dense hydrogen-rich environment \citep[e.g.][]{deng04,dild12}, a hybrid of the SNIa thermonuclear class, yet showing SNIIn features (notably the presence of hydrogen), often referred to in the literature as Ia-CSM \citep[e.g.][]{silv13}. The WD within the system explodes as an SNIa when it reaches the carbon ignition point and then interacts with a dense CSM producing type IIn-like emission lines (e.g. SN2002ic; \citealt{hamu03}, although see \citealt{bene06} for a possible massive star origin for these systems). The origin of this dense CSM is still unclear as is the percentage of the currently classified SNIIn which may actually belong to this group. \citet{silv13} studied SNIIn detected by the Palomar Transient Factory (PTF; \citealt{rau09, law09}) and found that $\sim$10 per cent of explosions classified as SNIIn were more likely to be thermonuclear supernovae, interacting with a dense CSM. They also find the SNIa-CSM in their sample to have a range of magnitudes in the {\it R}-band of --21.3 to --19, i.e. brighter than most CCSN events. Several SNIIn events have been reclassified as SNIa-CSM due to their similarity to previous explosions, most notably SN1997cy \citep{germ00,tura00}. However, this has been questioned by \citet{inse13} who intensively followed one of these events, SN2012ca, and found its nebular spectrum to be consistent with a core-collapse explosion. 

\subsection{Impostors}
Some explosions showing narrow emission lines, originally classified as SNIIn, have later been re-classed as SN `Impostors' when it has become clear that the progenitor stars have survived (e.g. SN1954J, \citealt{smit01}; SN2009ip, \citealt{berg09,mill09}). SN Impostors are thought to originate from the eruption of an LBV star, such as $\eta$ Carina, however, the eruption is non-terminal and hence not a true SN (e.g. \citealt{vand02}; \citealt{maun06}). Usually these events are much fainter than true SN explosions and so are distinct from the SNIIn group, however, the range in photometric properties of both the SNIIn and Impostor classes are so diverse that this may not always be the case (see \citealt{kiew12,tadd13} for the wide range of absolute magnitudes and decline rates of the SNIIn class). 

\subsubsection{2008S}
When SN2008S was discovered \citep{arbo08} it was classified as a SNIIn due to the narrow Balmer emission lines \citep{stan08} though some of the spectral features seen and the faint peak magnitude of the explosion led to some speculation that the event was actually a SN Impostor \citep{stee08}. Given its proximity, it was hoped that the progenitor star might have been directly detectable, but pre-explosion images obtained on the Large Binocular Telescope found nothing (upper limits M$_{U}$$>$--4.8, M$_{B}$$>$--4.3, M$_{V}$$>$--3.8; \citealt{prie08}). A point source was detected at the location of SN2008S in archival infrared {\it Spitzer} observations \citep{prie08}. The detection implied that the progenitor was a dust enshrouded $\sim$10~M$_{\odot}$ star, much lower than assumed SNIIn progenitor masses. The literature now agrees that this event was a SN Impostor rather than a true SN eruption \citep[e.g.][]{bond09,smith09}. It has come to define a group of SN Impostors which have lower masses than generally expected from LBV eruptions \citep{thom09}, which are usually accepted to be the progenitors of SN Impostors, and are often heavily dust enshrouded \citep{smith11c}.

\subsubsection{2009ip}
SN2009ip was incorrectly classified as a SN during its 2009 outburst; the progenitor star was known to have had previous S Dor-like outbursts (see \citealt{smit13} for an account of the star's pre-discovery variability), and was detected on pre-explosion images, taken 10 years prior to the 2009 discovery, as a massive LBV star (50-80 M$_{\odot}$; \citealt{smith10}). The star had another outburst in 2010 \citep{drak10}, and in 2012 had an outburst followed by re-brightening \citep{marg12}. It is still debated whether this most recent explosion is the transition into a true CCSN explosion. \citet{maue13} presented photometric and spectroscopic follow up of the event, which showed a SNII-like broad P Cygni profile in the Balmer lines, spectral lines with velocities typical of SN explosions and a peak magnitude of $\sim$--18. However \citet{fras13} find no evidence for nucleosynthesised material in late time spectra and the extensive multi-wavelength photometric and spectroscopic follow up of the event presented in \citet{marg13} shows the latest explosion to be consistent with another outburst of the LBV progenitor. \citet{smit13}, however, interpret the near infrared excess seen in the 2012 outburst of SN2009ip as the recent outburst propagating through previous episodic LBV outbursts which have deposited a dense CSM, and they conclude that the 2012 outburst was a true SN explosion. The work of \citet{maue14} also argues that the latest explosion was a true SN with the ejecta having $\gtrsim$10$^{51}$ ergs of kinetic energy, which is hard to reconcile with a progenitor having survived the explosion. The environment of SN2009ip is very unusual, as the star is located outside the main disk of the galaxy.

\subsection{This Paper}
The well-studied SNIIn explosions to date tend to be biased towards high-luminosity or `unusual' events \citep{kiew12}, which also means that any implications drawn from these events on the progenitors of the whole class could be flawed. Recently \citet{tadd13} presented the results of the follow up of five type IIn SNe from the Carnegie Supernova Project. Most of these were from targeted surveys and hence from bright, nearby spiral galaxies, with high extinction in only two cases, a highly inclined galaxy, and a SN close to the central regions of its host. \citet{tadd13} conclude that from their mass-loss equations, LBVs are still the most likely progenitor of this group.

In this paper, we conduct a study of the environments of CSM-interacting transients, in the form of type IIn SNe and SN Impostors. These events were selected by their presence in the Asiago SN Catalogue with an SNIIn classification, or through a literature search for Impostors. Events were not selected according to the amount of interaction observed, or by the types of galaxies in which they are present. We only require the host galaxies to have recession velocities less than 6000~kms$^{-1}$ and to have major to minor axis ratios less than 4:1, in order that our host galaxy analysis techniques are robust (see \citealt{habe12} for more detail). We present a new analysis of the radial distribution within their host galaxies of the resulting sample of interacting transients, and updated results for their respective association with H$\alpha$ emission. The total sample of events studied is 26 probable IIn and 13 probable impostors, although we can only analyse the detailed environment of 37 of these as explained below. We also carry out a robust analysis of the selection effects involved in both SNIIn and Impostor studies.

In Section 2 we present an analysis of the host galaxies within this sample, along with the positional information of the interacting transients in terms of the {\it R}-band and H$\alpha$ emission. Section 3 will analyse the association of the transients with star formation, as traced by H$\alpha$ emission. In Section 4 the selection effects within the samples of SNIIn and SN Impostors are explored. Section 5 discusses the individual events contained within these samples and conclusions are drawn in Section 6.

Throughout this paper comparisons will be made between the SNIIn and Impostor, or interacting transient class, and the SNIIP and SNIc classes (all CCSN subtype  samples are presented in \citealt{habe12,ande12} and SNIa samples in Anderson et al., in prep.). These comparison samples include SNe in all star-forming host galaxies (i.e. not just in undisturbed hosts, see \citealt{habe12}), however, it is emphasised that no events have been selected due to the interacting nature of their hosts, with all host galaxy classifications carried out after the SN analysis. The SNIIP and SNIc sub-types fall at the extremes of the currently understood mass sequence of CCSNe progenitors. The masses of SNIIP progenitors have been well established through direct detection methods (see \citealt{smar09} for a review) to lie within 8 and 20 M$_{\odot}$. Although the true mass range for type Ic SNe is unknown, the class as a whole are thought to result from the explosions of progenitors with much higher Zero-Age Main Sequence masses, even if these progenitors are within binary systems, both from their association with star formation \citep[e.g.][]{ande12,kunc13}, and through high mass stellar evolution models \citep[e.g.][]{geor12}.

\section{Host Galaxy Analysis}
The sample of interacting transients that will be analysed within this paper are presented in Table 1, including the reference from which the classification of the transient has been taken. The host galaxies are presented in Table 2, including the Hubble type and recession velocity\footnote{ Both taken from the NASA Extragalactic Database; ned.ipac.caltech.edu}, the distance, absolute magnitude, and inferred global and local (to the SN) metallicity. In total there are 39 events, and for each host galaxy we have \textit{R}-band and H$\alpha$ images, taken over a series of observing runs between 2001 and 2012. The details of the data reduction processes used are presented in \citet{ande12} for all of the observations taken prior to 2012. For the four host galaxies observed in 2012 by the Liverpool Telescope (LT), initial CCD reduction was done by the automated pipeline and an image subtraction pipeline built around the ISIS package \citep{alar00} was used to obtain continuum-subtracted H$\alpha$ images. 

The H$\alpha$ maps for each of the interacting-transient host galaxies can be seen in Figure 1, with the transients' positions marked with red circles. 

\begin{table*}
  \small
\begin{minipage}{160mm}
 \centering
 \caption{Transient sample, where host, RA and Dec are taken from the Asiago SN catalogue. The reference for the transient classification is given in the final column.}
   \begin{tabular}{lclccr}
    \hline 
    Event & Type & Host & RA$_{SN}$ (J2000) & Dec$_{SN}$ (J2000) & Classification Reference \\ 
    \hline 
     1954J & IMP & NGC~2403 & 07:36:55.20 & +65:37:54.0 & \citet{smit01}\\
     1961V & IMP & NGC~1058 & 02:43:36.42 & +37:20:43.58 & \citet{fili95}\\
     1987B & IIn & NGC~5850 & 15:07:02.97 & +01:30:13.19 & \citet{schl96}\\
     1987F & IIn & NGC~4615 & 12:41:38.99 & +26:04:22.40 & \citet{schl90}\\
     1993N & IIn & UGC~5695 & 10:29:46.20 & +13:01:14.00 & \citet{fili93IAUC}\\
     V1 & IMP & NGC~2366 & 07:28:43.37 & +69:11:23.9 & \citet{peti06,smith11}\\
     1994W & IIn & NGC~4041 & 12:02:10.89 & +62:08:32.35 & \citet{cumm94}\\
     1994Y & IIn & NGC~5371 & 13:55:36.86 & +40:27:53.17 & \citet{cloc94}\\
     1994ak & IIn & NGC~2782 & 09:14:01.47 & +40:06:21.50 & \citet{garn95}\\
     1995N & IIn & MCG-02-38-17 & 14:49:28.27 & -10:10:15.40 & \citet{poll95}\\
     1996bu & IIn & NGC~3631 & 11:20:59.30 & +53:12:08.40 & \citet{naka96}\\
     1996cr & IIn & ESO097-G13 & 14:13:10.01 & -65:20:44.40 & \citet{baue07}\\
     1997bs & IMP & NGC~3627 & 11:20:14.25 & +12:58:19.6 & \citet{vand99}\\
     1997eg & IIn & NGC~5012 & 13:11:36.73 & +22:55:29.40 & \citet{fili97iauc}\\
     1999bw & IMP & NGC~3198 & 10:19:46.81 & +45:31:35.0 & \citet{fili99}\\
     1999el & IIn & NGC~6951 & 20:37:17.72 & +66:06:11.50 & \citet{cao99}\\
     1999gb & IIn & NGC~2532 & 08:10:13.70 & +33:57:29.80 & \citet{jha99}\\
     2000P & IIn & NGC~4965 & 13:07:10.53 & -28:14:02.50 & \citet{jha00}\\
     2000cl & IIn & NGC~3318 & 10:37:16.07 & -41:37:47.80 & \citet{stat01}\\ 
     2001ac & IMP & NGC~3504 & 11:03:15.37 & +27:58:29.5 & \citet{math01}\\
     2001fa & IIn & NGC~673 & 01:48:22.22 & +11:31:34.4 & \citet{fili01iau}\\
     2002A & IIn & UGC~3804 & 07:22:36.14 & +71:35:41.50 & \citet{bene02}\\
     2002bu & IMP & NGC~4242 & 12:17:37.18 & +45:38:47.4 & \citet{smith11}\\
     2002fj & IIn & NGC~2642 & 08:40:45.10 & -04:07:38.50 & \citet{hamu02iauc}\\
     2002kg & IMP & NGC~2403 & 07:37:01.83 & +65:34:29.3 & \citet{schw03}\\
     2003G & IIn & IC~208 & 02:08:28.13 & +06:23:51.9 & \citet{hamu03iau}\\
     2003dv & IIn & UGC~9638 & 14:58:04.92 & +58:52:49.90 & \citet{kota03}\\
     2003gm & IMP & NGC~5334 & 13:52:51.72 & -01:06:39.2 & \citet{pata03}\\
     2003lo & IIn & NGC~1376 & 03:37:05.12 & -05:02:17.30 & \citet{math04}\\
     2005db & IIn & NGC~214 & 00:41:26.79 & +25:29:51.6 & \citet{blan05}\\
     2005gl & IIn & NGC~266 & 00:49:50.02 & +32:16:56.8 & \citet{blanc05}\\
     2005ip & IIn & NGC~2906 & 09:32:06.42 & +08:26:44.40 & \citet{smit09}\\
     2006am & IIn & NGC~5630 & 14:27:37.24 & +41:15:35.40 & \citet{quim06}\\
     2006bv & IMP & UGC~7848 & 12:41:01.55 & +63:31:11.6 & \citet{smith11}\\
     2006fp & IMP & UGC~12182 & 22:45:41.13 & +73:09:47.8 & \citet{blon06}\\
     2008J & IIn & MCG-02-07-33 & 02:34:24.20 & -10:50:38.5 & \citet{stri08}\\
     2008S & IMP & NGC~6946 & 20:34:45.35 & +60:05:57.8 & \citet{smith11}\\
     2010dn & IMP & NGC~3184 & 10:18:19.89 & +41:26:28.8 & \citet{smith11}\\ 
     2010jl & IIn & UGC~5189A & 09:42:53.33 & +09:29:41.8 & \citet{bene10}\\
     \hline
   \end{tabular}
 \end{minipage} 
\end{table*}

\begin{table*}
  \small
  \begin{minipage}{160mm}
    \centering
    \caption{The host galaxy information for each interacting transient in this sample. The recession velocity is given in units of kms$^{-1}$ and the distance in Mpc. The methods for calculating the inferred global and local metallicities are described in the text.} 
    \begin{tabular}{lcllccccc}
      \hline
      Event & Type & Host & Hubble Type & V$_{r}$ & Distance & Abs B-band Mag\footnote{From the Uppsala General Catalogue of Galaxies (UGC; \citealt{nils73}) where possible, otherwise from the Third Reference Catalogue of Bright Galaxies (RC3; \citealt{deva95})} & Galaxy Metallicity\footnote{Metallicities given in terms of 12+log(O/H) and calculated using \citet{trem04}.} & Local Metallicity\\
      \hline
      1954J & IMP & NGC~2403 & SABcd & 130 & 3.58 & --18.47 & 8.65 & ---\\  
      1961V & IMP & NGC~1058 & SAc & 518 & 7.67 & --17.62 & 8.50 & 8.30 \\ 
      1987B & IIn & NGC~5850 & SBb & 2556 & 37.58 & --19.28 & 8.80 & 8.02 \\ 
      1987F & IIn & NGC~4615 & Scd & 4716 & 69.35 & --20.41 & 9.01 & 8.32\\ 
      1993N & IIn & UGC~5695 & S? & 2940 & 43.24 & --18.28 & 8.62 & 8.31 \\
      V1 & IMP & NGC~2366 & IBm & 80 & 3.57 & --16.16 & 8.23 & ---\\
      1994W & IIn & NGC~4041 & SAbc & 1234 & 22.70 & --20.18 & 8.97 & 8.61 \\ 
      1994Y & IIn & NGC~5371 & SABbc & 2558 & 37.62 & --21.38 & 9.19 & 8.94\\ 
      1994ak & IIn & NGC~2782 & SABa & 2543 & 37.40 & --20.56 & 9.04 & 8.61 \\
      1995N & IIn & MCG-02-38-17 & IBm pec & 1856 & 27.29 & --17.68 & 8.51 & ---\\
      1996bu & IIn & NGC~3631 & SAc & 1156 & 13.10 & --19.59 & 8.86 & 8.28\\
      1996cr & IIn & ESO097-G13 & SAb & 434 & 4.21 & --16.02 & 8.20 & ---\\
      1997bs & IMP & NGC~3627 & SABb & 727 & 10.01 & --21.10 & 9.14 & 9.03\\ 
      1997eg & IIn & NGC~5012 & SABc & 2619 & 38.51 & --19.33 & 8.81 & 8.53\\
      1999bw & IMP & NGC~3198 & SBc & 663 & 13.99 & --20.03 & 8.94 & 8.56\\ 
      1999el & IIn & NGC~6951 & SABbc & 1424 & 22.56 & --19.47 & 8.84 & 8.17\\ 
      1999gb & IIn & NGC~2532 & SABc & 5252 & 77.24 & --21.54 & 9.22 & 8.95\\ 
      2000P & IIn & NGC~4965 & SABd & 2261 & 33.25 & --19.86 & 8.91 & 8.64\\  
      2000cl & IIn & NGC~3318 & SABb & 2775 & 40.81 & --20.86 & 9.10 & 9.03\\
      2001ac & IMP & NGC~3504 & SABab & 1534 & 20.05 & --20.01 & 8.94 & 8.54\\ 
      2001fa & IIn & NGC~673 & SABc & 5182 & 76.21 & --21.11 & 9.14 & 8.94\\
      2002A & IIn & UGC~3804 & Scd & 2887 & 42.46 & --20.14 & 8.96 & 8.48\\
      2002bu & IMP & NGC~4242 & SABdm & 506 & 7.90 & --17.59 & 8.49 & 7.49\\
      2002fj & IIn & NGC~2642 & SBbc & 4345 & 63.90 & --20.68 & 9.06 & 8.73\\
      2002kg & IMP & NGC~2403 & SABcd & 130 & 3.58 & --18.47 & 8.65 & 8.51\\ 
      2003G & IIn & IC~208 & SAbc & 3524 & 51.82 & --19.37 & 8.82 & 8.61\\ 
      2003dv & IIn & UGC~9638 & Im & 2771 & 33.40 & --17.12 & 8.40 & ---\\
      2003gm & IMP & NGC~5334 & SBc & 1386 & 32.62 & --18.87 & 8.73 & 8.42\\ 
      2003lo & IIn & NGC~1376 & SAcd & 4153 & 61.07 & --21.04 & 9.13 & 8.88\\
      2005db & IIn & NGC~214 & SABc & 4537 & 66.72 & --21.12 & 9.15 & 8.92\\  
      2005gl & IIn & NGC~266 & SBab & 4661 & 68.54 & --21.58 & 9.23 & 9.09\\ 
      2005ip & IIn & NGC~2906 & Scd & 2140 & 31.47 & --19.39 & 8.83 & 8.33\\
      2006am & IIn & NGC~5630 & Sdm & 2655 & 39.04 & --19.36 & 8.82 & 8.70\\
      2006bv & IMP & UGC~7848 & SABcd & 2513 & 36.96 & --18.04 & 8.58 & 8.35\\
      2006fp & IMP & UGC~12182 & S & 1534 & 21.91 & --16.20 & 8.24 & 7.64\\
      2008J & IIn & MCG-02-07-33 & SBbc & 4759 & 69.99 & --20.16 & 8.97 & 8.89\\
      2008S & IMP & NGC~6946 & SABcd & 48 & 5.96 & --18.38 & 8.64 & 8.06\\ 
      2010dn & IMP & NGC~3184 & SABcd & 592 & 11.95 & --19.99 & 8.94 & ---\\
      2010jl & IIn & UGC~5189A & Irr & 3207 & 47.16 & --19.57 & 8.86 & --- \\
      \hline
    \end{tabular}
  \end{minipage}
\end{table*}

\begin{figure*}
  \begin{minipage}{160mm}
    \begin{center}
      \renewcommand{\thesubfigure}{\thefigure.\arabic{subfigure}}
      \caption{H$\alpha$ observations of a sample of the SNIIn and Impostor host galaxies, with the SN positions marked with red circles. Each galaxy is displayed with North to the top, and East to the left.}
     \subfigure[NGC~2403: SN1954J (left); SN2002kg (right)]{\includegraphics[width=0.3\textwidth,height=0.15\textheight]{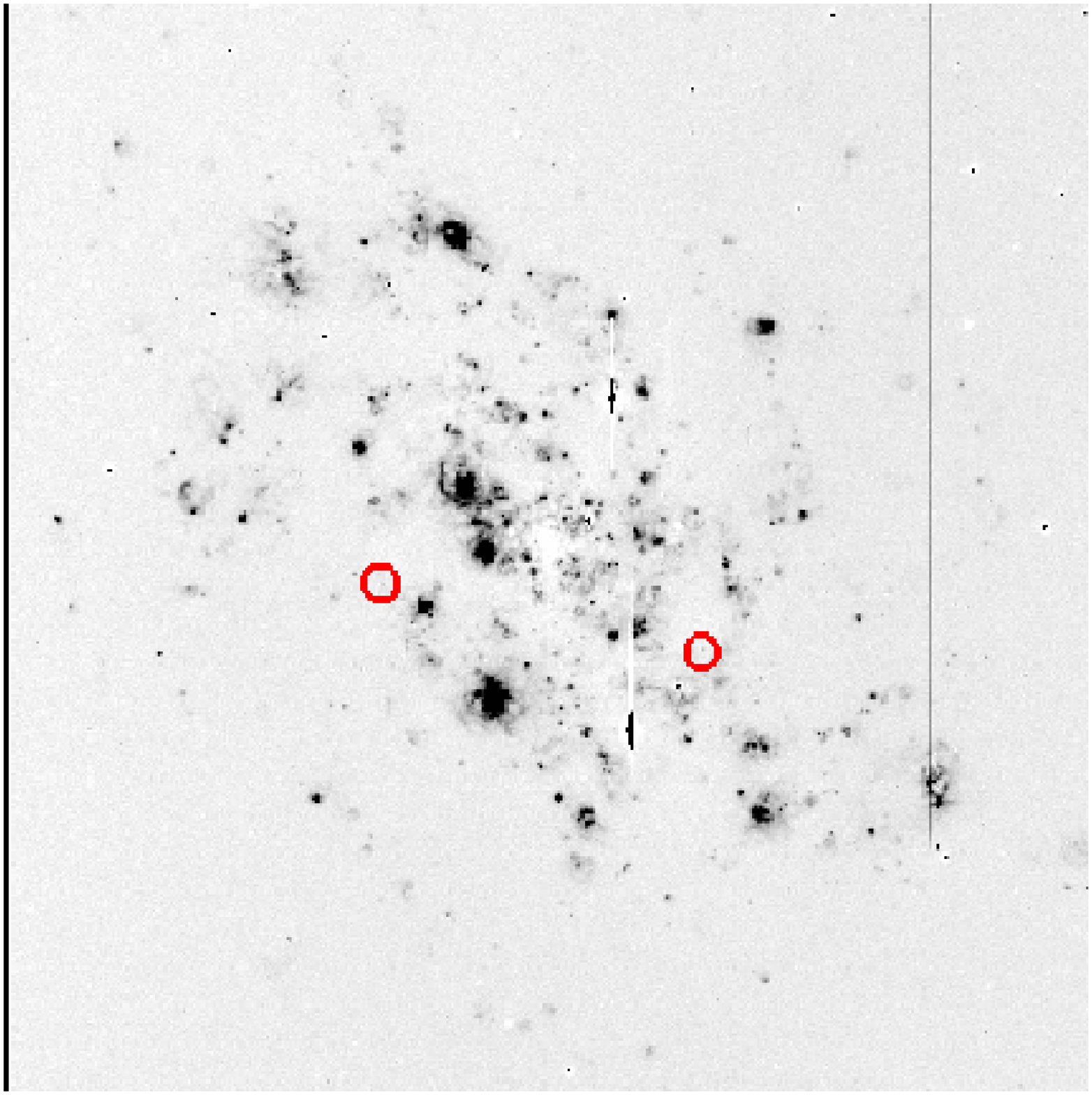}}  
     \subfigure[NGC~1058: SN1961V]{\includegraphics[width=0.3\textwidth,height=0.15\textheight]{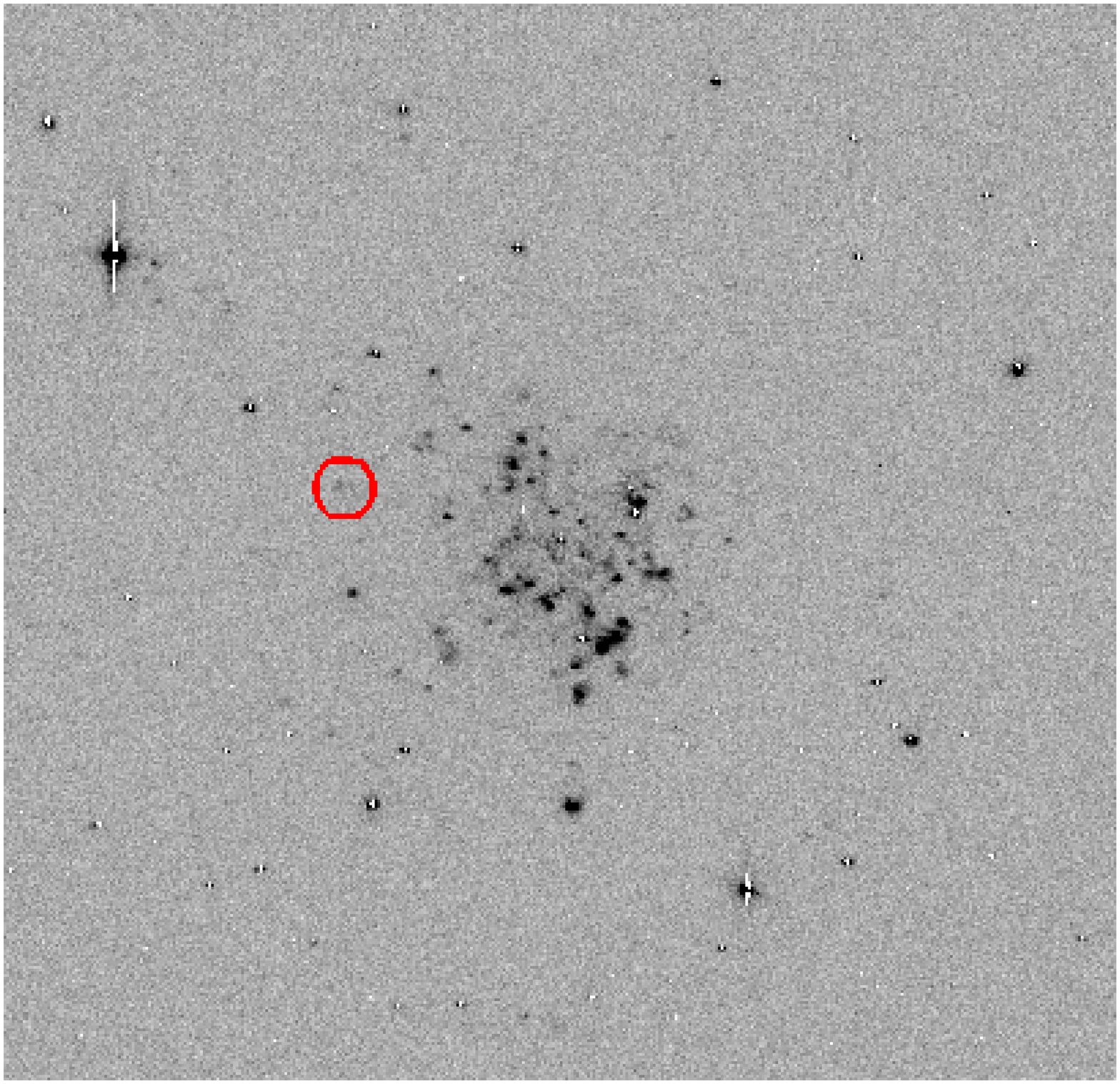}}   
     \subfigure[NGC~5850: SN1987B]{\includegraphics[width=0.3\textwidth,height=0.15\textheight]{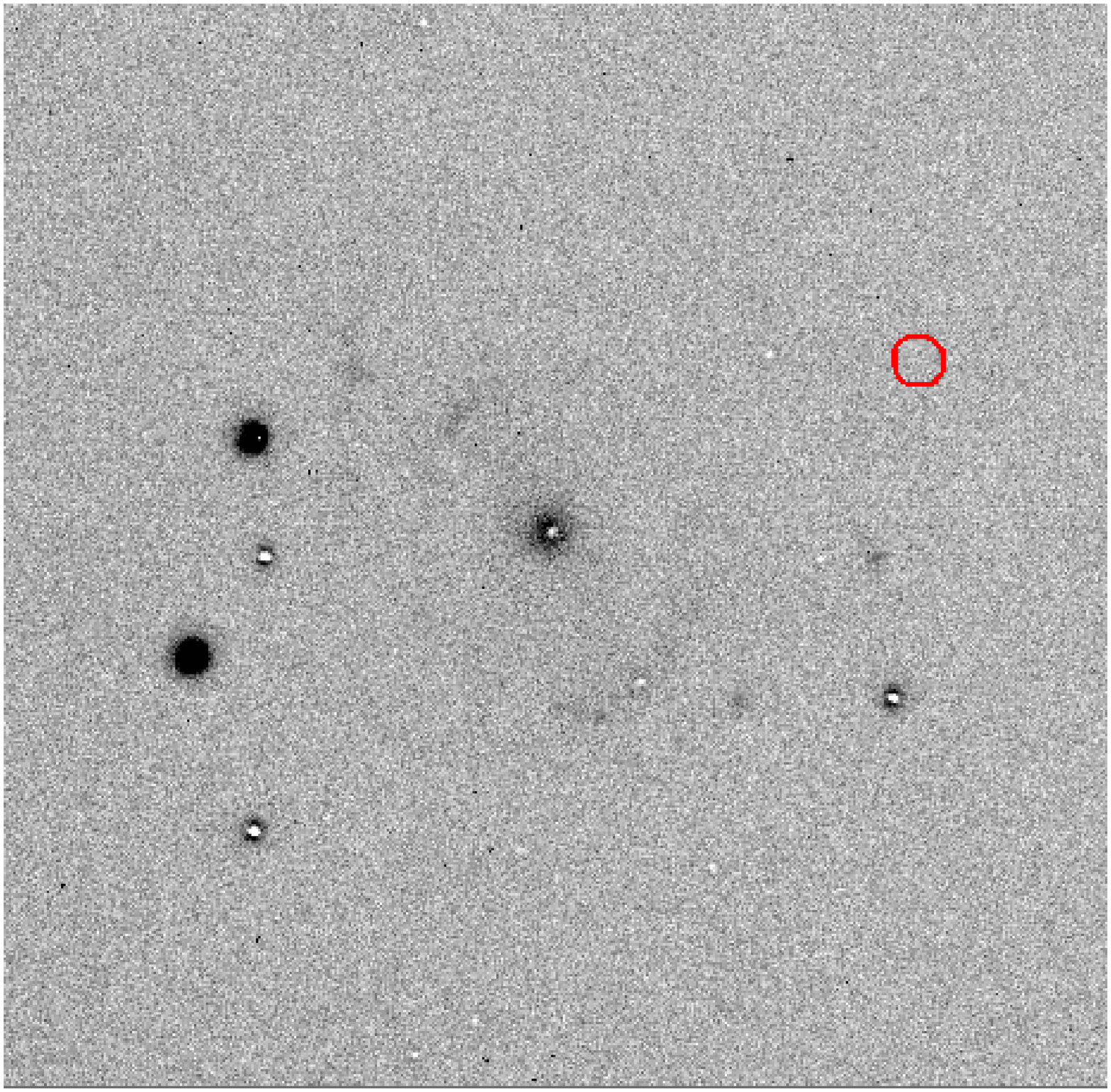}}\\   
     \subfigure[NGC~4615: SN1987F]{\includegraphics[width=0.3\textwidth,height=0.15\textheight]{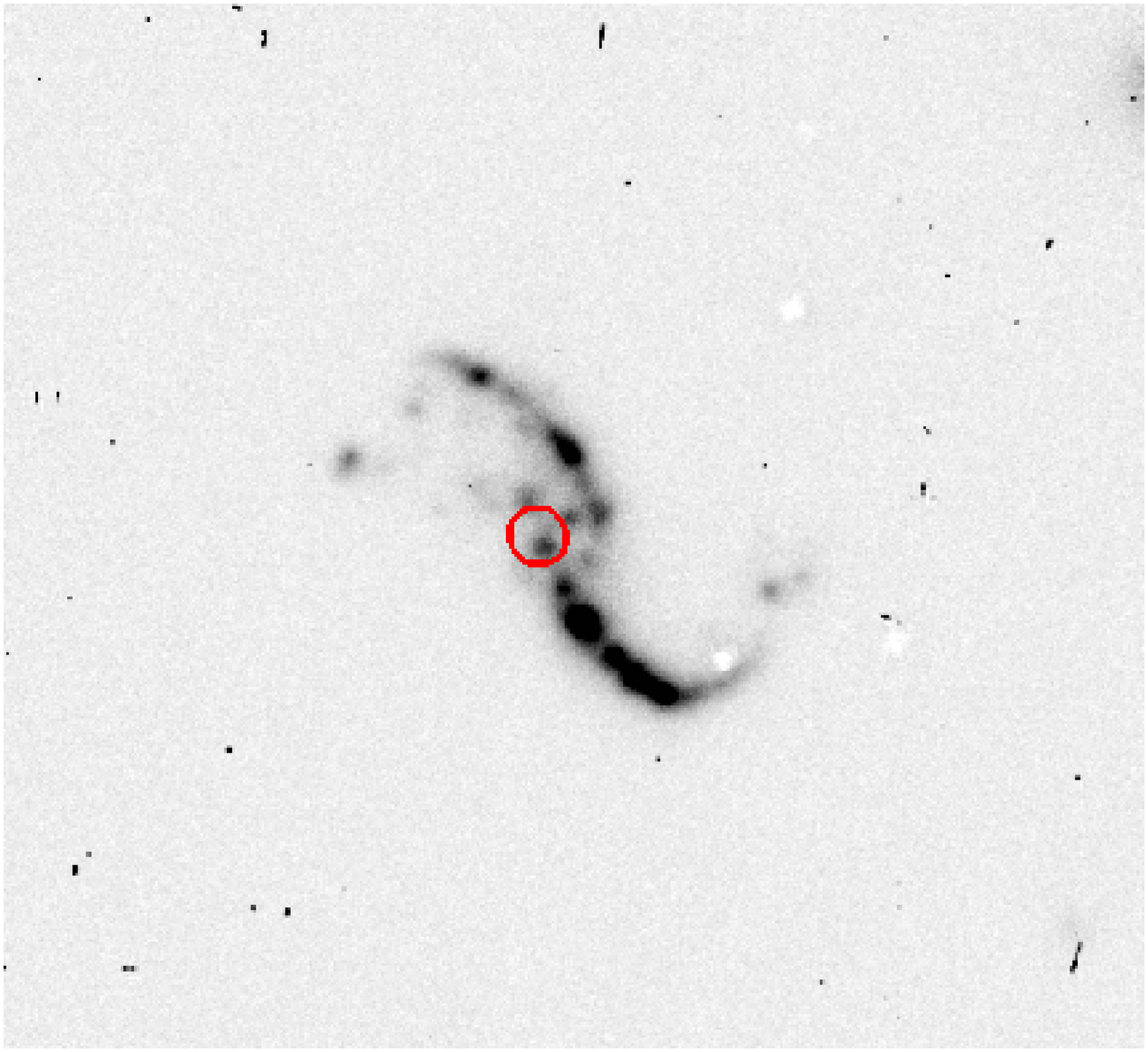}}   
     \subfigure[UGC~5695: SN1993N]{\includegraphics[width=0.3\textwidth,height=0.15\textheight]{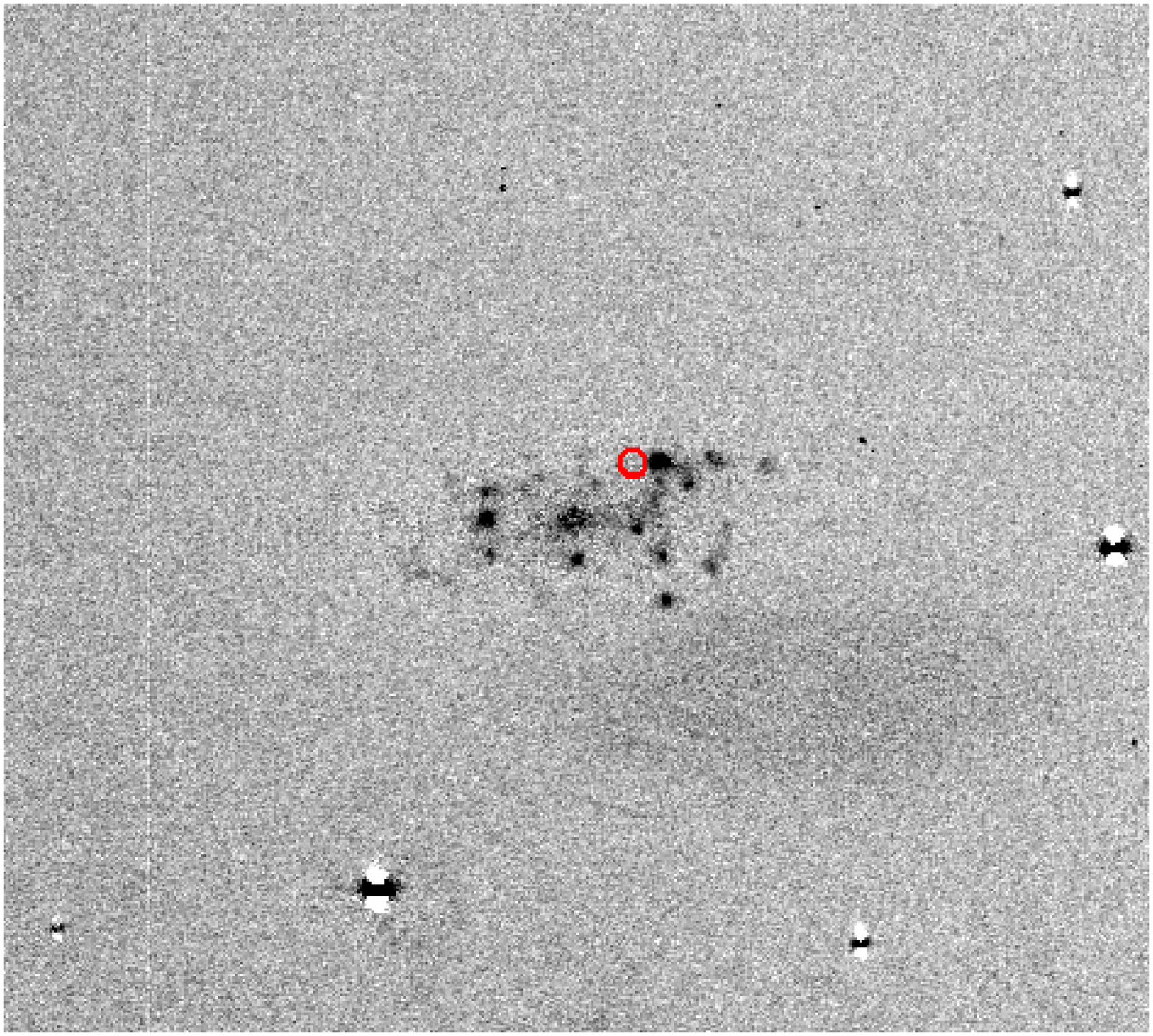}}  
     \subfigure[NGC~4041: SN1994W]{\includegraphics[width=0.3\textwidth,height=0.15\textheight]{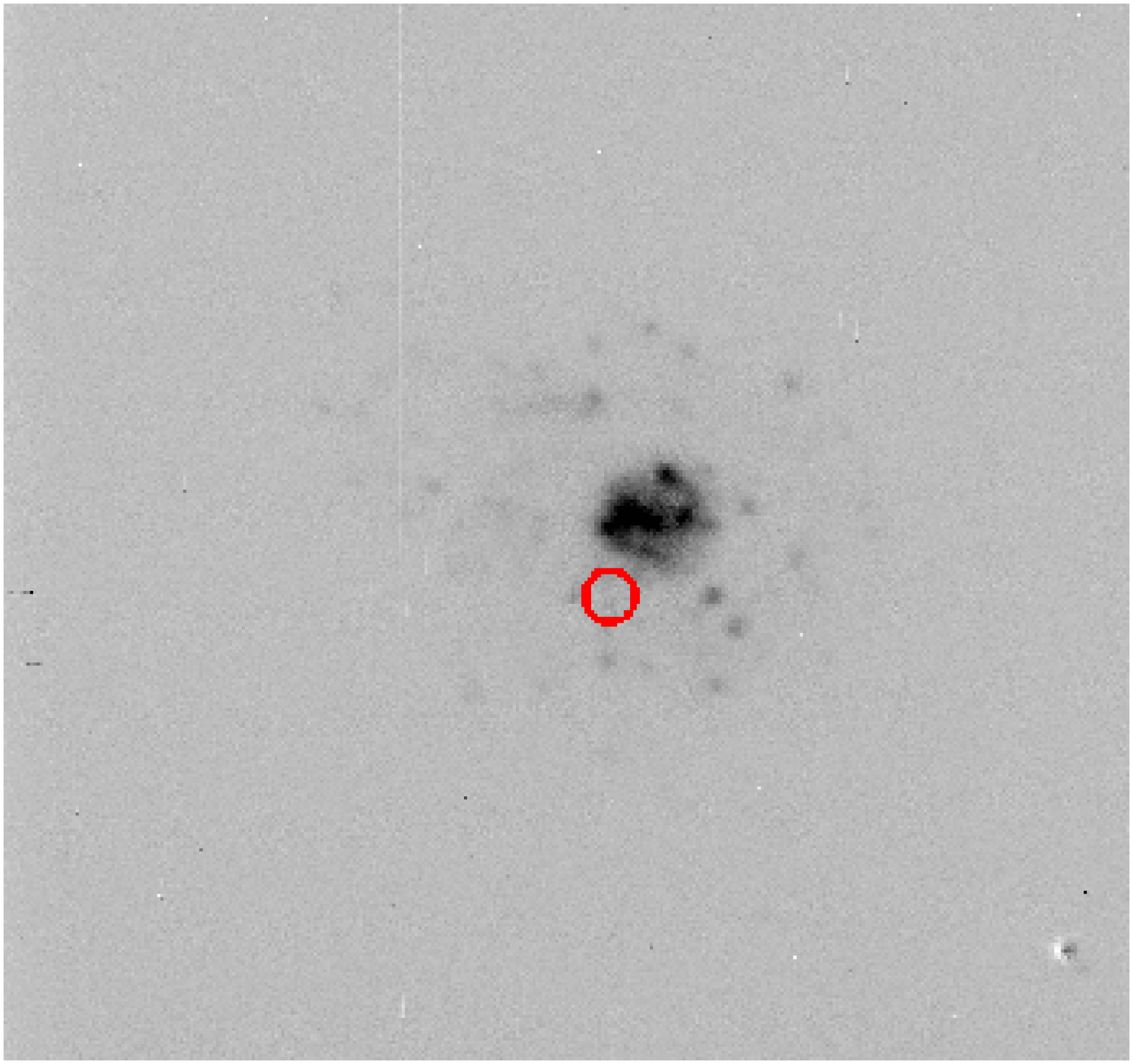}}\\   
     \subfigure[NGC~5371: SN1994Y]{\includegraphics[width=0.3\textwidth,height=0.15\textheight]{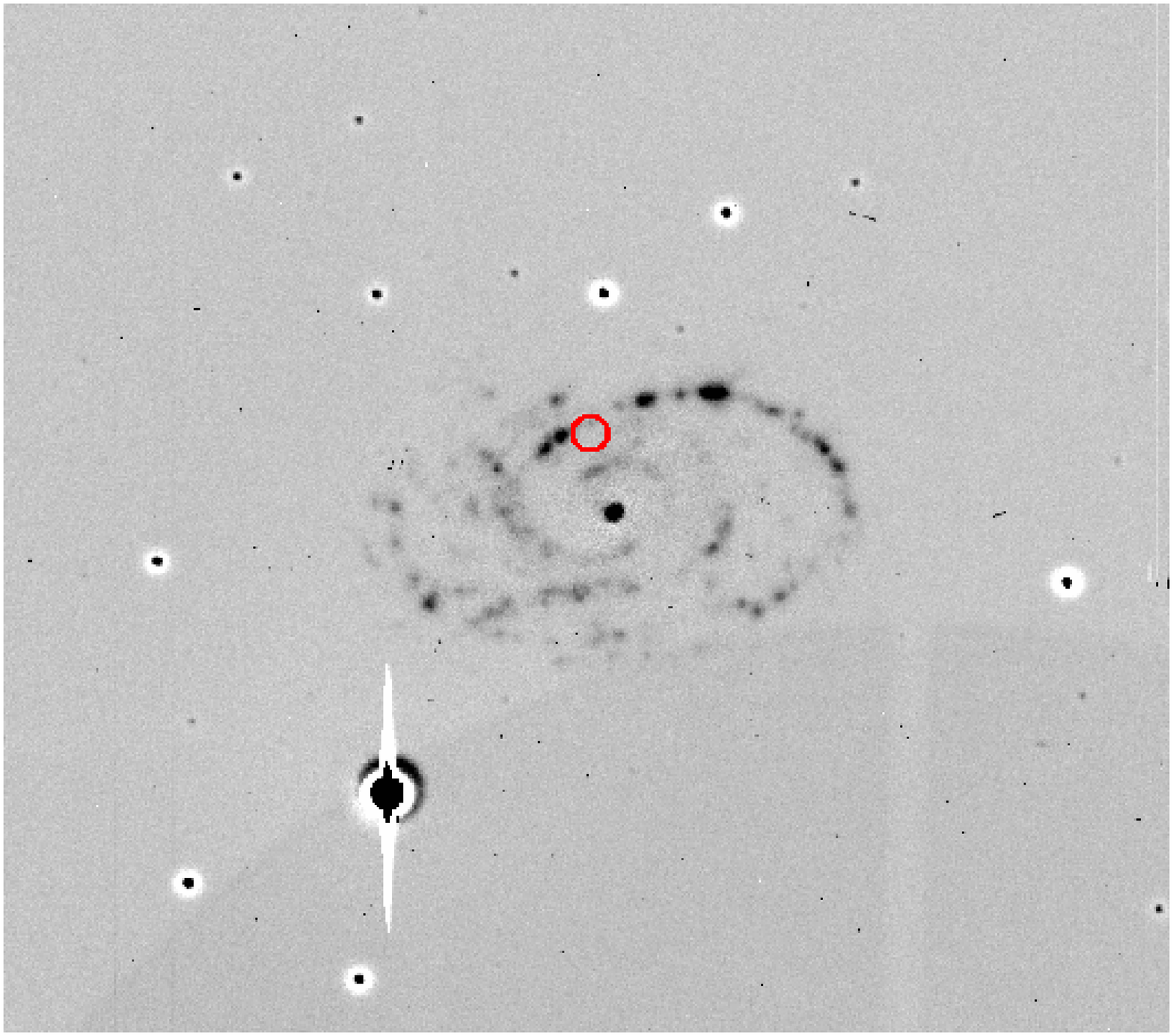}}   
     \subfigure[NGC~2782: SN1994ak]{\includegraphics[width=0.3\textwidth,height=0.15\textheight]{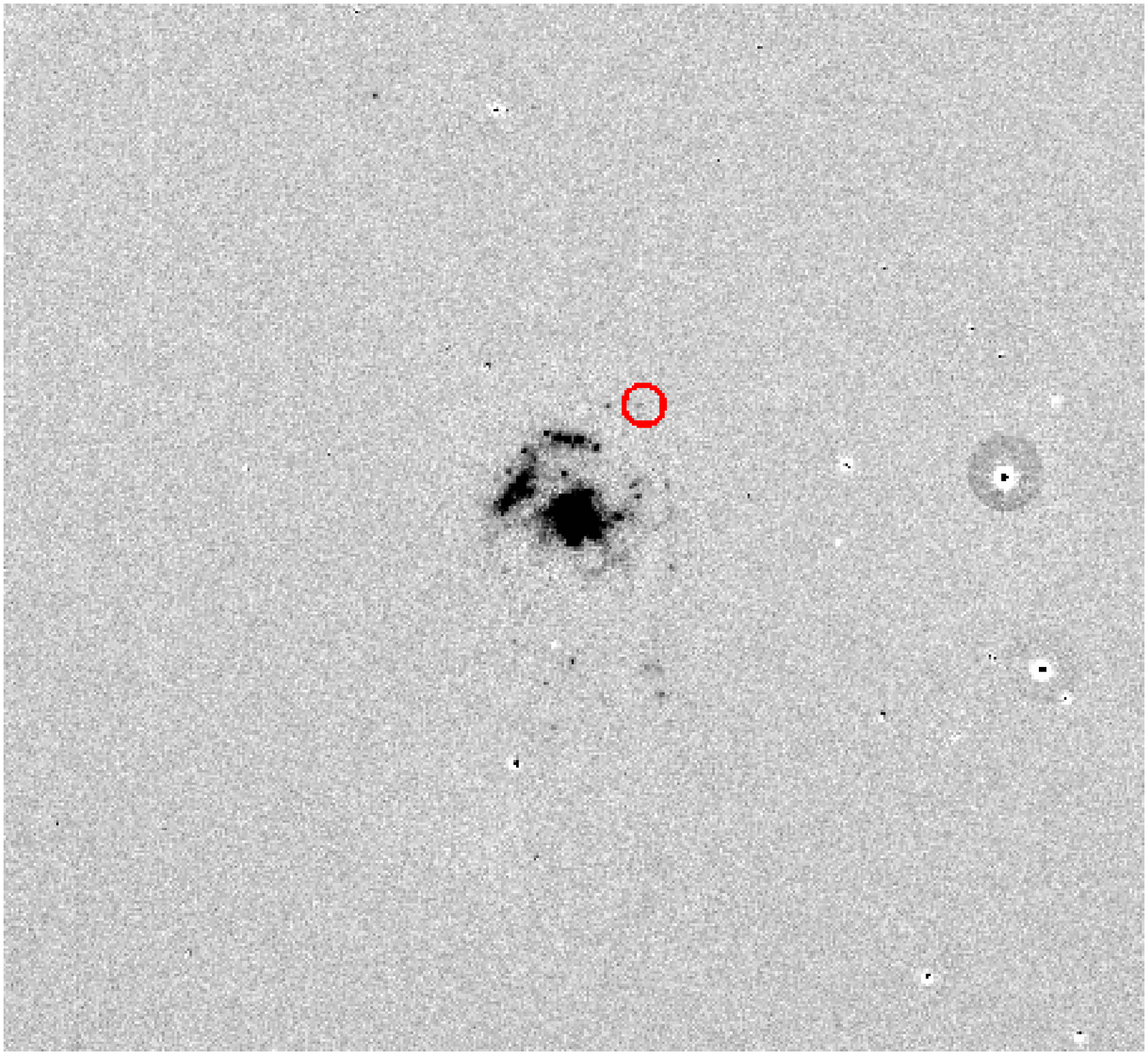}}   
     \subfigure[MCG-02-38-17: SN1995N]{\includegraphics[width=0.3\textwidth,height=0.15\textheight]{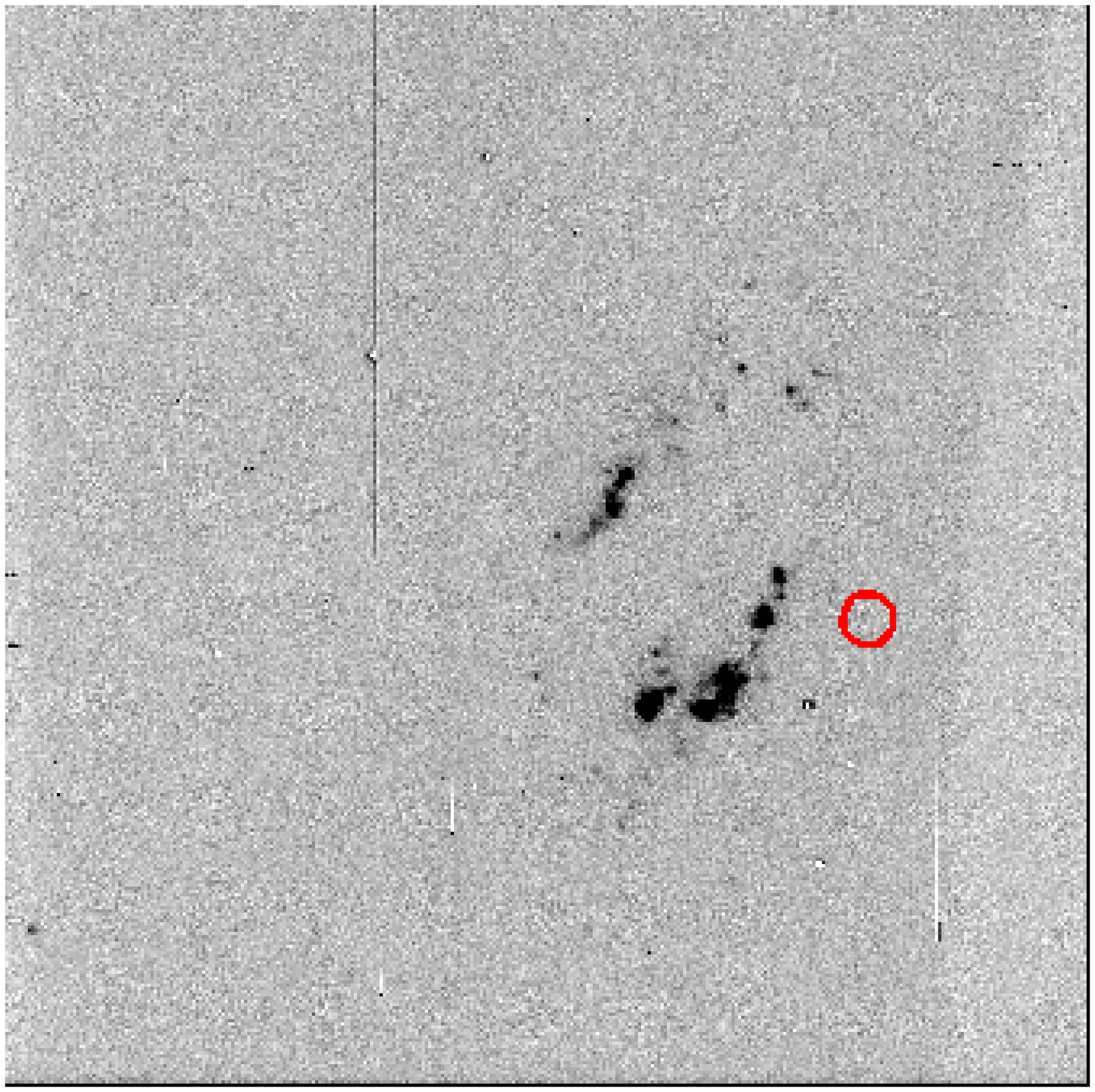}}   
     \subfigure[NGC~3631: SN1996bu]{\includegraphics[width=0.3\textwidth,height=0.15\textheight]{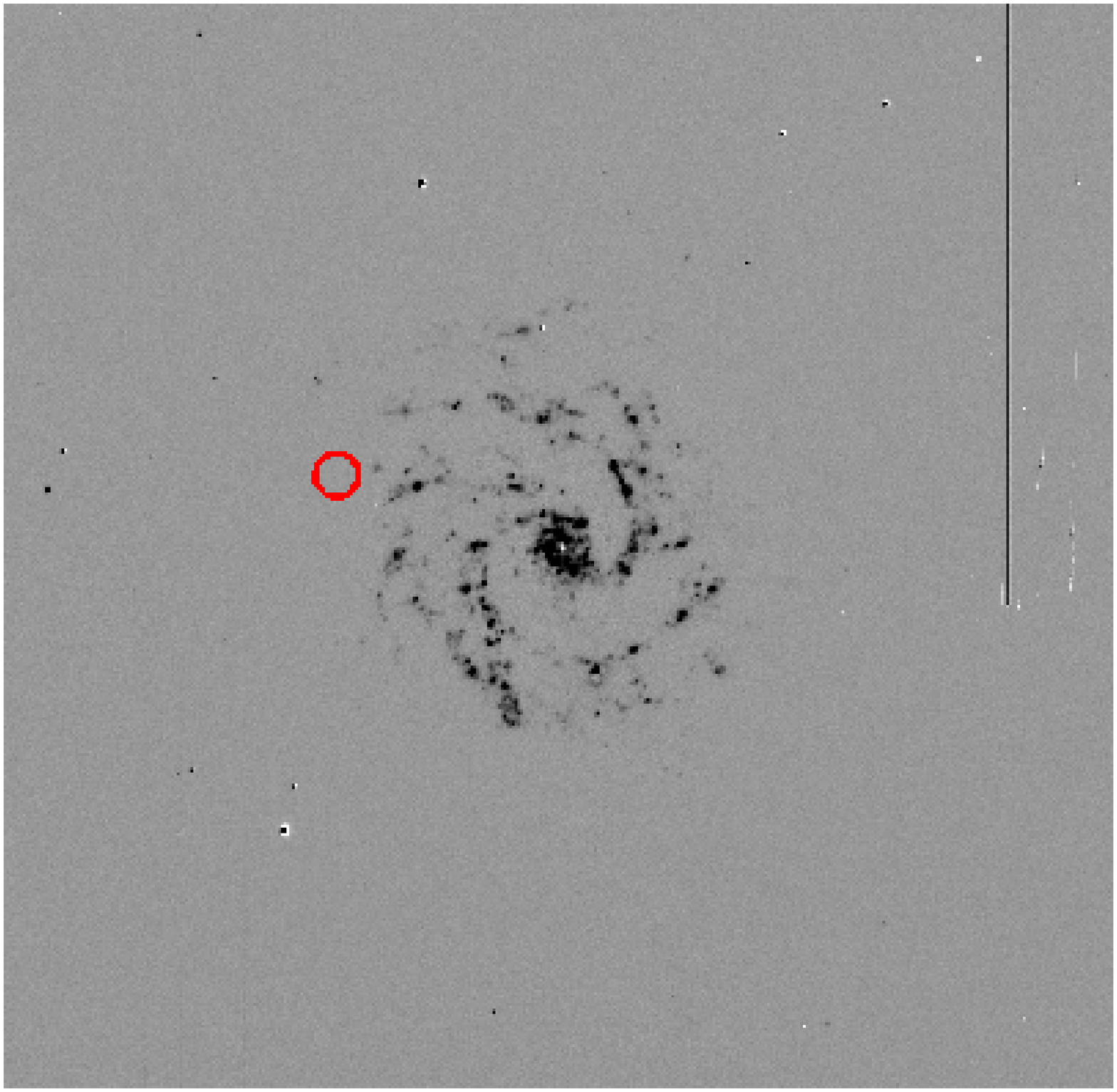}}   
     \subfigure[ESO97-G13: SN1996cr]{\includegraphics[width=0.3\textwidth,height=0.15\textheight]{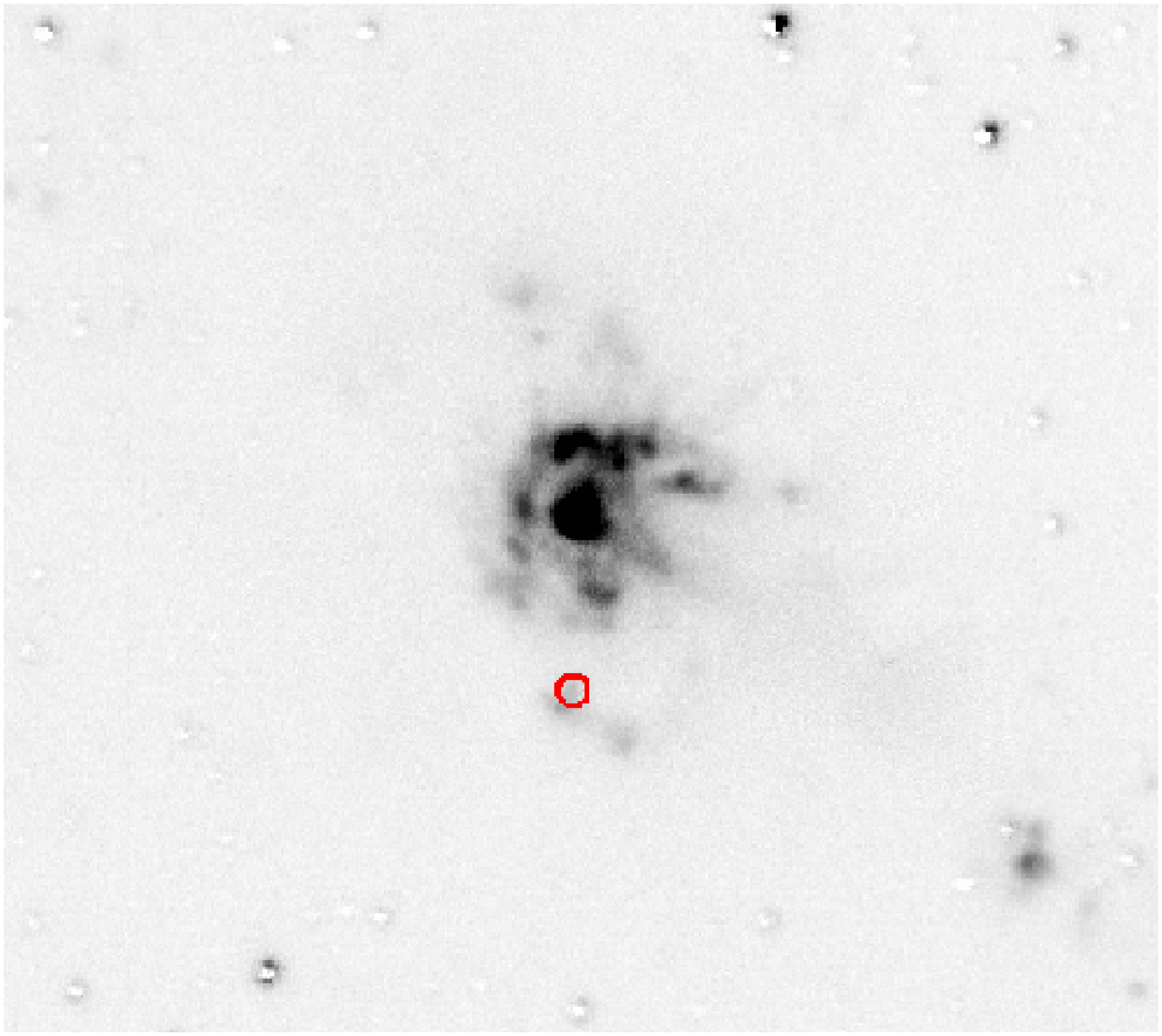}}   
     \subfigure[NGC~3627: SN1997bs]{\includegraphics[width=0.3\textwidth,height=0.15\textheight]{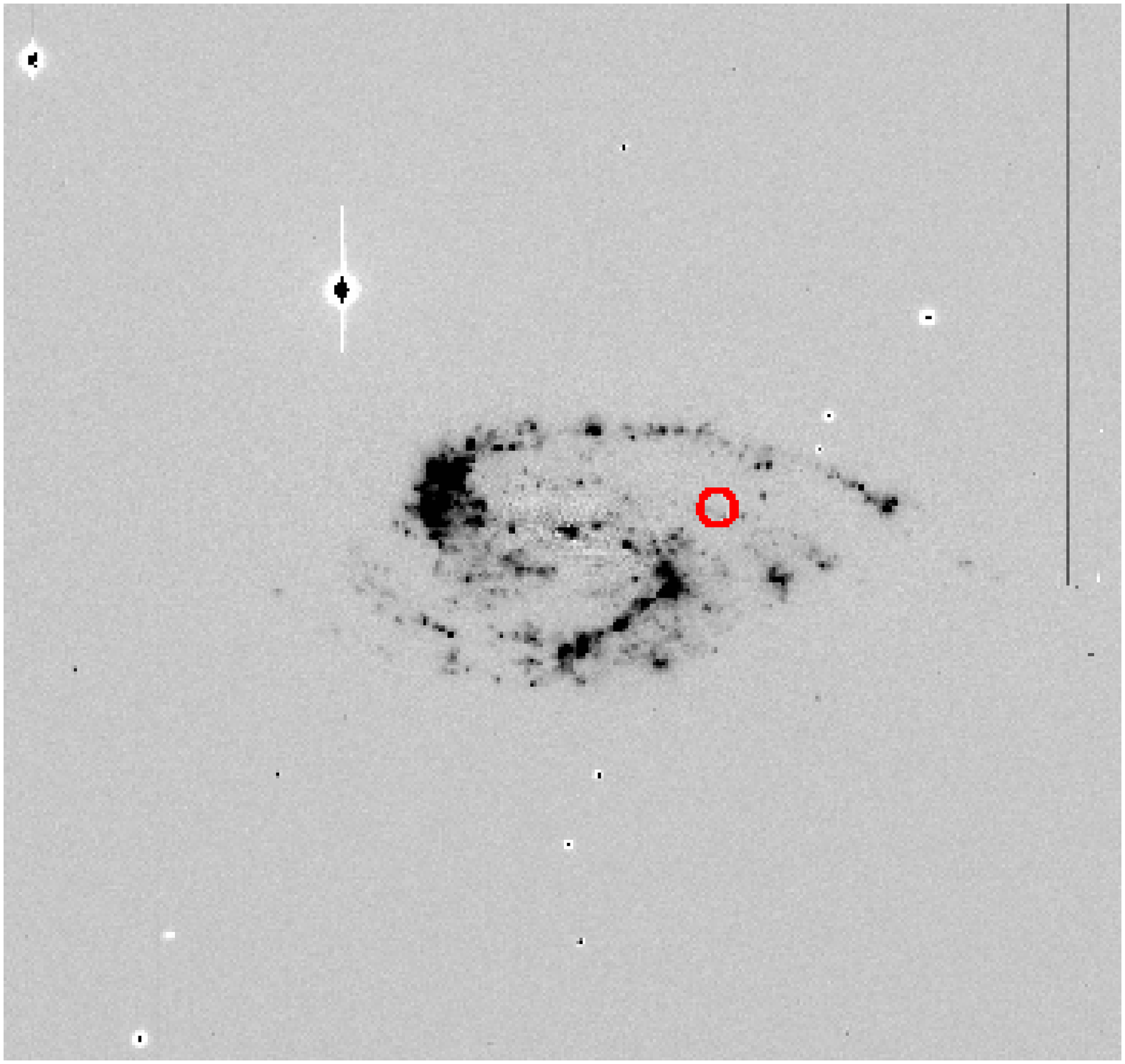}}   
     \subfigure[NGC~5012: SN1997eg]{\includegraphics[width=0.3\textwidth,height=0.15\textheight]{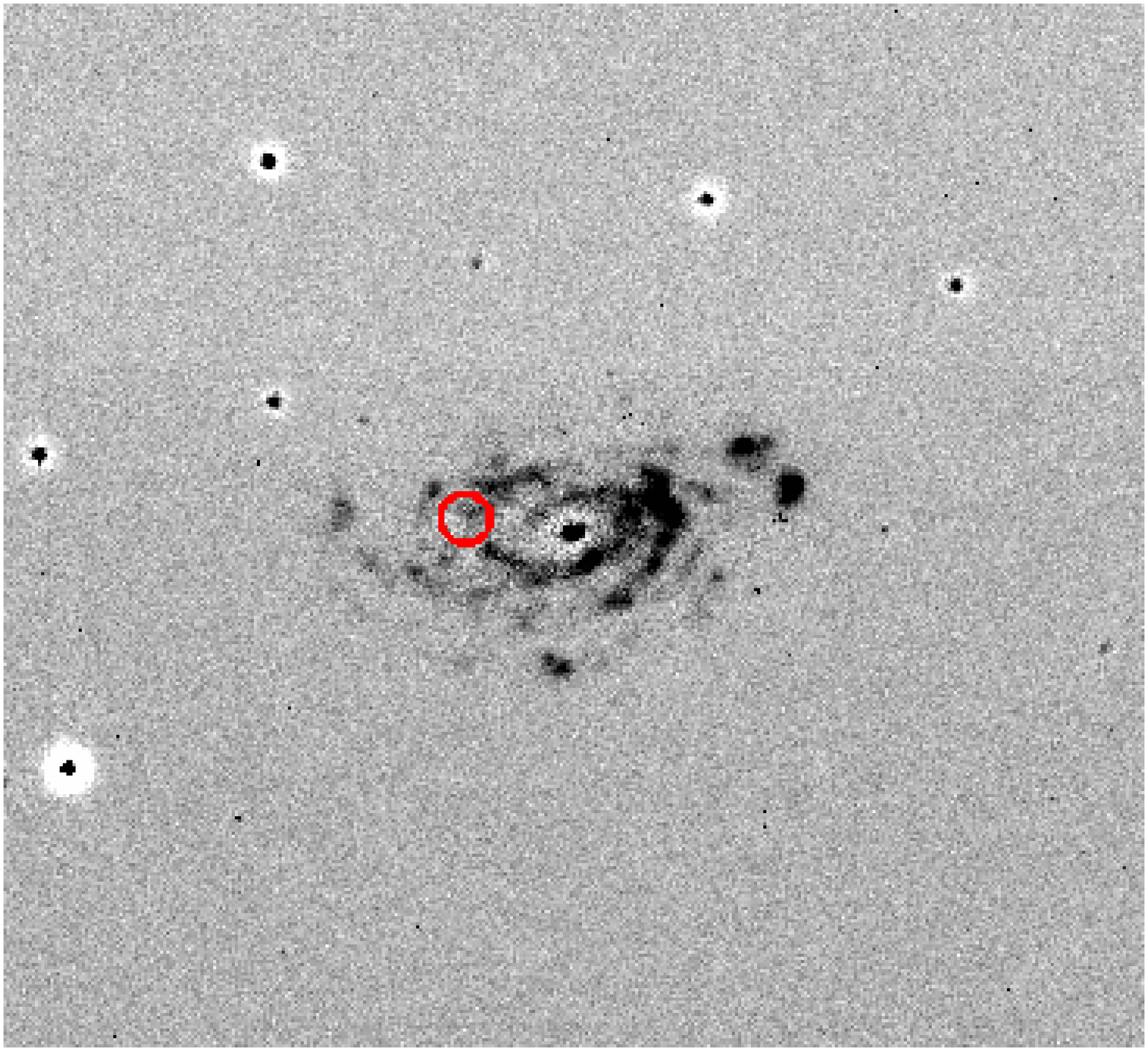}}   
     \subfigure[NGC~3198: SN1999bw]{\includegraphics[width=0.3\textwidth,height=0.15\textheight]{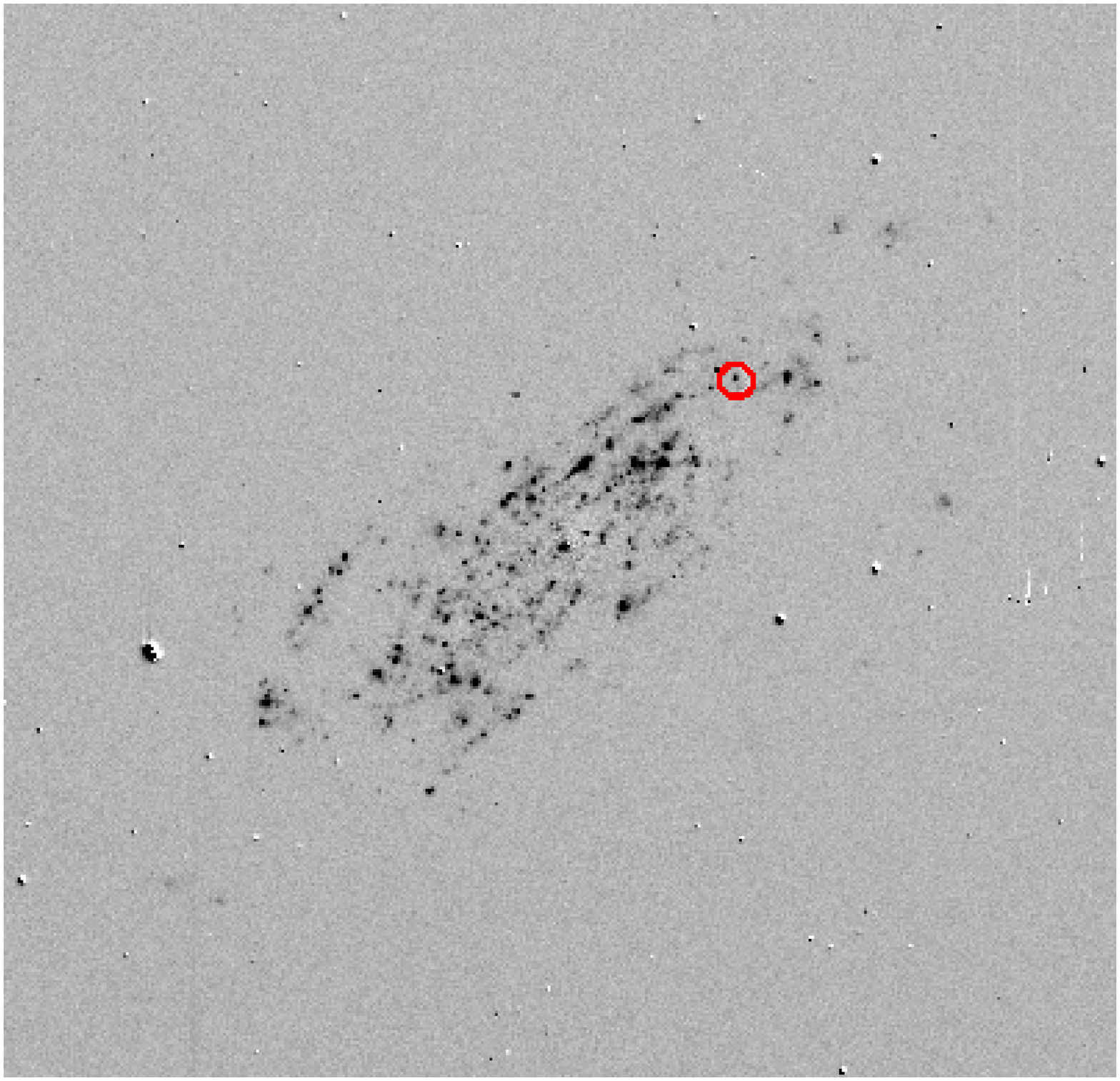}}   
     \subfigure[NGC~6951: SN1999el]{\includegraphics[width=0.3\textwidth,height=0.15\textheight]{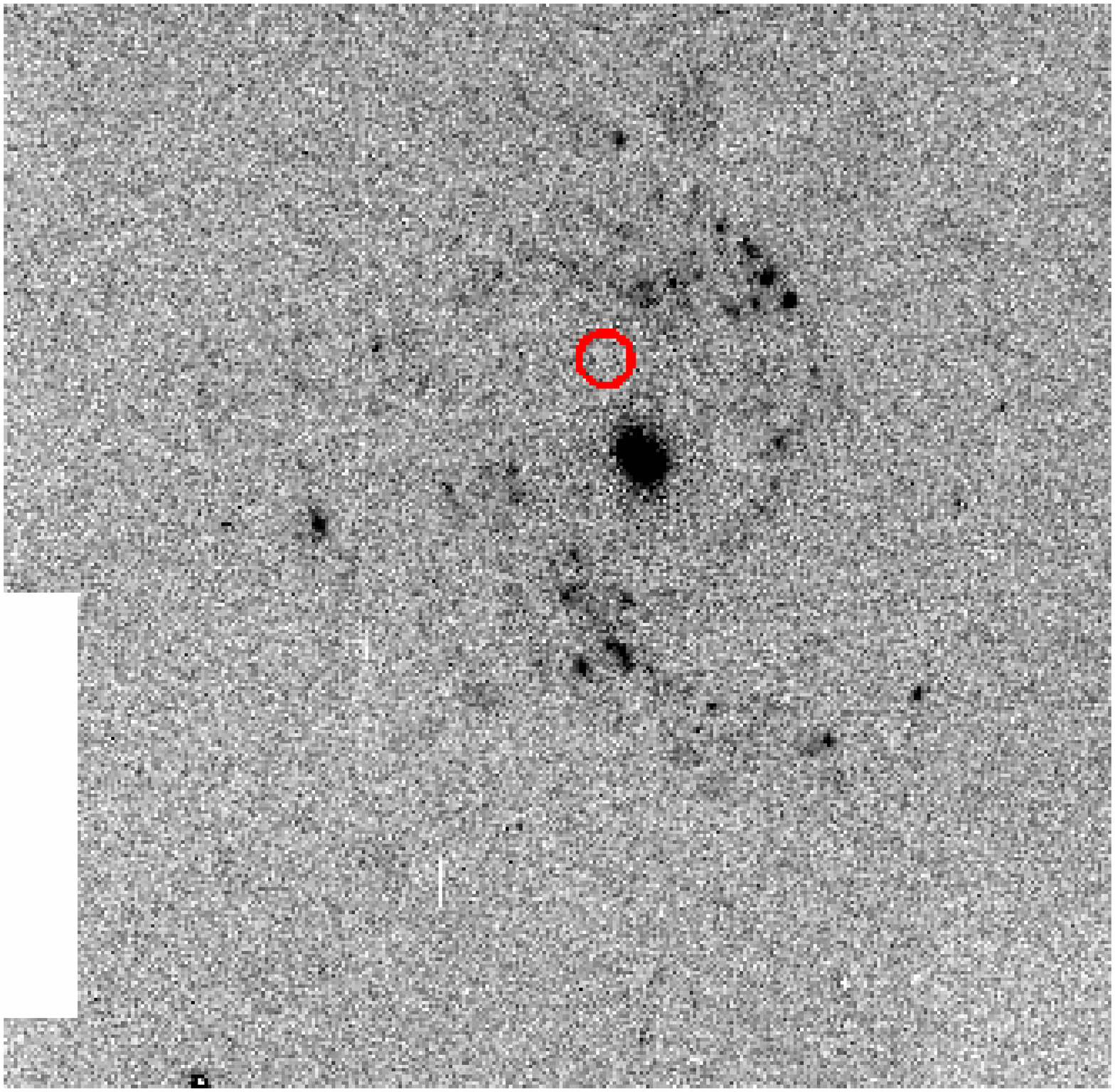}} \\  
    \label{fig:1}
     \end{center}
  \end{minipage}
\end{figure*}

\begin{figure*}
  \begin{minipage}{160mm}
    \begin{center}
      \contcaption{H$\alpha$ observations of a sample of the SNIIn and Impostor host galaxies, with the SN positions marked with red circles.}
      \renewcommand{\thesubfigure}{\thefigure.\arabic{subfigure}}
      \subfigure[NGC~2532: SN1999gb]{\includegraphics[width=0.3\textwidth,height=0.15\textheight]{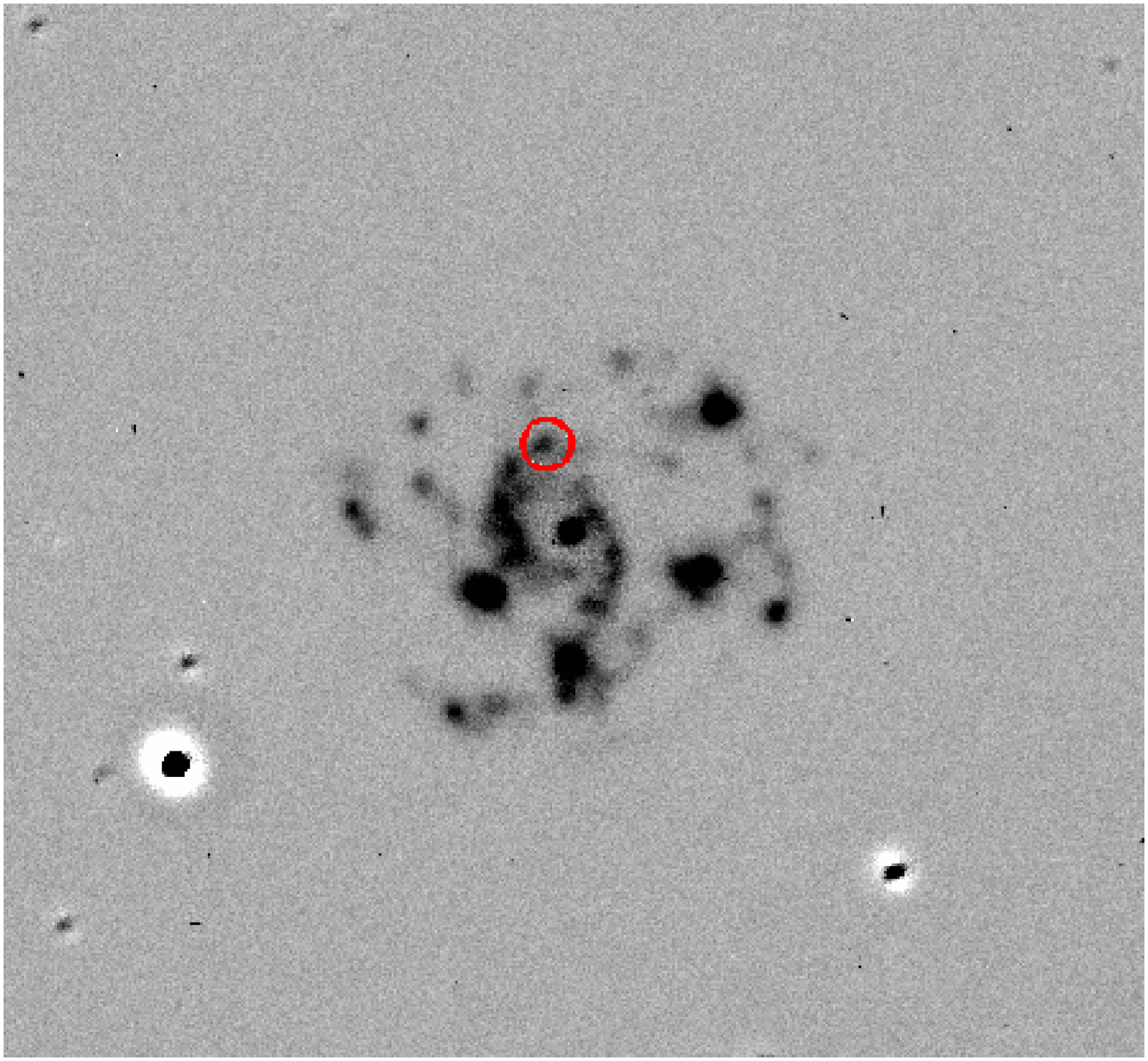}}   
     \subfigure[NGC~4965: SN2000P]{\includegraphics[width=0.3\textwidth,height=0.15\textheight]{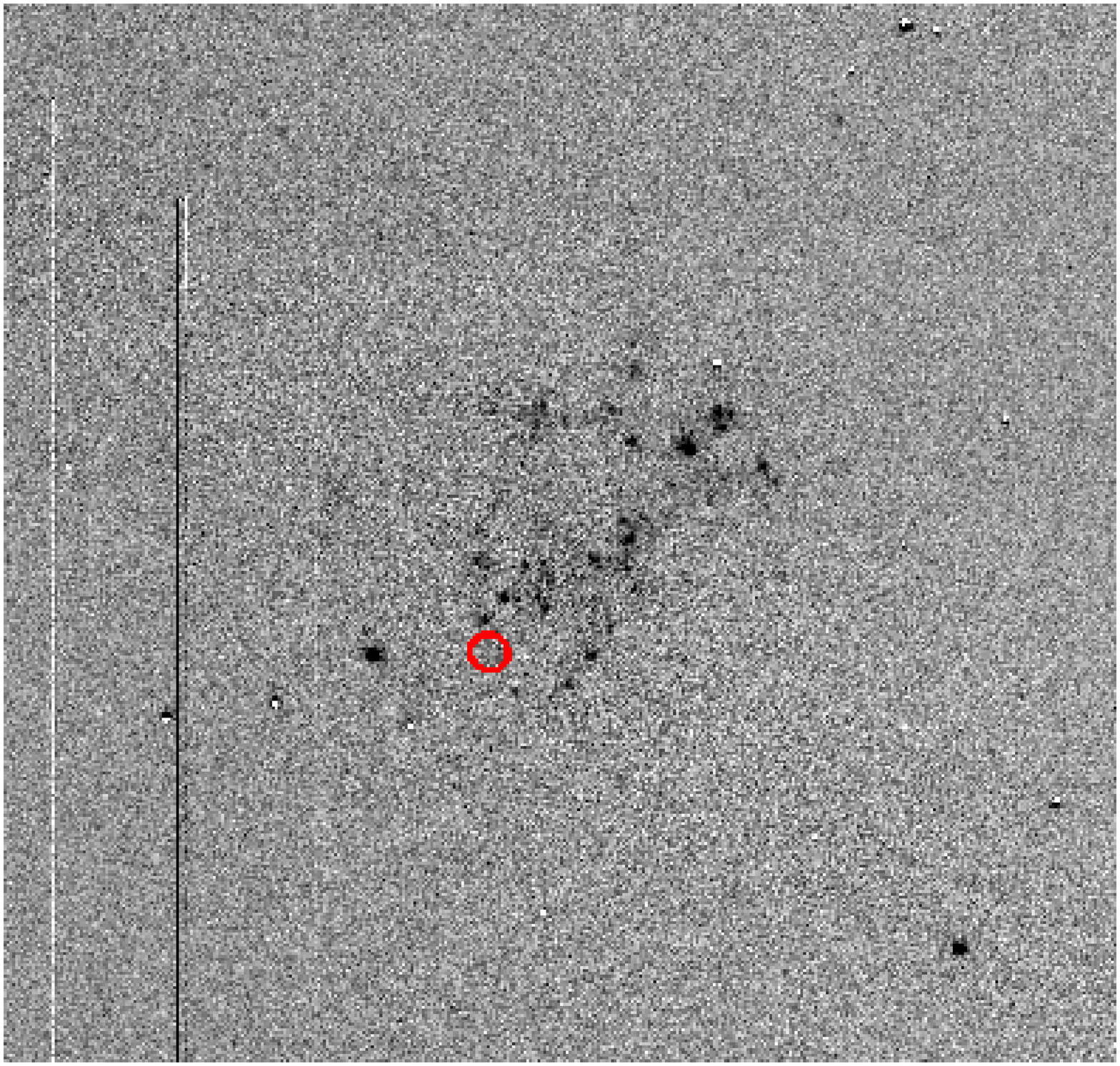}}   
     \subfigure[NGC~3318: SN2000cl]{\includegraphics[width=0.3\textwidth,height=0.15\textheight]{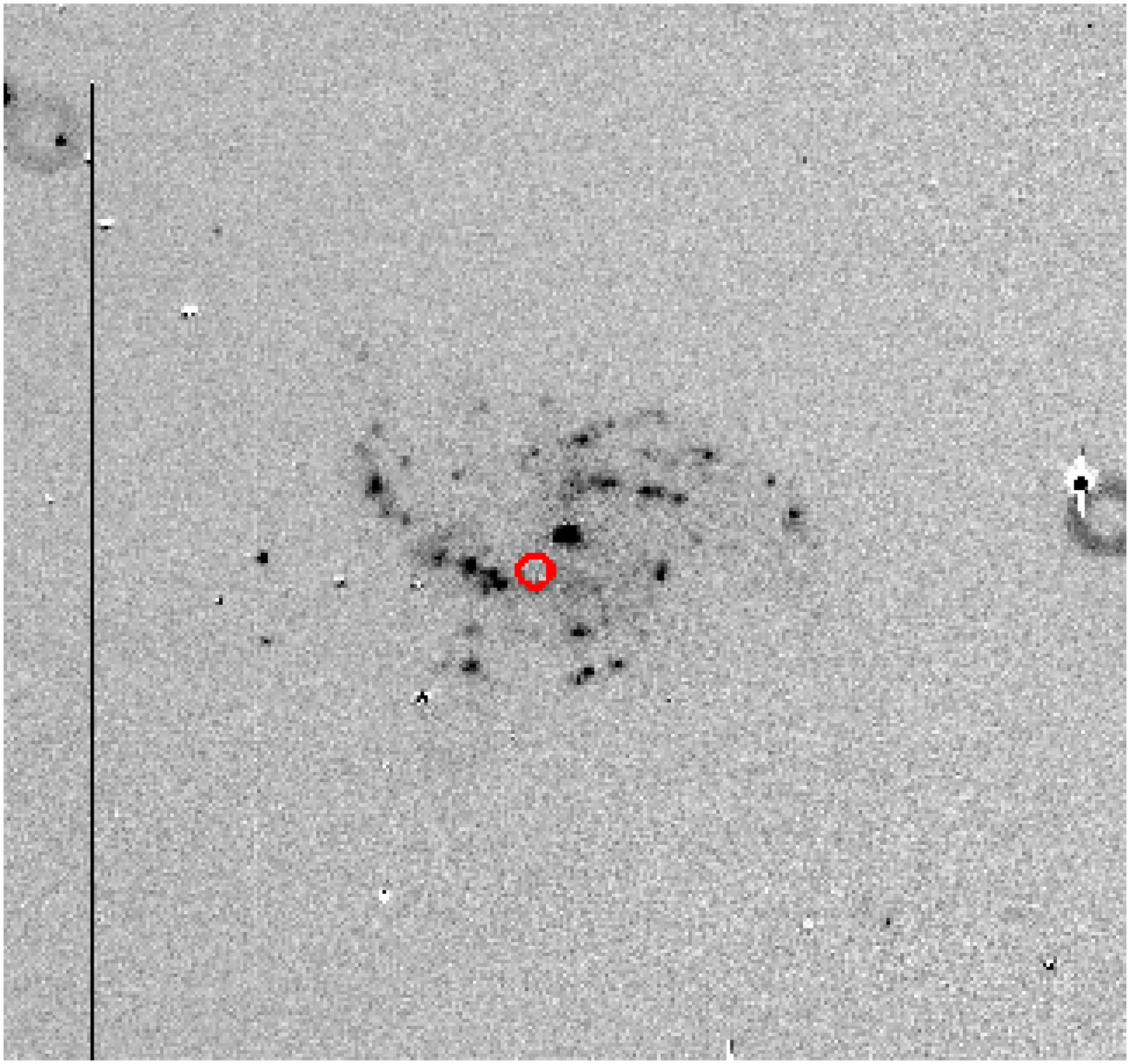}}   
     \subfigure[NGC~3504: SN2001ac]{\includegraphics[width=0.3\textwidth,height=0.15\textheight]{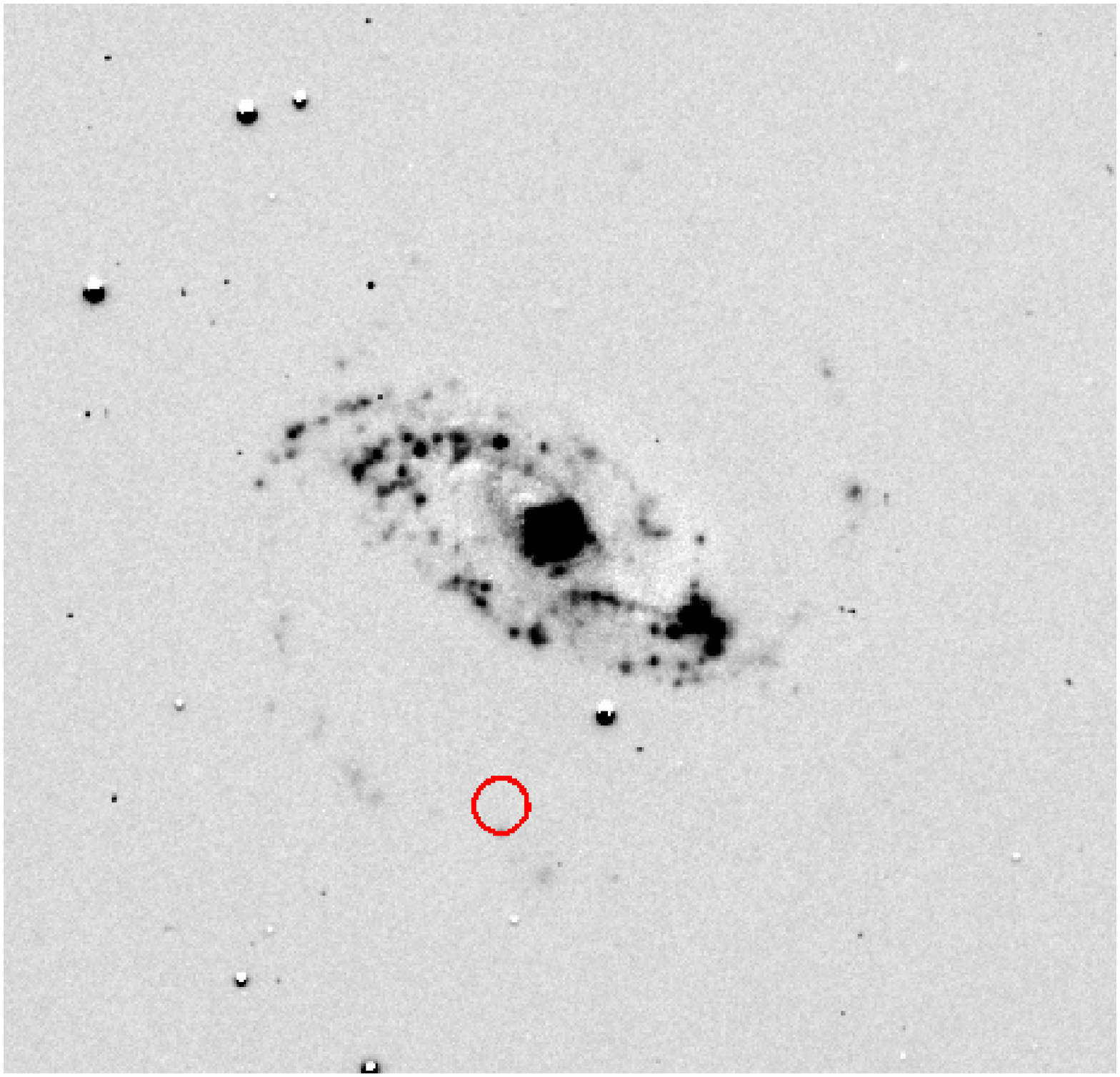}}   
     \subfigure[NGC~673: SN2001fa]{\includegraphics[width=0.3\textwidth,height=0.15\textheight]{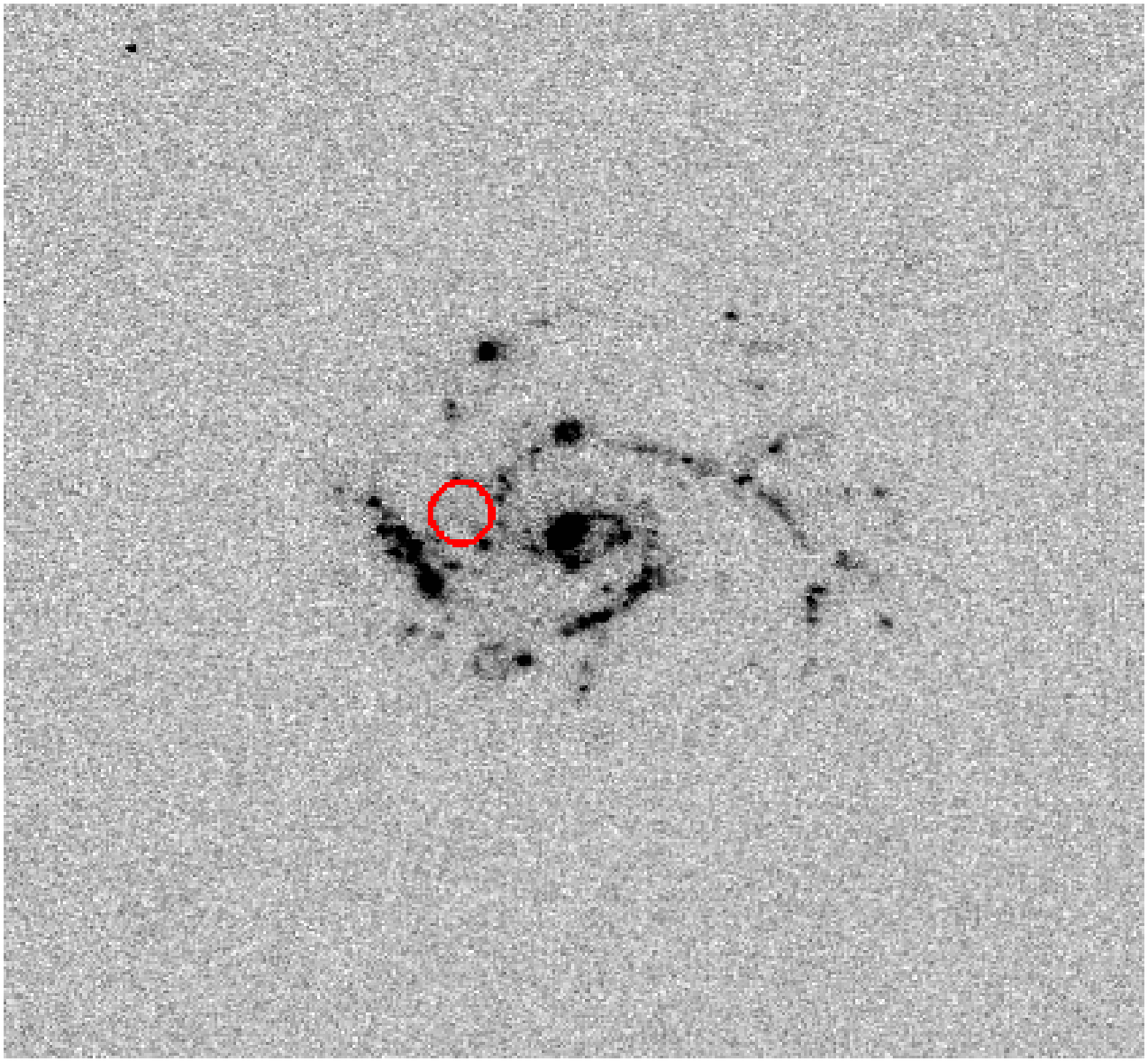}}   
     \subfigure[UGC~3804: SN2002A]{\includegraphics[width=0.3\textwidth,height=0.15\textheight]{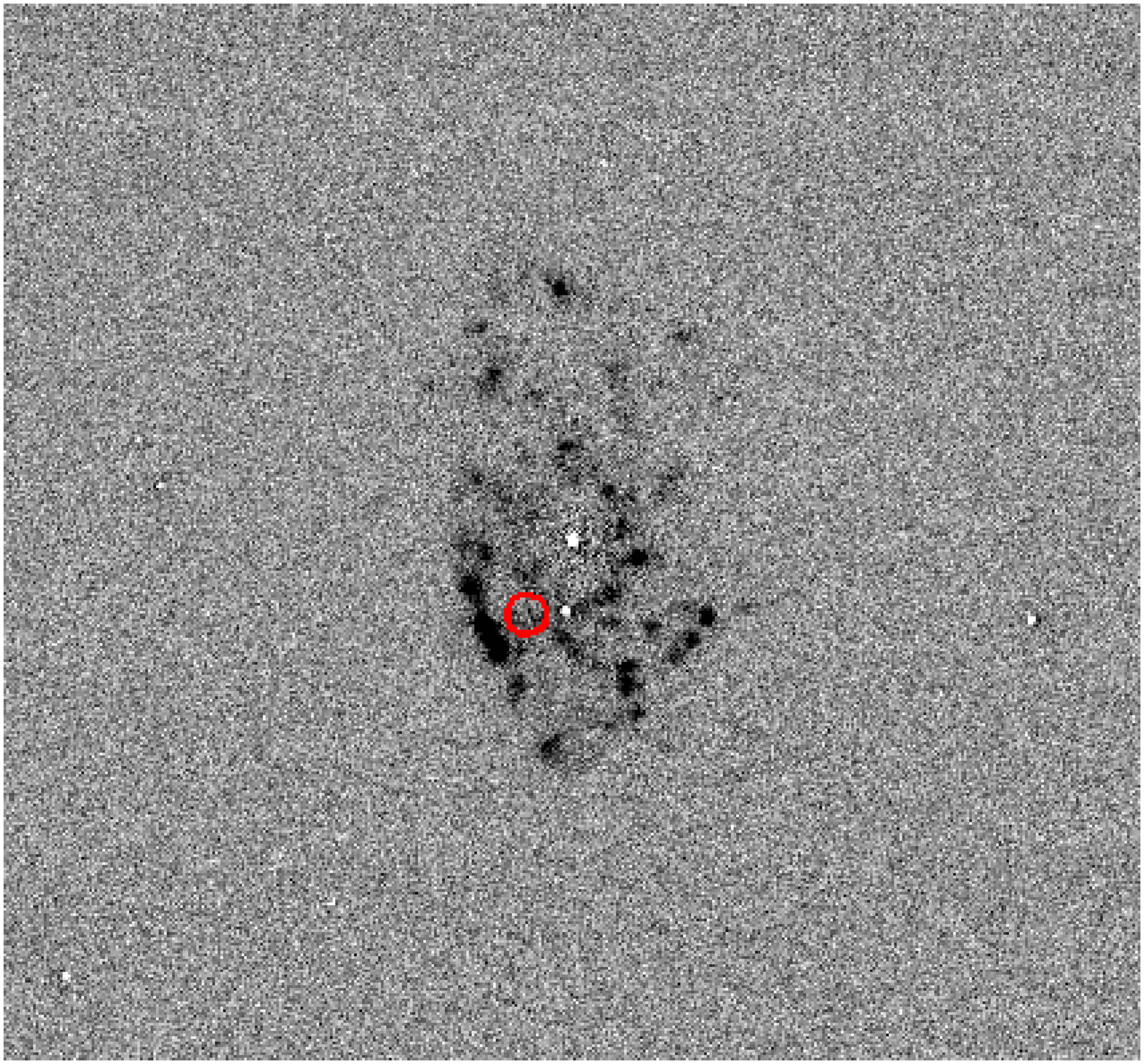}}   
     \subfigure[NGC~4242: SN2002bu]{\includegraphics[width=0.3\textwidth,height=0.15\textheight]{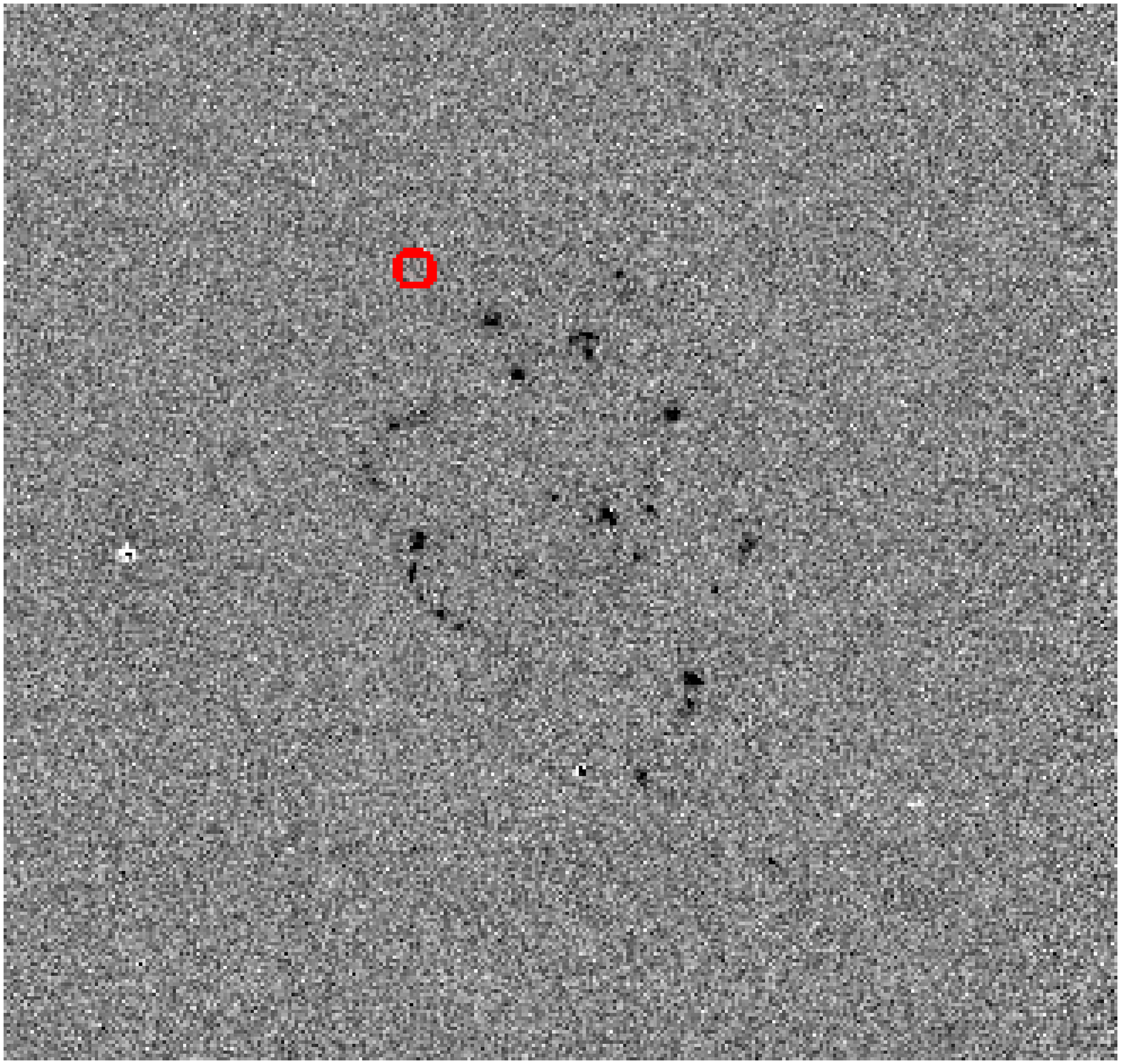}}   
     \subfigure[NGC~2642: SN2002fj]{\includegraphics[width=0.3\textwidth,height=0.15\textheight]{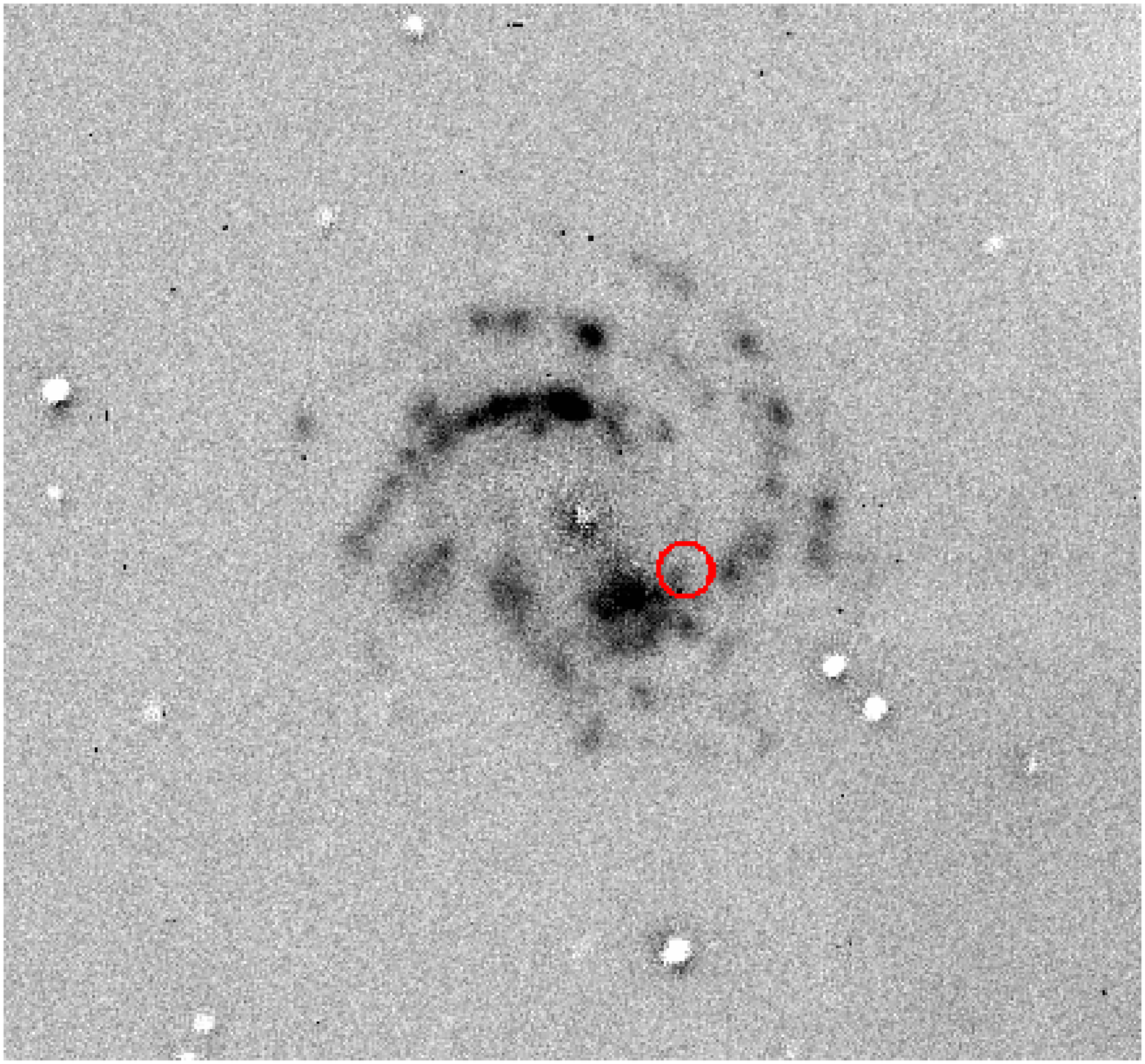}}   
     \subfigure[IC~208: SN2003G]{\includegraphics[width=0.3\textwidth,height=0.15\textheight]{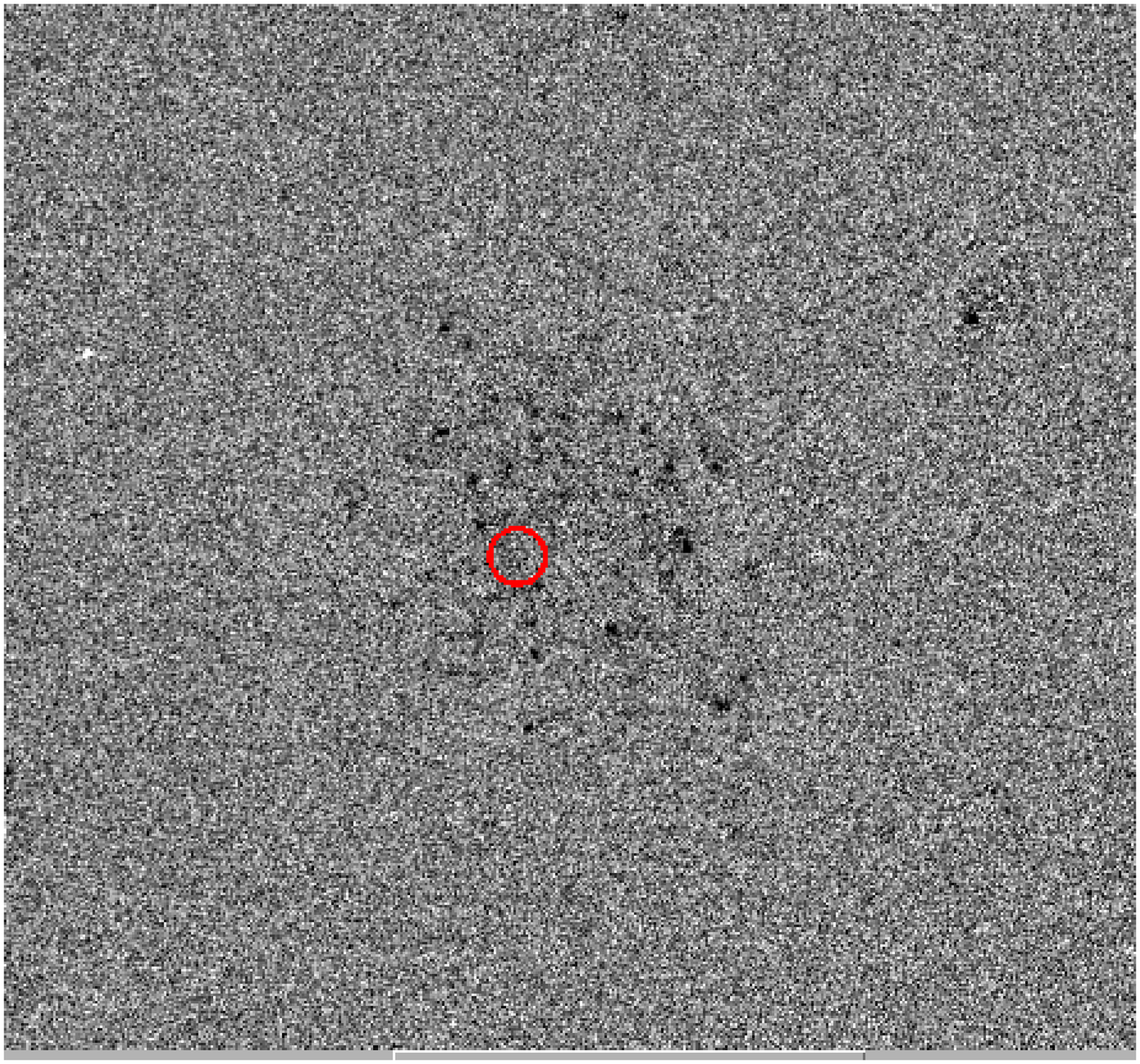}}   
     \subfigure[UGC~9638: SN2003dv]{\includegraphics[width=0.3\textwidth,height=0.15\textheight]{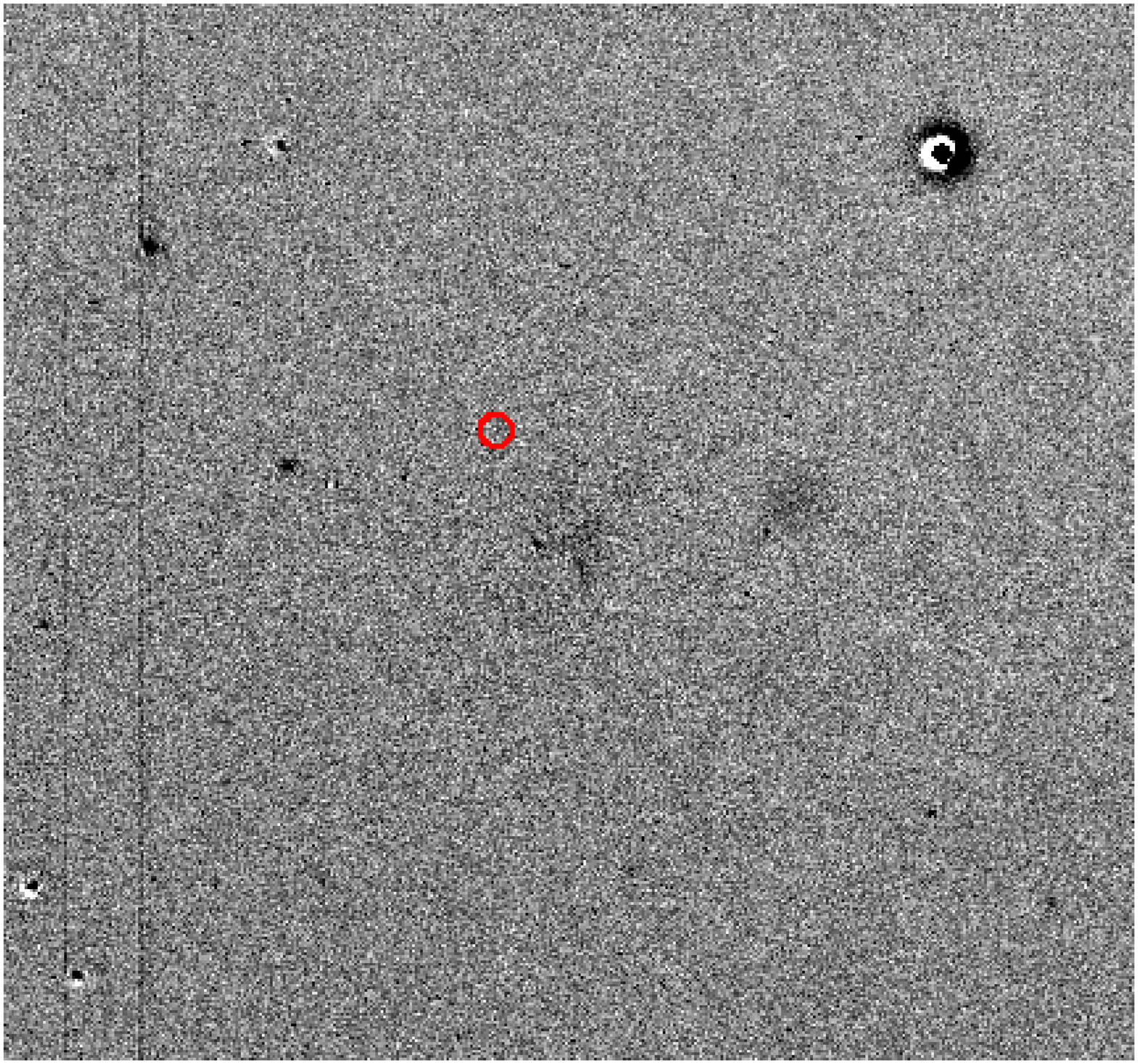}}   
     \subfigure[NGC~5334: SN2003gm]{\includegraphics[width=0.3\textwidth,height=0.15\textheight]{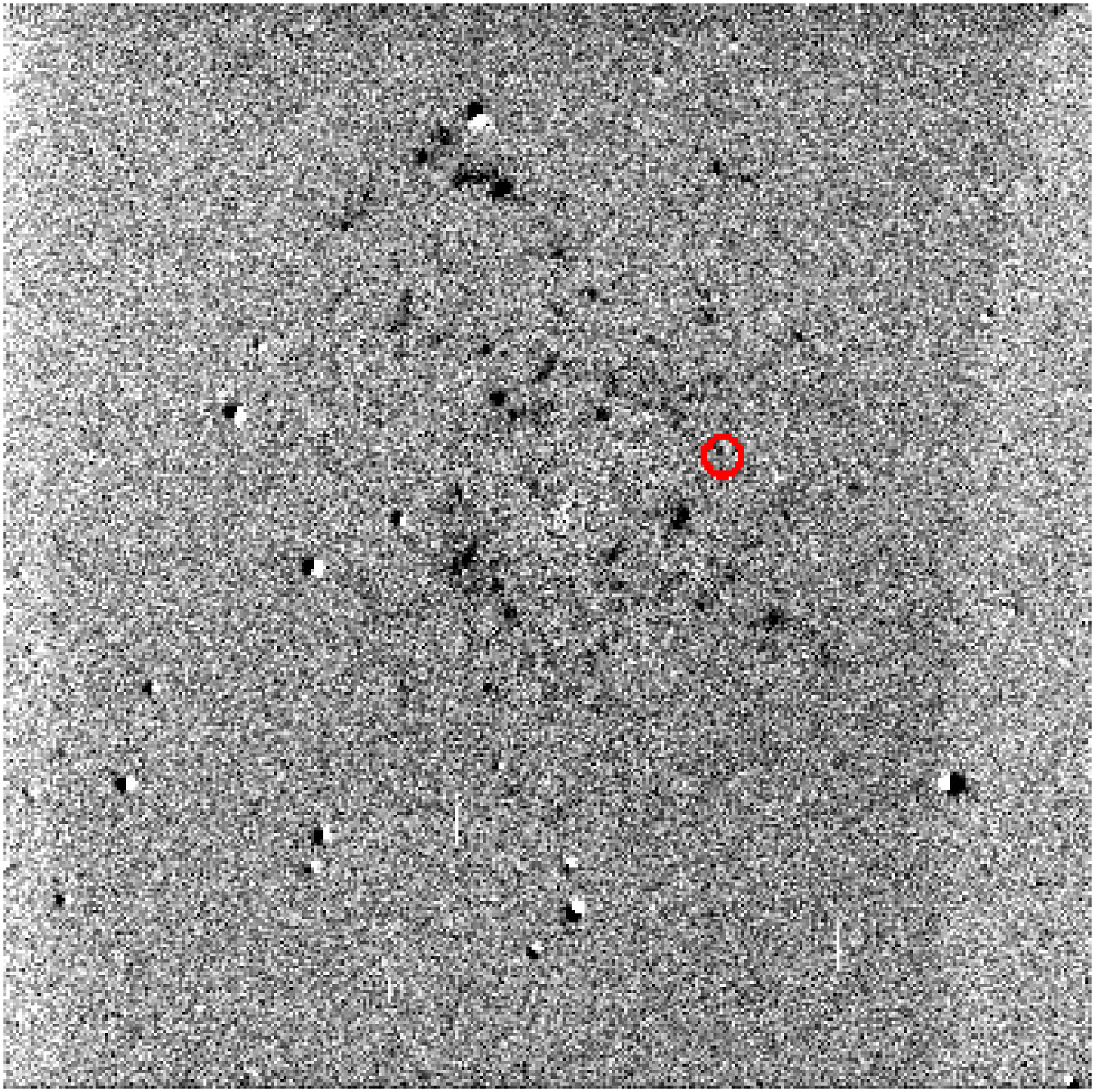}}   
     \subfigure[NGC~1376: SN2003lo]{\includegraphics[width=0.3\textwidth,height=0.15\textheight]{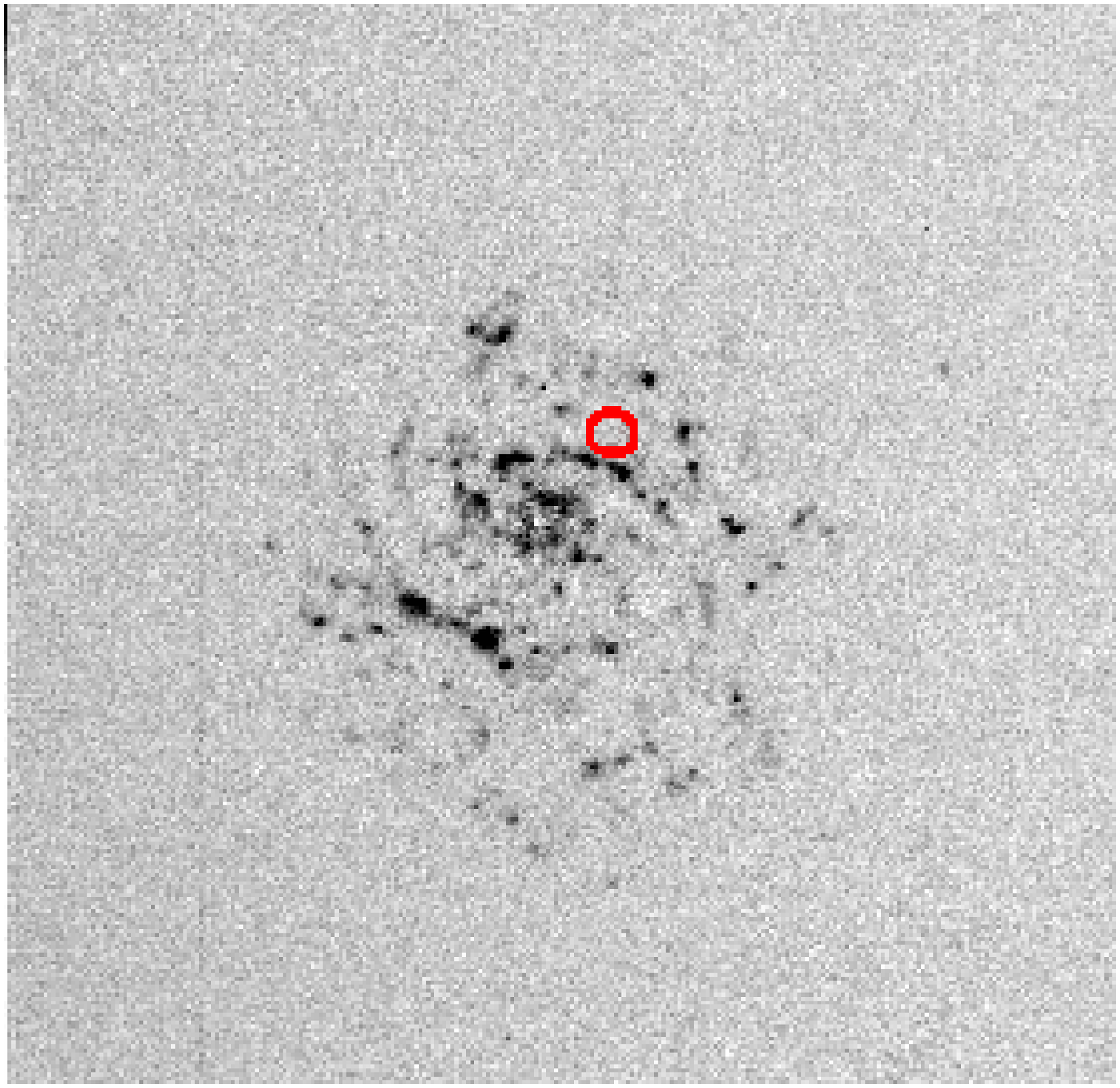}}   
     \subfigure[NGC~214: SN2005db]{\includegraphics[width=0.3\textwidth,height=0.15\textheight]{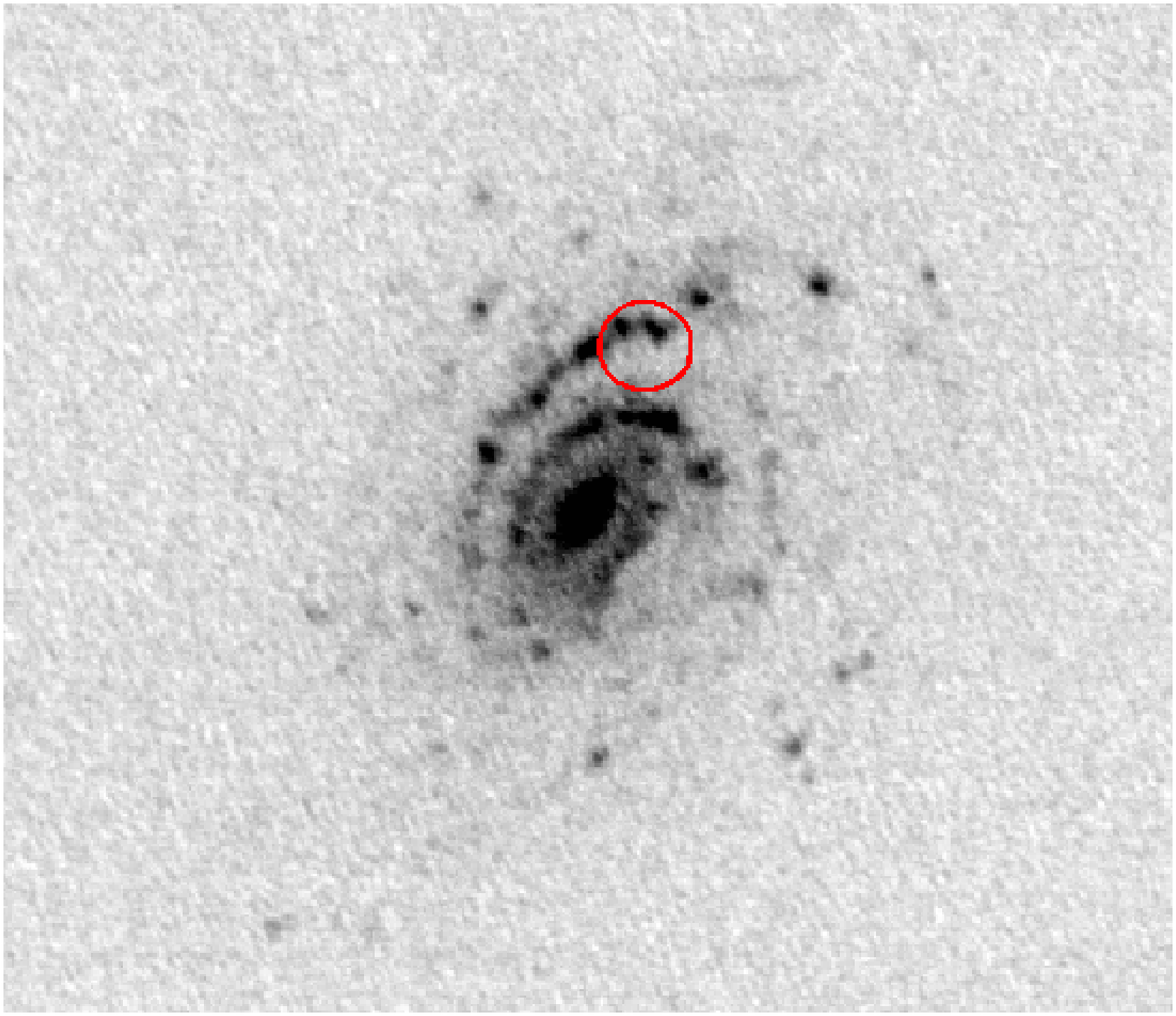}}   
     \subfigure[NGC~266: SN2005gl]{\includegraphics[width=0.3\textwidth,height=0.15\textheight]{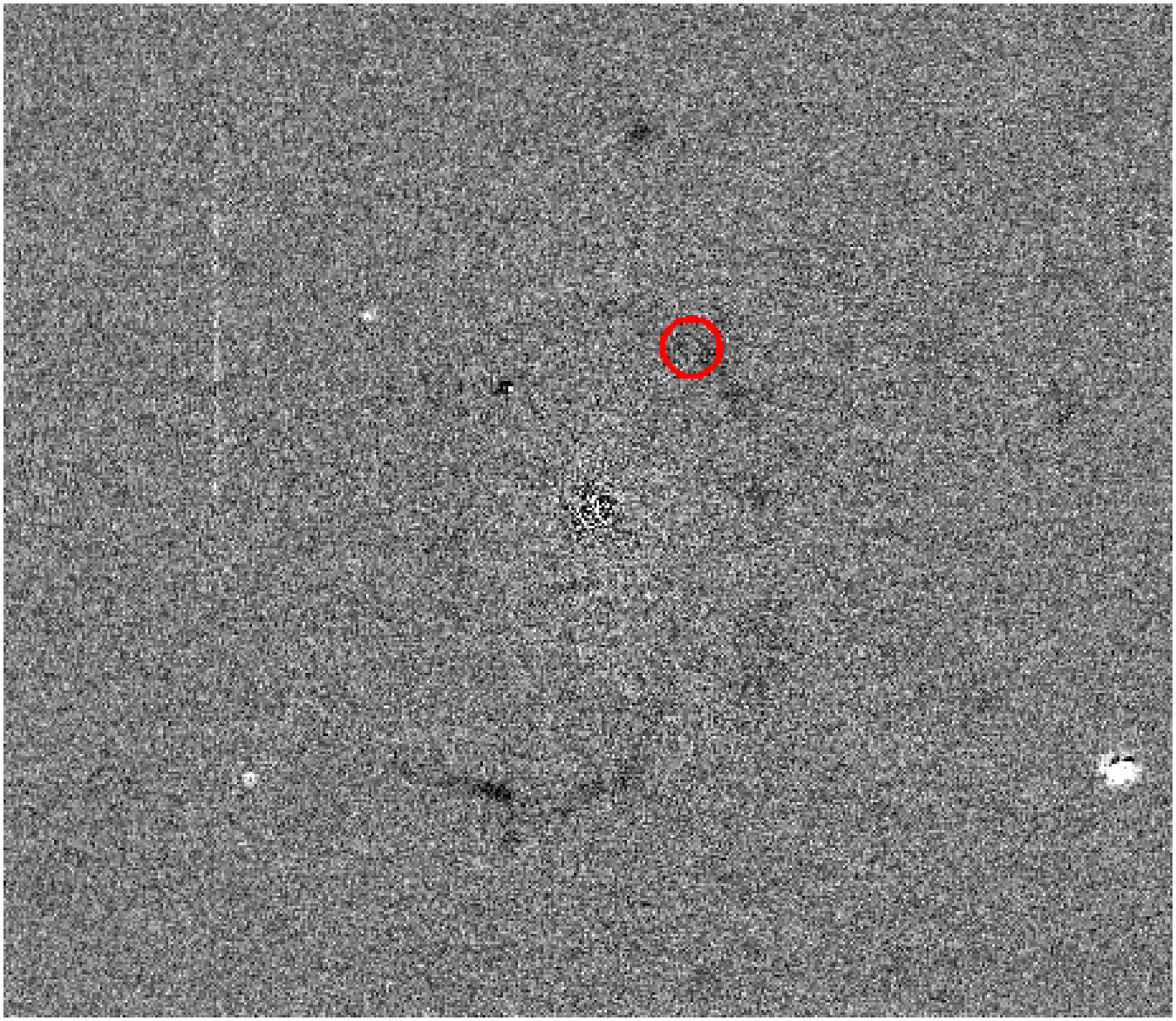}}  
     \subfigure[NGC~2906: SN2005ip]{\includegraphics[width=0.3\textwidth,height=0.15\textheight]{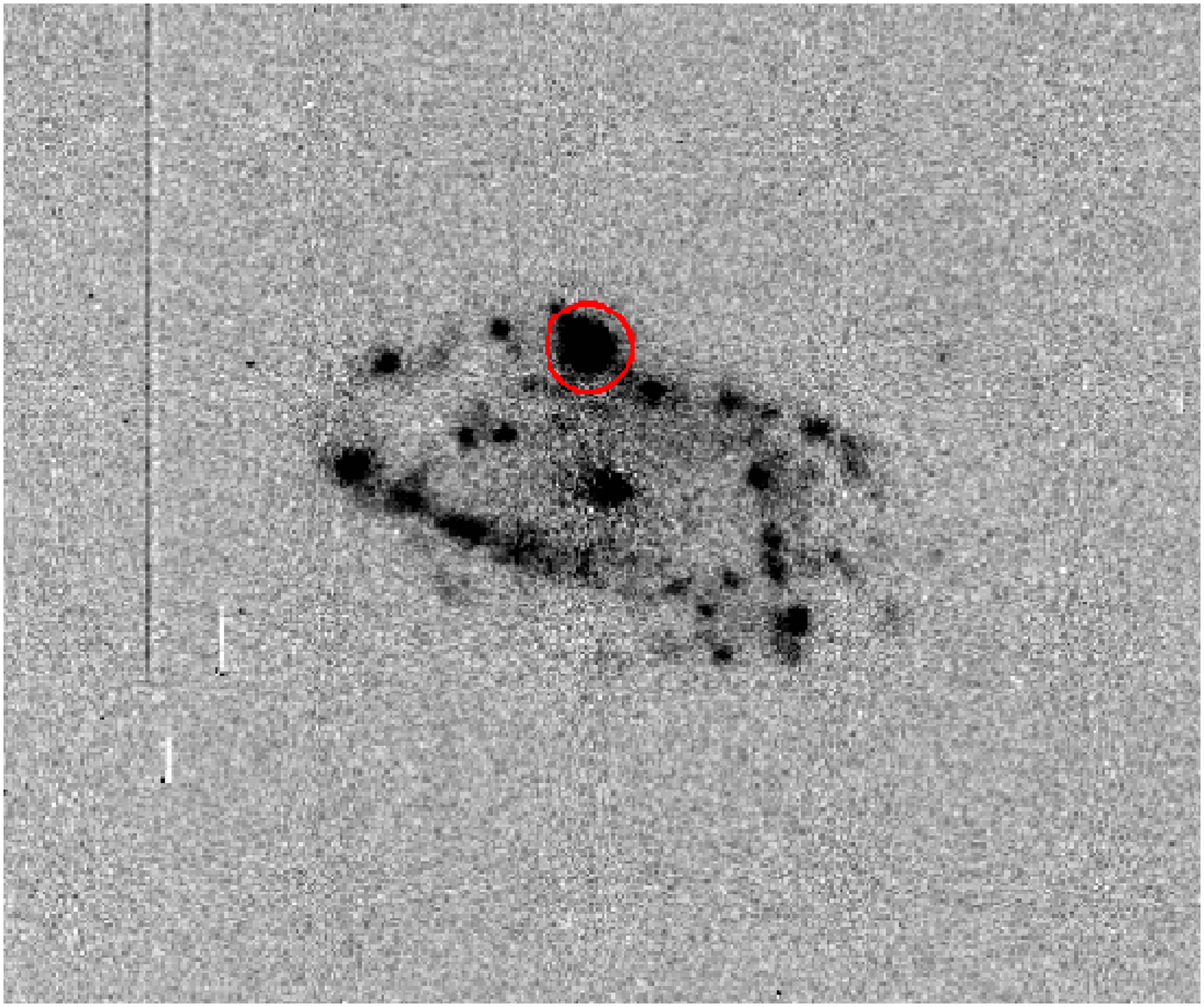}}  \\
    \end{center}
  \end{minipage}
\end{figure*}

\begin{figure*}
  \begin{minipage}{160mm}
    \begin{center}
     \contcaption{H$\alpha$ observations of a sample of the SNIIn and Impostor host galaxies, with the SN positions marked with red circles.}
    \renewcommand{\thesubfigure}{\thefigure.\arabic{subfigure}}
     \subfigure[NGC~5630: SN2006am]{\includegraphics[width=0.3\textwidth,height=0.15\textheight]{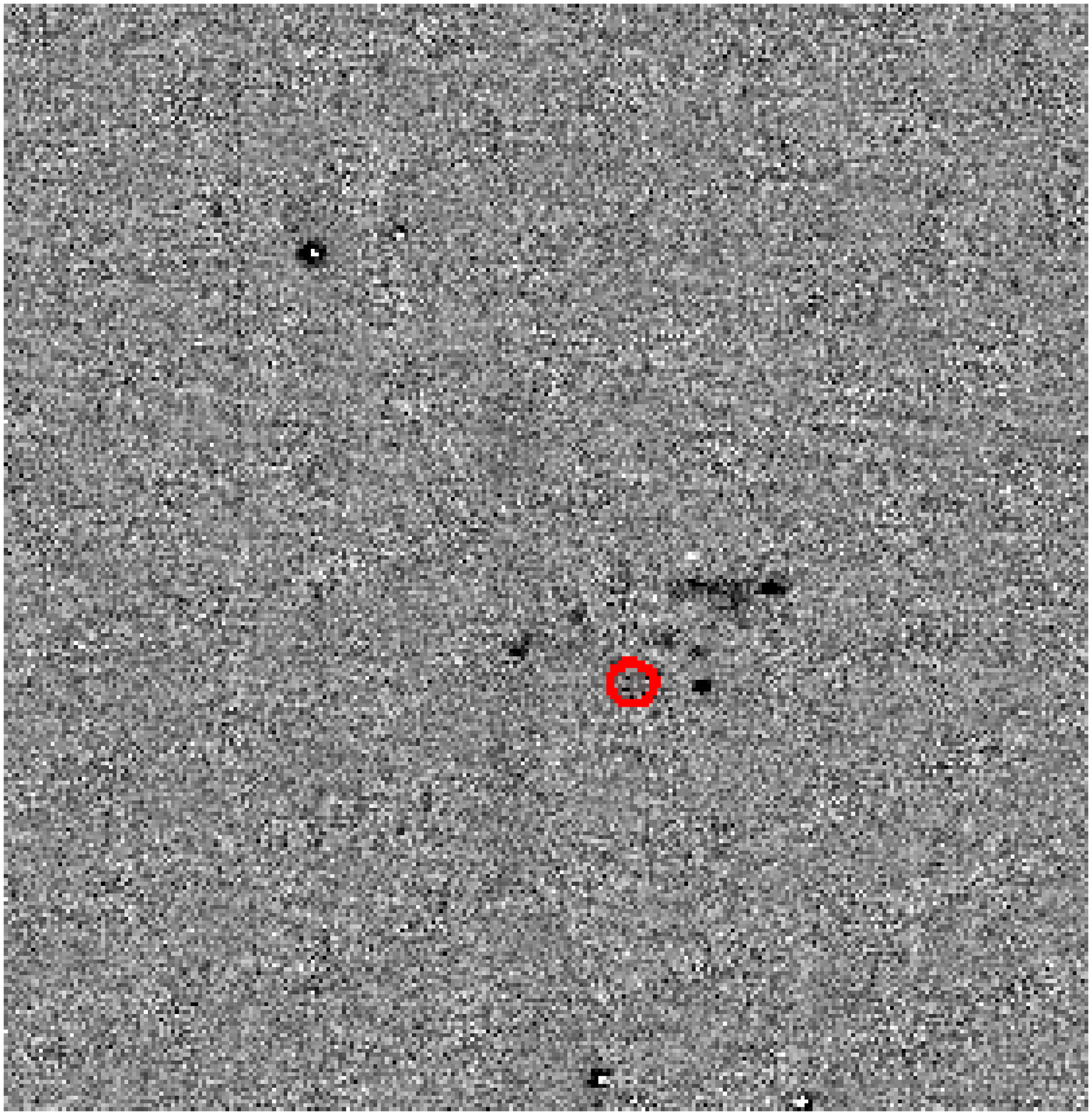}}   
     \subfigure[UGC~7848: SN2006bv]{\includegraphics[width=0.3\textwidth,height=0.15\textheight]{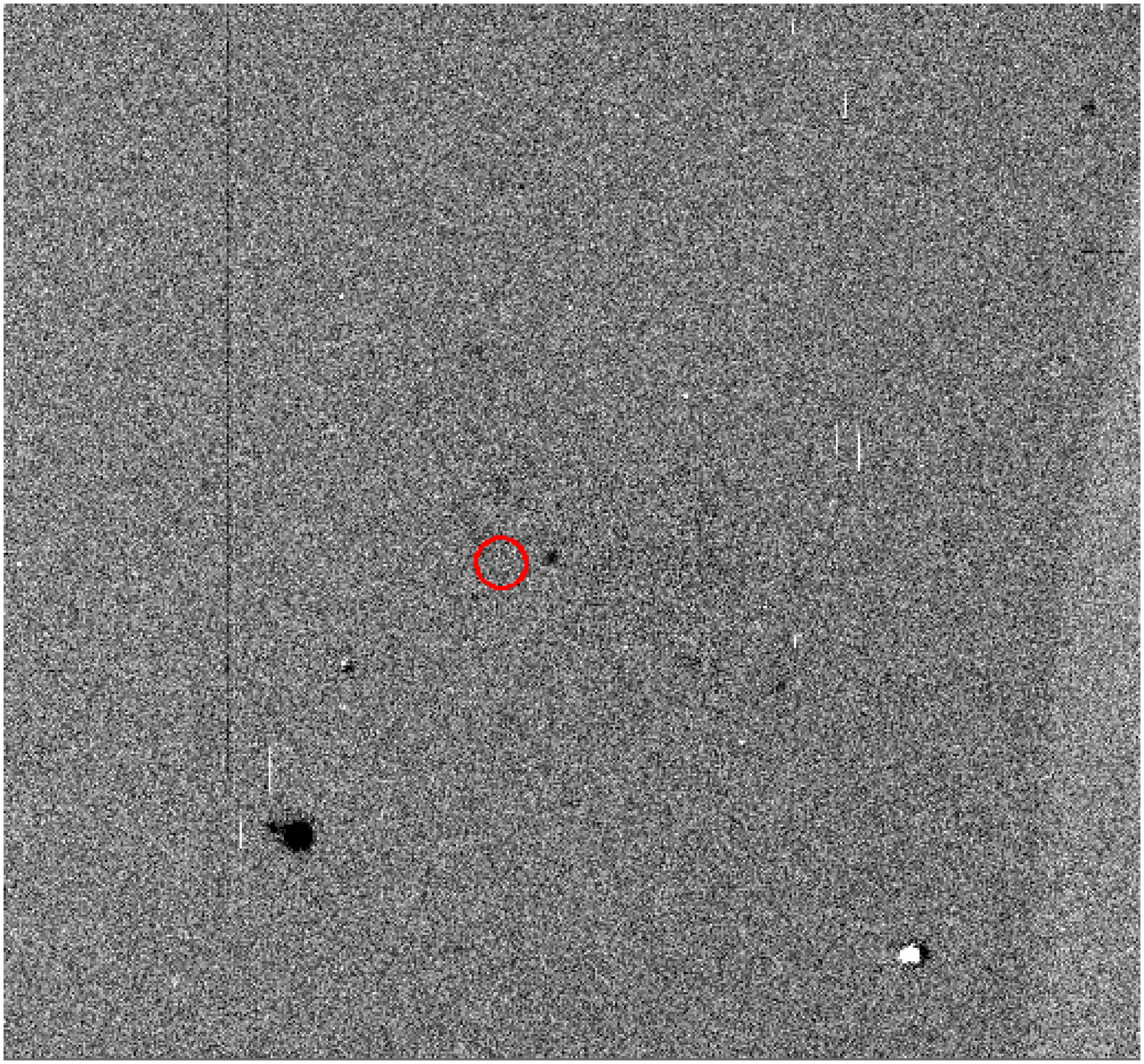}}   
     \subfigure[UGC~12182: SN2006fp]{\includegraphics[width=0.3\textwidth,height=0.15\textheight]{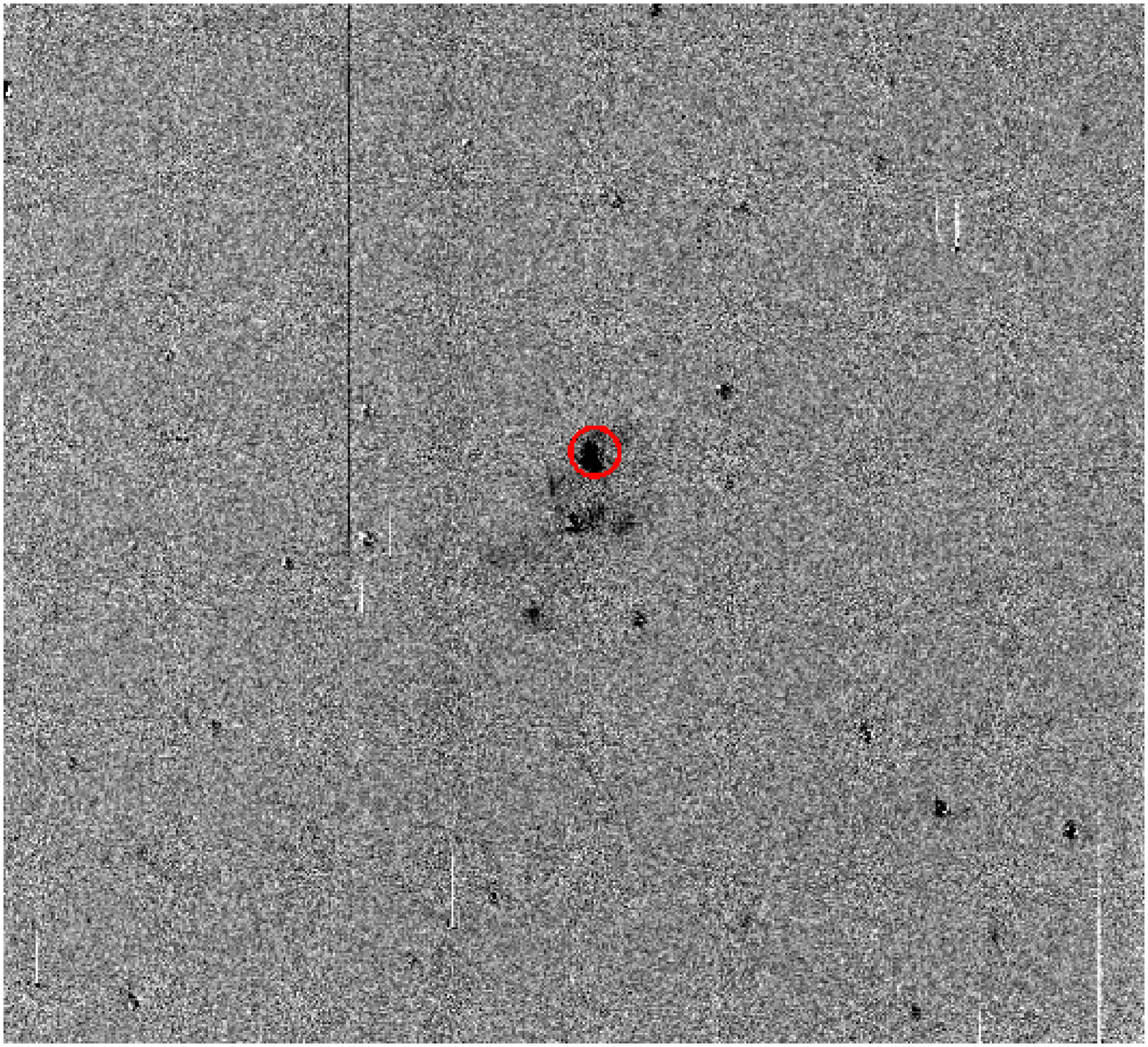}}   
     \subfigure[MCG-02-07-33: SN2008J]{\includegraphics[width=0.3\textwidth,height=0.15\textheight]{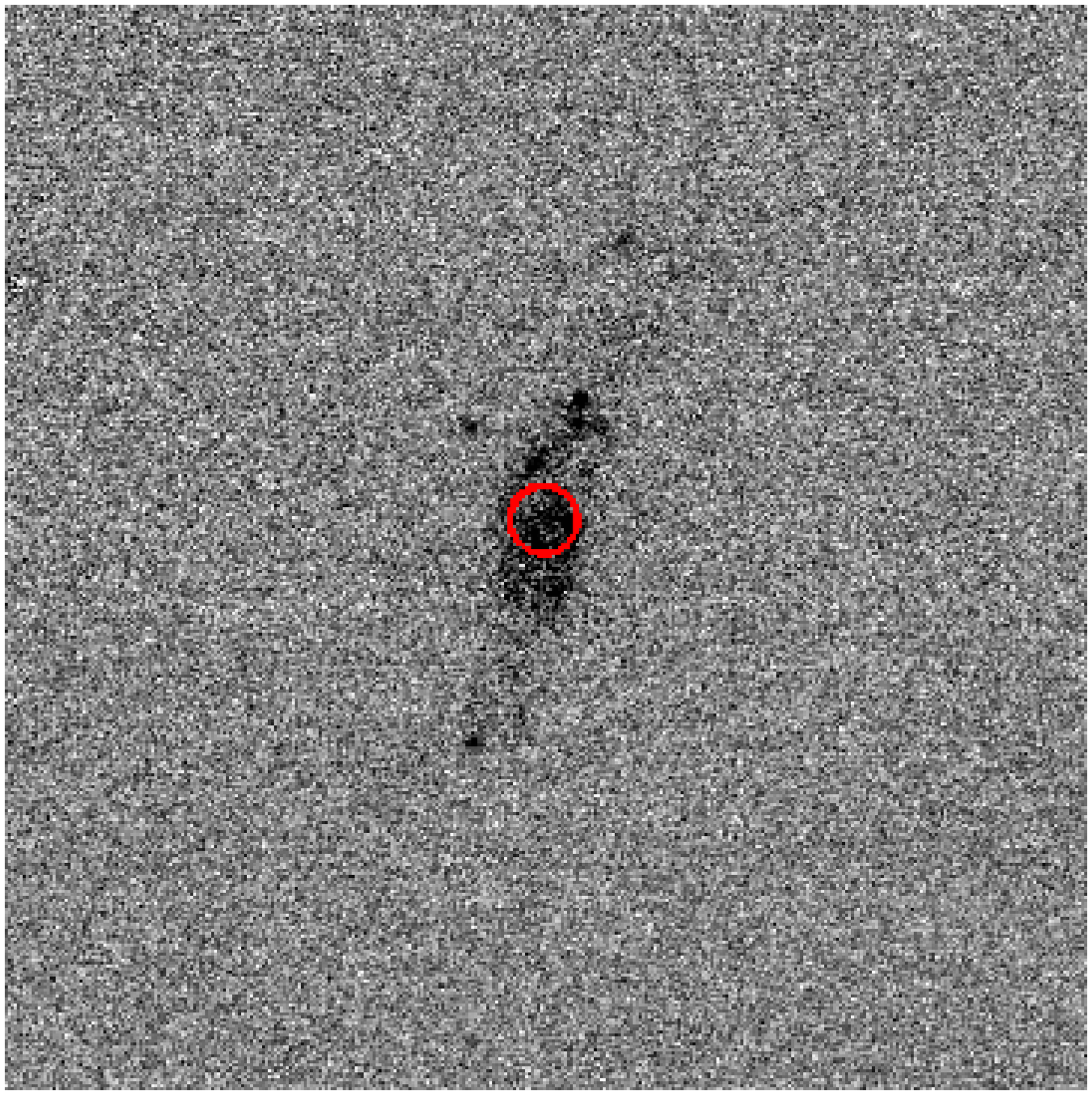}}   
     \subfigure[NGC~6946: SN2008S]{\includegraphics[width=0.3\textwidth,height=0.15\textheight]{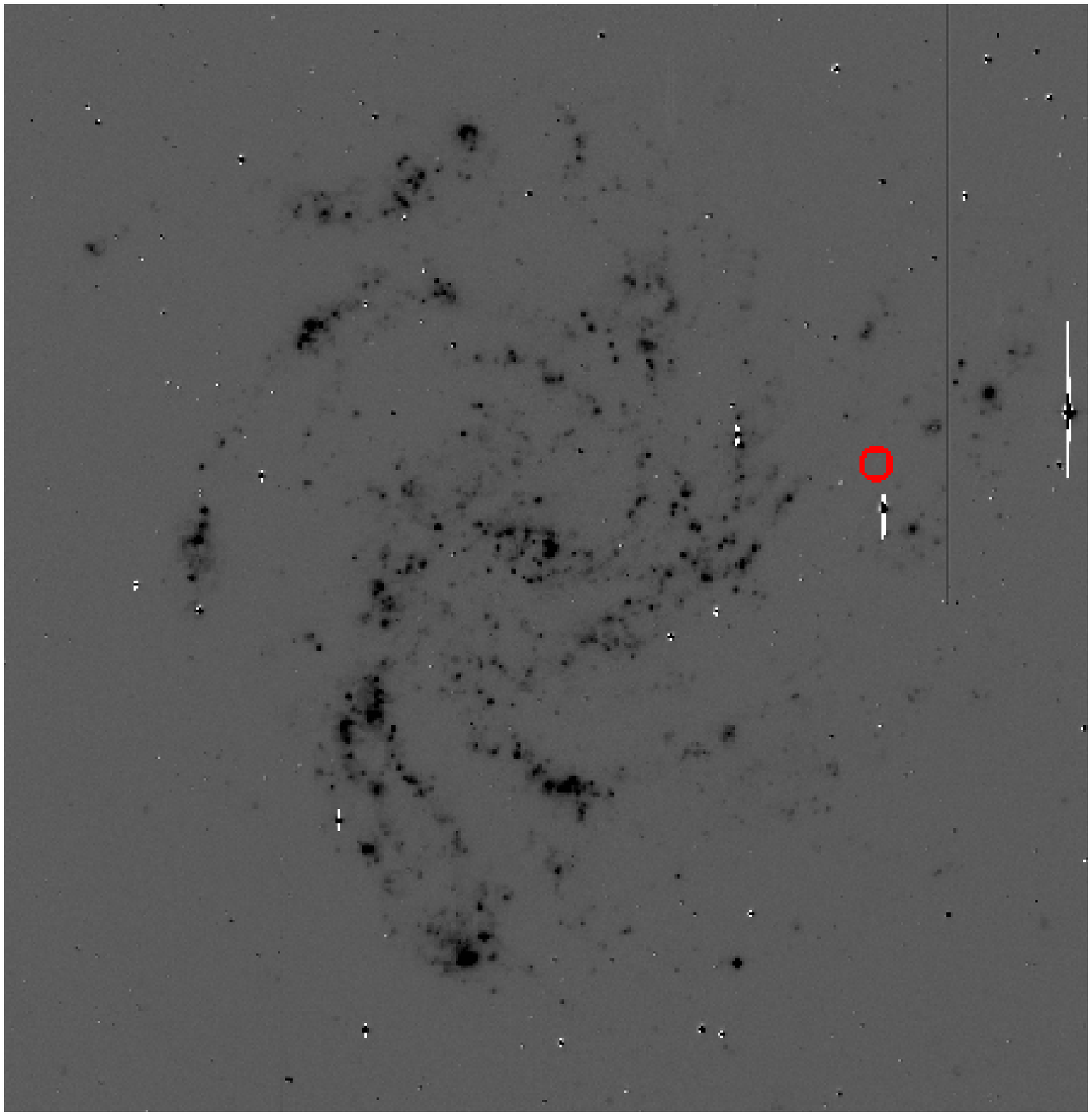}}   
     \subfigure[NGC~3184: SN2010dn]{\includegraphics[width=0.3\textwidth,height=0.15\textheight]{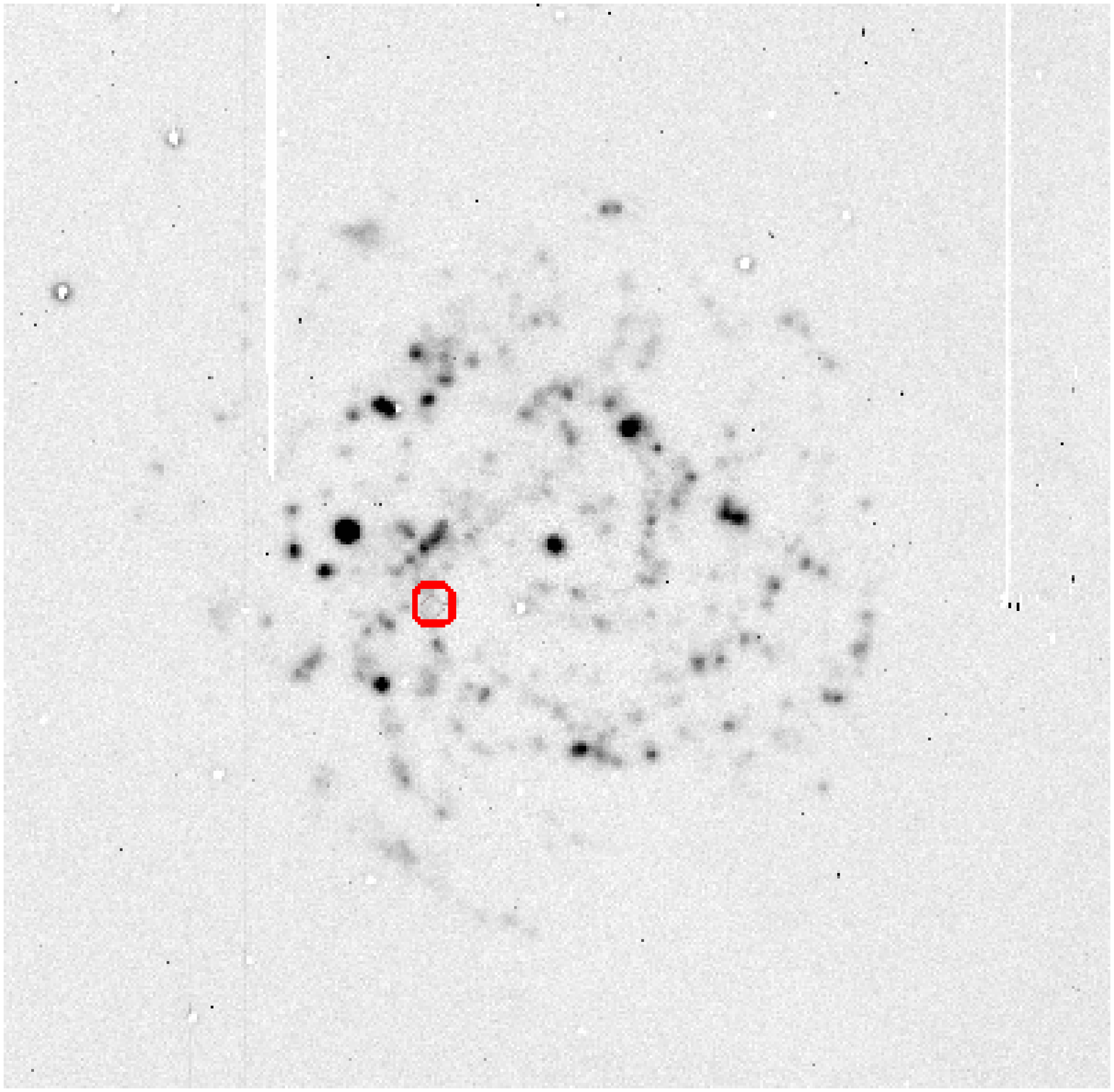}}   
     \subfigure[UGC~5189A: SN2010jl]{\includegraphics[width=0.3\textwidth,height=0.15\textheight]{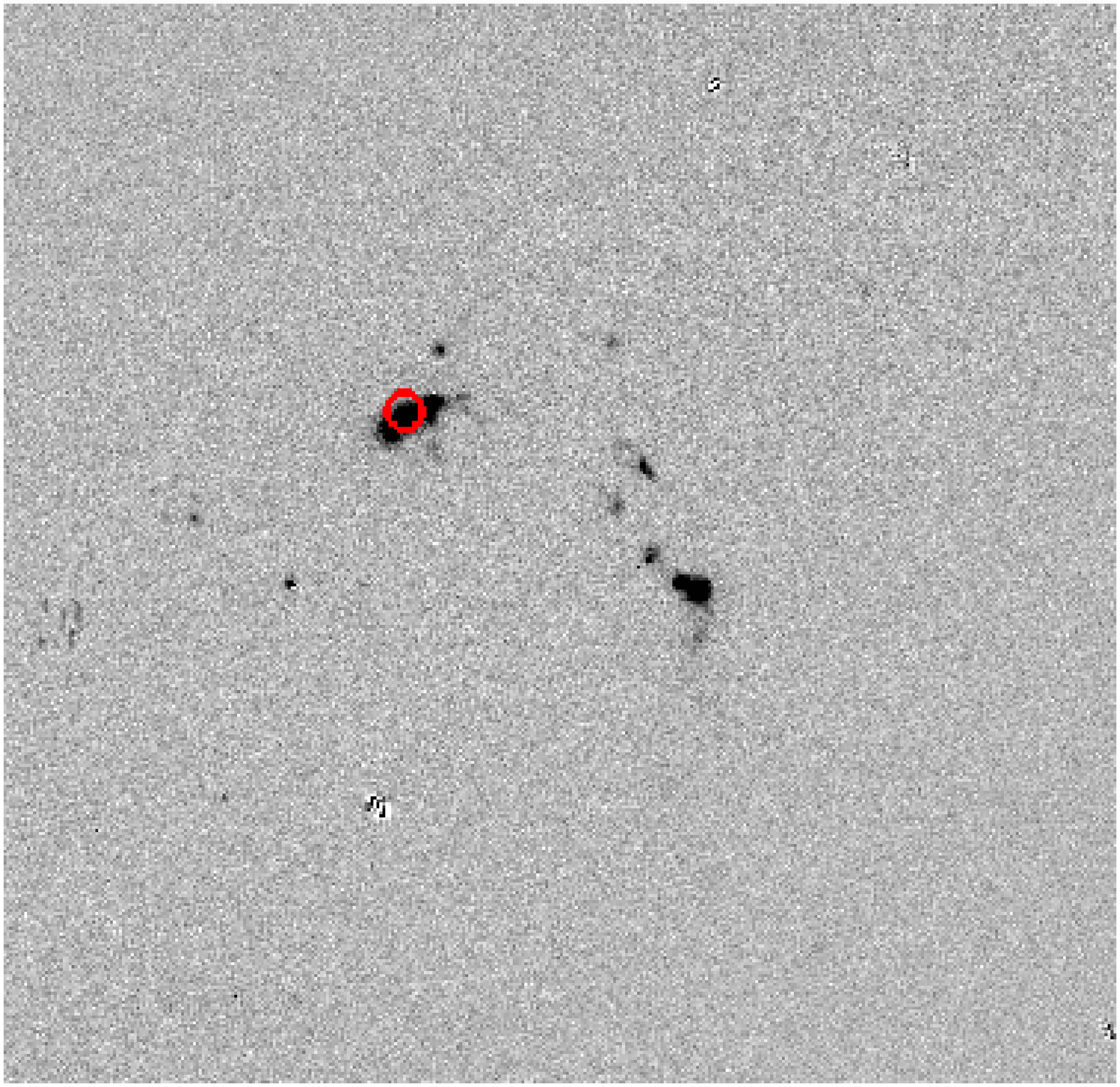}} 
     \subfigure[NGC~2366: V1]{\includegraphics[width=0.3\textwidth,height=0.15\textheight]{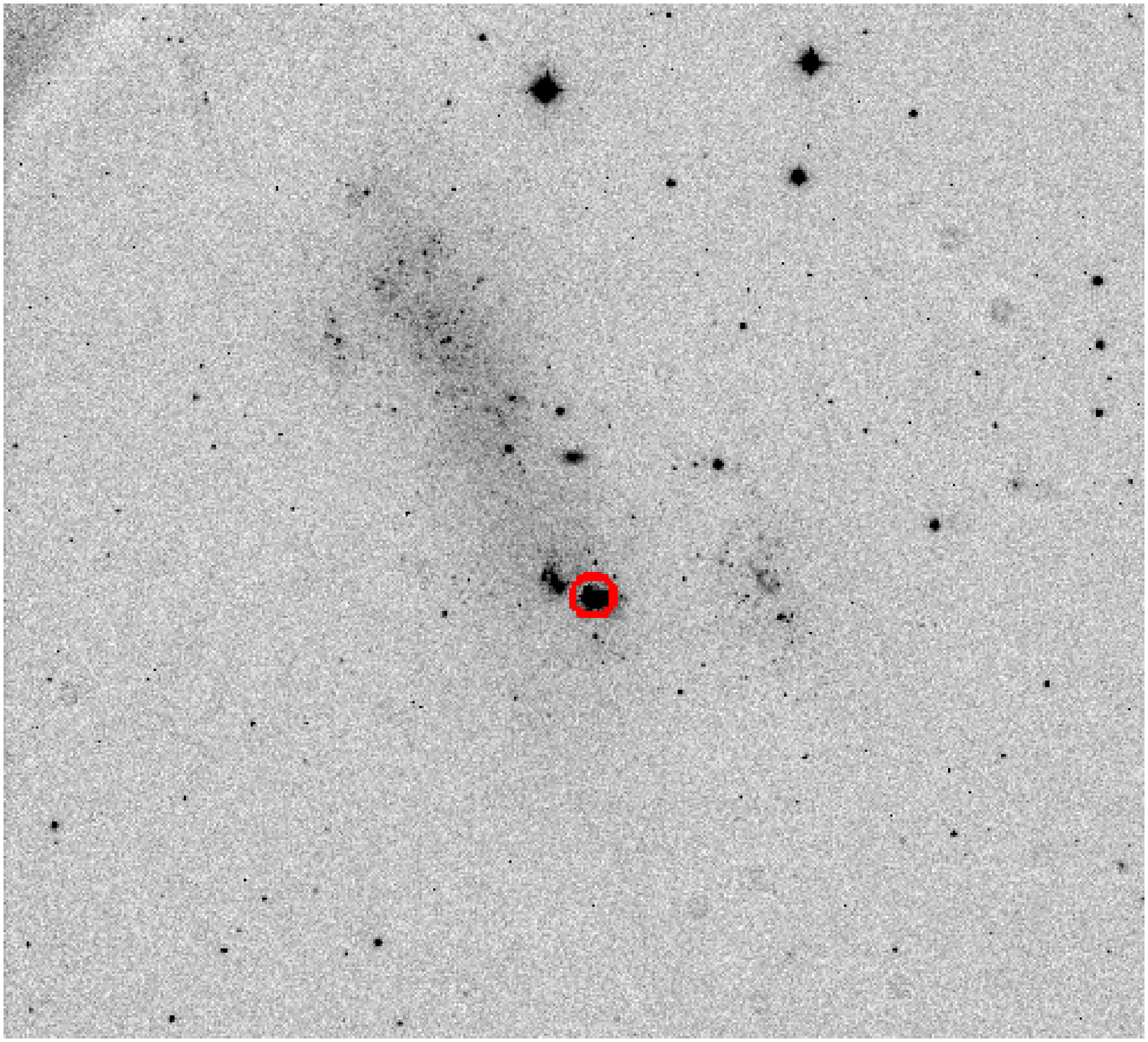}}\\
    \end{center}
  \end{minipage}
\end{figure*}

The host galaxy classifications of the interacting transient group encompass Hubble types from Sab to Im, but no ellipticals, lenticulars or very early-type spirals (S0a or Sa) are present in the group. Figure 2 shows the cumulative distribution of SNIIn, Impostor, SNIIP and SNIc host galaxy T-types, compared with the contributions of each host type to the star formation (SF) density in the local Universe, taken from \citet{jame08}. The distribution of interacting transient hosts follows the SF density in the local Universe. This analysis is in disagreement with the small number of SNIIn host galaxies presented in \citet{li11}, where the authors claim that SNIIn occur preferentially in less massive, late-type spirals, compared to SNIIP. In our analysis, the two distributions are statistically indistinguishable. The only outlier is the SNIc host distribution, which is offset to earlier type hosts compared to SNIIP, SNIIn and Impostor hosts. No such offset is found in \citet{li11}. This result is not, however, statistically significant, which may reflect the small sample sizes in some of the groups (i.e. only 12 Impostors). If real, this offset may indicate that SNIc preferentially occur in higher metallicity environments as claimed previously \citep[e.g.][]{arca10,modj11}, although this was not found by other recent analyses \citep[e.g.][]{ande10,habe12}.

The average host galaxy $B$-band absolute magnitudes for the various SN subtypes are presented in Table 3, and calculated using the apparent magnitudes given in the Third Reference Catalogue of bright galaxies (RC3; \citealt{deVau91}). These were then converted into absolute magnitudes using the distance modulus, where the distance was taken from redshift independent methods, given in NED, for hosts within 2000kms$^{-1}$, or using the recession velocity for hosts beyond 2000kms$^{-1}$, and taking H$_{0}$=68 kms$^{-1}$Mpc$^{-1}$ \citep[e.g.][]{planck13}.

\begin{table}
    \small
    \centering
    \caption{Average $B$-band host galaxy absolute magnitudes}
    \begin{tabular}{l c}
      \hline
      Type & M$_{B}$\\
      \hline
      SNIIn & --19.81 $\pm$ 0.27\\
      Impostor & --18.73 $\pm$ 0.39\\
      Interacting & --19.47 $\pm$ 0.27\\
      SNIIP & --19.58 $\pm$ 0.18\\
      SNIc & --19.54 $\pm$ 0.30\\
      \hline
    \end{tabular}
\end{table}

The Lick Observatory Supernova Search (LOSS; \citealt{li11}) found marginal evidence that SNIIn were from less luminous hosts than SNIIP, while this study finds the samples to be consistent using a Kolmogorov-Smirnov (KS) test. This test establishes the probability that two populations are statistically similar, and produces two values which will be used throughout this analysis, P and D. P is the probability that the two populations are drawn from the same parent distribution, and D is the maximum distance between the populations on a normalised cumulative distribution plot. Here the results indicate that the probability, P, of the two samples being drawn from the same parent population is 0.592 (D=0.180). Both studies find that the SNIIP host galaxy absolute magnitudes are consistent with those of the host galaxies of SNIc (here P=0.361, D=0.184). We also find that the SNIIn host galaxies are completely consistent with the SNIc host population (P=0.933, D=0.128). However, it is interesting to note that the host galaxies of the SN Impostor class are less luminous than any of the comparison samples; this is statistically significant in terms of the SNIc (P=0.016, D=0.513) and SNIIn (P=0.021, D=0.467) samples, and even compared to the SNIIP host sample, there is only a $\sim$6 per cent chance of the distributions being drawn from the same parent population. 

\begin{figure}
\includegraphics[height=85mm,angle=270]{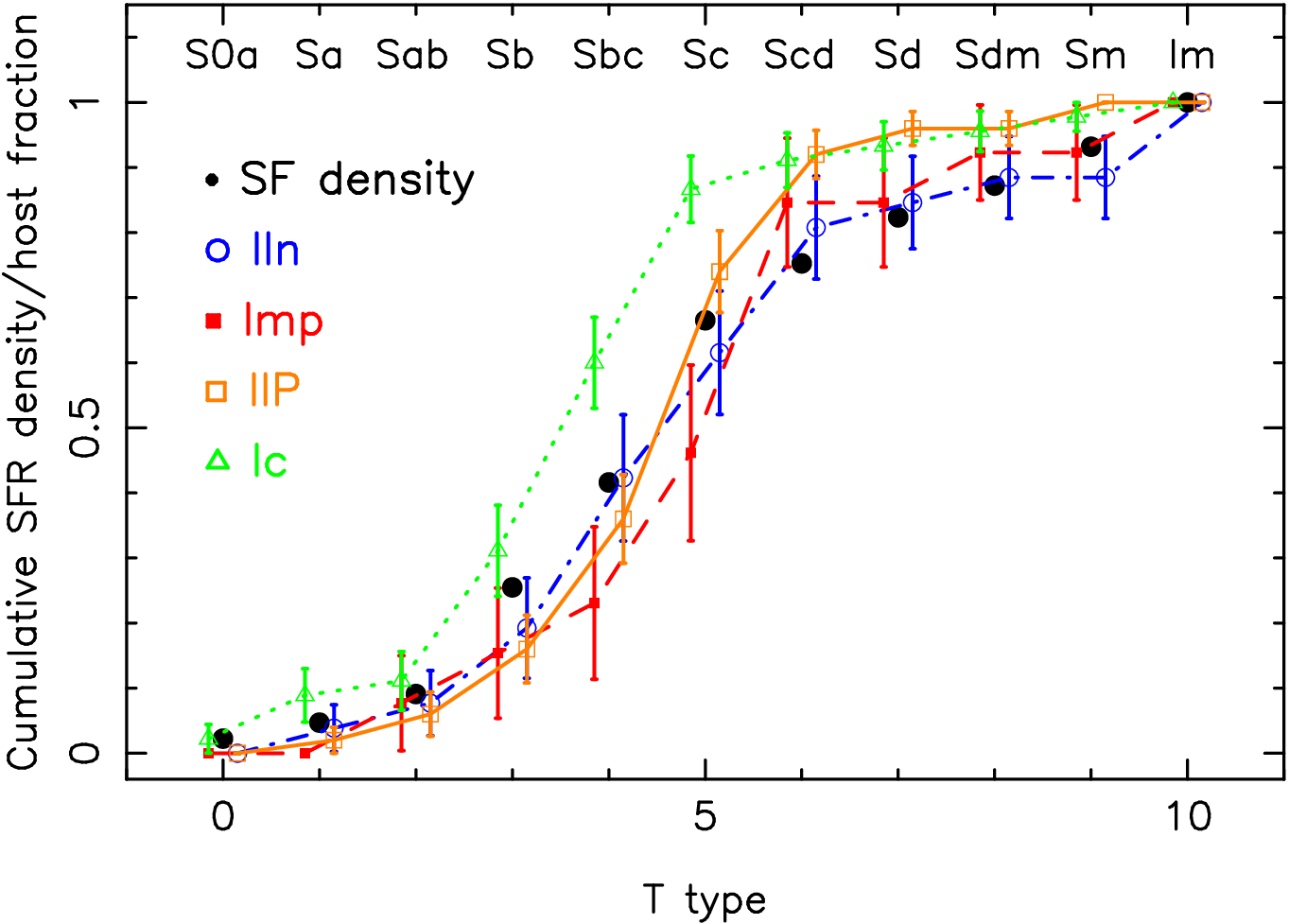}
\caption{
Distribution of the Hubble types of SNeIIn (blue) and Impostor (red) host galaxies compared with the contributions of each host type to the star formation density in the local Universe (black dots), and the host galaxies of SNIIP (orange) and SNIc (green). 
}
\label{fig:tsfr_IIn}
\end{figure}

\subsection{Metallicity}
Using the host galaxy absolute magnitudes it is possible to estimate the global metallicity of the host using equation 2 in \citet{trem04} (hereafter T04):

\begin{equation}
12+log(O/H) = -0.185(\pm0.001)M_{b}+5.238(\pm0.018)
\end{equation}

The host metallicities for each of the SNe in our interacting transient sample are presented in Table 2, based upon T04, and the average host metallicity for the various SN subtypes is shown in Table 4. Where possible the T04 metallicities have been converted into the equivalent O3N2 diagnostic presented in Pettini \& Pagel (2004; hereafter PP04), in order to compare to local SN metallicities presented in the literature \citep[e.g.][]{ande10,modj11}. These conversions have been carried out using the equations presented in \citet{kewl08} which are only valid for metallicities between 8.05 and 8.90, therefore only $\sim$75 per cent of the local metallicities could be converted using this technique. The average O3N2 metallicity given in Table 4 is therefore converted directly from the average T04 value.\\

{\begin{table}
    \small
    \centering
    \caption{Average Estimated Host Metallicity}
    \begin{tabular}{l l c c c}
      \hline
      Type & Number &  \multicolumn{3}{c}{Average Host Metallicity} \\
      \hline
      & & T04 & O3N2 & Error \\
      \hline
      SNIIn & 26 & 8.90 & 8.97 & 0.17 \\
      Impostor & 13 & 8.67 & 8.76 & 0.23 \\
      Interacting & 38 & 8.84 & 8.90 & 0.14 \\
      SNIIP & 50 & 8.86 & 9.01 & 0.13 \\
      SNIc & 45 & 8.85 & 9.06 & 0.13 \\
      \hline
    \end{tabular}
  \end{table}
}

We are able to compare these estimates of host galaxy metallicities, with the measured bulge metallicities taken using the Intermediate Dispersion Spectrograph (IDS) on the Isaac Newton Telescope (INT) during an observing run in November 2010, for six of the IIn host galaxies in this sample.

Table 5 presents the measured metallicities in O3N2 (O3N2$_{meas.}$) for SNe host galaxies in this sample, converted into T04 (T04$_{con.}$) for direct comparison to the estimated T04 metallicities (T04$_{est.}$).
This analysis has showed that the host galaxies of SNIIn are consistent with the distribution of star formation between galaxy types in the Universe, indicating that there is no preference for less luminous, lower metallicity, hosts. There are however indications that the host galaxies of SN Impostors may be less luminous and therefore of lower metallicities than true SN host galaxies.
{\begin{table}
    \footnotesize
    \centering
    \caption{Measured SN host galaxy metallicities}
    \begin{tabular}{l l c c c}
      \hline
      SN & Host & O3N2$_{meas.}$ & T04$_{con.}$ & T04$_{est.}$ \\
      \hline
      1994ak & NGC2782 & 8.59 & 8.39 & 9.04\\
      1999gb & NGC2532 & 8.79 & 8.54 & 9.22\\
      2002A & UGC3804 & 9.14 & 8.83 & 8.96\\
      2003lo & NGC1376 & 8.66 & 8.44 & 9.13\\
      2005db & NGC214 & 8.85 & 8.59 & 9.15\\
      2005ip & NGC2906 & 9.14 & 8.83 & 8.83\\
      \hline
    \end{tabular}
  \end{table}
}

We do not currently have metallicity measurements taken locally to the SNe for most of these events, therefore in order to assess whether mean metallicities in the local regions of SNe vary substantially between subtypes, we must make some assumptions. Taking the global metallicities presented in Table 2, we can assume that these represent the nuclear metallicity in the host galaxy, and we can apply a typical metallicity gradient (e.g. those presented for the various Hubble T-types in \citealt{henr99}, which are adopted in this paper) out to a defined radius (such as the R$_{25}$ isophotal radius). Such assumptions have been applied by several authors \citep[e.g.][]{bois09}. R$_{25}$ is the standard isophotal radius where the {\it B}-band surface brightness is 25 mag arcsec$^{-2}$ \citep{deVau91}. By correcting the SN offset from the centre for galaxy inclination and converting this distance to a fraction of the R$_{25}$, we can estimate a local metallicity for the site of the SN. We were only able to do this for spiral hosts with known R$_{25}$ values, and as a result not all of the sample have an estimated local metallicity value, as can be seen in Table 2.

This technique yielded estimates of the average local metallicities for the various SN subtypes, where again the conversions to O3N2 are limited to those SNe with local values within the range 8.05-8.90, and therefore the O3N2 values presented in Table 6 are converted directly from the average T04 value. \\

{\begin{table}
    \small
    \centering
    \caption{Average estimated local metallicities for subtypes}
    \begin{tabular}{l l c c c}
      \hline
      Type & Number & \multicolumn{3}{c}{Average SN Metallicity}\\
      \hline
      & & T04 & O3N2 & Error \\
      \hline
      SNIIn & 22 & 8.636 & 8.894 & 0.201\\
      Impostor & 10 & 8.290 & 8.476 & 0.288\\
      Interacting & 32 & 8.528 & 8.772 & 0.170\\
      SNIIP & 47 & 8.526 & 8.770 & 0.135\\
      SNIc & 44 & 8.662 & 8.923 & 0.140\\
      \hline
    \end{tabular}
  \end{table}
}

These estimates appear higher than true, local metallicity measurements when compared to values in the literature which give an average SNII (dominated by SNIIP) value of 8.62$\pm$0.04 in \citet{ande10}, and an average for PTF detected SNII of 8.65$\pm$0.09 in \citet{stol13}, and an average SNIc value of 8.63$\pm$0.06 (averaged from the data presented in \citealt{ande10,modj11,lelo11,sand12}). It is therefore interesting to compare the estimates to local SNIIn metallicity measurements, for which we have three. These were again obtained using the IDS on the INT, La Palma, in November 2010, and are presented in Table 7.\\

{\begin{table}
    \small
    \centering
    \caption{Measured local SN site metallicities}
    \begin{tabular}{l c c c}
      \hline
      SN & O3N2$_{meas.}$ & T04$_{con.}$ & T04$_{est.}$ \\
      \hline
      1994ak & 8.83 & 8.58 & 8.61\\
      1999gb & 8.72 & 8.49 & 8.95\\
      2003lo & 8.66 & 8.44 & 8.88\\
      \hline
    \end{tabular}
  \end{table}
}
 While the absolute values of each estimated local metallicity measurement are possibly unreliable, we may still use the comparative values and address any difference between them. Whilst the samples of SNIIn and SNIc have similar estimated local metallicity values, we see a slightly lower value for the SNIIP hosts compared to SNIIn, though both are consistent within the large errors.

The large discrepancies between the converted T04 values presented here, and the estimated T04 values are not unexpected. For any given host galaxy magnitude the T04 data show a spread of $\sim$0.3-0.4 dex in metallicity (figure 4 of \citealt{trem04}). Combined with this the metallicity gradients presented in \citet{henr99} show a spread of $\sim$1.0 dex in metallicity at any given radius (lower left panel of figure 4a in \citealt{henr99}). The conversions presented in \citet{kewl08} also contain errors of $\lesssim$0.07 dex. We have attempted to combine all of these errors in our calculations which reflect the large uncertainties in such an analysis. 
 
When looking at the comparisons of the Impostor group with the other subtypes, with the errors taken into account, all are formally consistent. However, the mean local metallicity value of the Impostors appears to be smaller and may indicate that these events prefer low metallicity environments. This result should be followed up by measuring the true host environments of this class of transient, which is beyond the scope of this paper. The relative difference between the bulge and estimated local metallicity for the Impostor sample is interesting, although the errors involved in these estimates are large. It is therefore worthwhile to analyse in more detail the radial location of these events to see if they are really located in the outer regions of the host galaxies. We are able to use here a well tested statistic, known as Fr({\it R}) which has been developed to analyse the distribution of events within their host galaxies \citep{ande09,habe10}, and is described in detail in the next section.  

\subsection{Radial Analysis of Interacting Transients}

Our previous work has shown that useful information on the progenitor stars
of different supernova types is revealed by the distributions of their radial positions, relative to the underlying stellar populations, within their host galaxies. We compare the interacting transients presented in this paper with the distribution of both SNIIP and SNIc presented in \citet{habe12}, and have thus conducted a radial analysis for the present sample. For each SNIIn within the sample the radial position was calculated using the technique described in \citet{ande09} in terms of the fraction of the total galaxy {\it R}-band and H$\alpha$ emission within the position of the SN (defined as Fr({\it R}) and Fr(H$\alpha$) respectively). This gives each SN a value between 0 and 1, where 0 would indicate that the SN was in the centre of the host galaxy and 1 indicates an extreme outlying SN. Table 8 presents the results of the radial analysis which can also be seen in histogram format in Figures 3 and 4, showing the fractional {\it R}-band and H$\alpha$ values respectively. 

\begin{table*}
  \small
  \begin{minipage}{160mm}
 \centering
 \caption{Fractions of {\it R}-band and H$\alpha$ light within the locations of interacting transients in this sample, followed by the NCR index for each interacting transient within the sample. Also included are the dates of detection and of our observation, along with the telescope used.}
  \small
   \begin{tabular}{llccccrcrc}
    \hline 
    Event & Type & Fr({\it R}) & Fr({\it H$\alpha$}) & NCR$_{H\alpha}$ & NCR$_{UV}$ & Discovery & Discovery Mag$_{R}$ & Observation & Telescope\\ 
    \hline 
     1987B & IIn & 0.951 & 1.000 & 0.000 & 0.000 & 24-02-1987 & --18.7 & 07-02-2008 & INT\\    
     1987F & IIn & 0.489 & 0.333 & 0.352 & 0.541 & 23-04-1987 & --19.7 & 07-02-2008 & INT\\    
     1993N & IIn & 0.512 & 0.261 & 0.000 & 0.000 & 15-04-1993 & --15.7 & 09-02-2010 & ESO 2.2\\ 
     1994W & IIn & 0.491 & 0.541 & 0.795 & 0.679 & 30-07-1994 & --18.8 & 09-11-2008 & LT\\    
     1994Y & IIn & 0.355 & 0.212 & 0.000 & 0.331 & 12-08-1994 & --19.1 & 07-02-2008 & INT\\    
     1994ak & IIn & 0.725 & 0.977 & 0.000 & 0.311 & 24-12-1994 & --16.6 & 05-02-2008 & INT\\   
     1995N & IIn & 0.612 & 0.822 & 0.001 & 0.000 & 05-05-1995 & --15.0 & 15-03-2008 & LT\\
     1996bu & IIn & 0.923 & 0.993 & 0.000 & 0.000 & 14-11-1996 & --13.3 & 06-02-2008 & INT\\   
     1996cr & IIn & 0.881 & 0.909 & 0.575 & -- & 16-03-2007 & --10.2 & 07-02-2010 & ESO 2.2\\ 
     1997eg & IIn & 0.503 & 0.449 & 0.338 & 0.418 & 04-12-1997 & --17.3 & 07-02-2008 & INT\\   
     1999el & IIn & 0.320 & 0.259 & 0.048 & 0.232 & 20-10-1999 & --17.2 & 10-08-2009 & LT\\   
     1999gb & IIn & 0.485 & 0.443 & 0.676 & 0.489 & 22-11-1999 & --18.3 & 06-02-2008 & INT\\   
     2000P & IIn & 0.375 & 0.336 & 0.000 & 0.620 & 08-03-2000 & --18.5 & 05-02-2010 & ESO 2.2\\ 
     2000cl & IIn & 0.226 & 0.157 & 0.312 & 0.613 & 26-05-2000 & --18.3 & 04-02-2010 & ESO 2.2\\ 
     2001fa & IIn & 0.302 & 0.311 & 0.147 & 0.376 & 18-10-2001 & --17.5 & 22-10-2012 & LT\\
     2002A & IIn & 0.419 & 0.253 & 0.401 & 0.803 & 01-01-2002 & --15.7 & 22-01-2001 & JKT\\    
     2002fj & IIn & 0.488 & 0.372 & 0.558 & 0.841 & 12-09-2002 & --18.3 & 06-02-2008 & INT\\   
     2003G & IIn & 0.209 & 0.102 & 0.000 & 0.000 & 08-01-2003 & --17.4 & 22-10-2012 & LT\\      
     2003dv & IIn & 0.926 & 0.780 & 0.000 & 0.301 & 22-04-2003 & --16.4 & 04-04-2009 & INT\\   
     2003lo & IIn & 0.398 & 0.384 & 0.000 & 0.293 & 31-12-2003 & --16.7 & 08-02-2010 & ESO 2.2\\ 
     2005db & IIn & 0.531 & 0.533 & 0.398 & 0.657 & 19-07-2005 & --16.8 & 20-10-2012 & LT\\    
     2005gl & IIn & 0.519 & -- & 0.000 & 0.995 & 05-10-2005 & --17.2 & 21-10-2012 & LT\\       
     2005ip & IIn & 0.399 & 0.528 & -- & -- & 05-11-2005 & --17.0 & 17-01-2008 & INT\\      
     2006am & IIn & 0.604 & 0.617 & 0.000 & 0.445 & 22-02-2006 & --14.5 & 11-02-2007 & LT\\   
     2008J & IIn & 0.087 & 0.069 & 0.807 & 0.770 & 15-01-2008 & --18.8 & 22-10-2012 & LT\\
     \hline                                        
     Mean & & 0.508 & 0.485 & 0.225 & 0.422 && --16.9 &&\\
     $\sigma$ & & 0.045 & 0.059 & 0.058 & 0.062 && 0.41 &&\\
     \hline
     1954J & IMP &  0.680 & 0.681 & 0.187 & 0.738 & ?-10-1954 & --12.3 & 20-03-2005 & INT\\
     1961V & IMP & 0.968 & 0.931 & 0.363 & 0.000 & 11-12-1961 & --16.9 & 19-11-2000 & JKT\\
     1997bs & IMP & 0.362 & 0.348 & 0.023 & 0.328 & 15-04-1997 & --13.2 & 06-02-2008 & INT\\
     1999bw & IMP & 0.745 & 0.755 & 0.000 & 0.466 & 20-04-1999 & --12.9 & 06-02-2008 & INT\\
     2001ac & IMP & 0.826 & 0.992 & 0.000 & 0.000 & 12-03-2001 & --13.3 & 29-03-2005 & INT\\
     2002bu & IMP & 0.896 & 0.930 & 0.000 & 0.000 & 28-03-2002 & --14.0 & 08-05-2000 & JKT\\
     2002kg & IMP & 0.378 & 0.146 & 0.055 & 0.654 & 26-10-2002 & --8.77 & 20-03-2005 & INT\\
     2003gm & IMP & 0.536 & 0.352 & 0.000 & 0.468 & 06-07-2003 & --15.6 & 15-05-2009 & LT \\
     2006bv & IMP & 0.579 & 1.000 & 0.000 & -- & 28-04-2006 & --15.0 & 05-03-2009 & LT\\
     2006fp & IMP & 1.000 & 1.000 & 0.965 & 0.000 & 17-09-2006 & --14.1 & 04-08-2008 & LT\\
     2008S & IMP & 0.632 & 0.563 & 0.000 & 0.031 & 01-02-2008 & --11.9 & 30-03-2005 & INT\\
     2010dn & IMP & 0.224 & 0.198 & 0.000 & 0.762 & 31-05-2010 & --12.9 & 06-02-2008 & INT\\
     NGC2366-V1 & IMP & --- & --- & --- & --- & 1993 & --6.9 & 12-03-2003 & JKT\\
     \hline
     Mean & & 0.652 & 0.658 & 0.133 & 0.313 & & --13.4 & &\\
     $\sigma$ & & 0.072 & 0.094 & 0.082 & 0.096 && 0.59 & &\\
     \hline
     Transient Mean & & 0.555 & 0.543 & 0.194 & 0.387 && --15.8 &&\\
     Transient $\sigma$ & & 0.039 & 0.052 & 0.047 & 0.052 && 0.43 &&\\
     \hline
     \hline
   \end{tabular} 
 \end{minipage}
\end{table*}

An interesting feature of Figure 3 is the lack of SNIIn in the central regions of host galaxies, with only one of the 26 occurring in the central 20\% of the host galaxy light. We also find a lack of SN Impostor detections in these central locations. 

\begin{figure}
 \centering
  \includegraphics[width=0.8\columnwidth,angle=0]{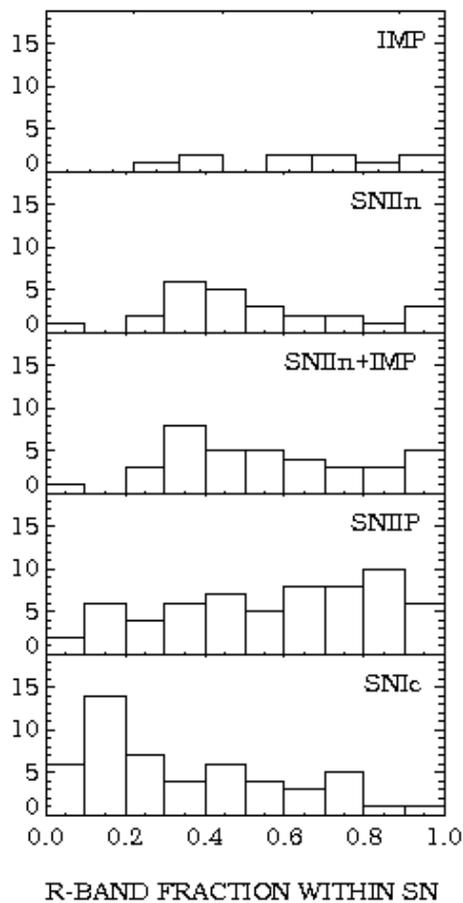}
  \caption{Distributions of Fr({\it R}) values for SNIIn and Impostors, compared with SNIIP and SNIc.}
\end{figure}
\begin{figure}
 \centering
  \includegraphics[width=0.76\columnwidth,angle=0]{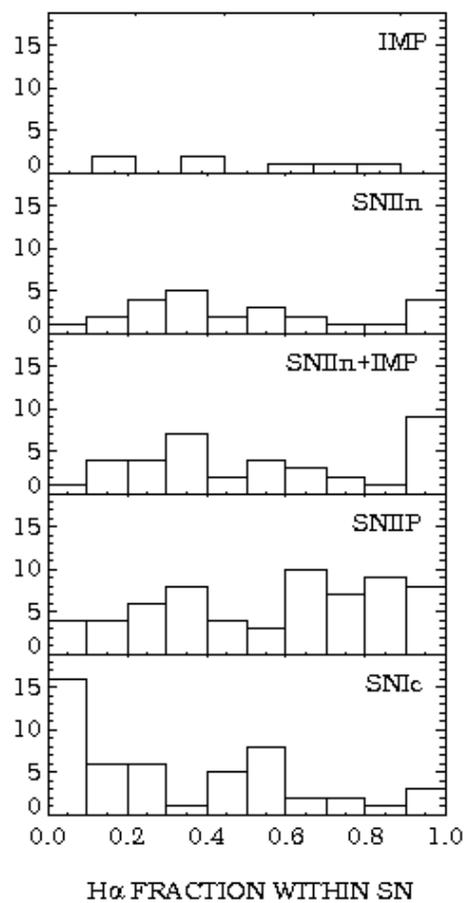}
  \caption{Distributions of Fr(H$\alpha$) values for SNIIn and Impostors, compared with SNIIP and SNIc.}
\end{figure}

Tables 9 and 10 show the probabilities of the distributions of SNIIn and Impostors being drawn from the same parent population as SNIIP and SNIc in terms of the Fr({\it R}) and Fr(H$\alpha$) values respectively, in the form of a KS test. Comparisons to other individual SN sub-groups are also shown where possible (sample size $>$ 10). The radial analysis of these various subtypes can be found in \citet{ande09,habe10,habe12}. The comparison to a sample of SNIa (Anderson at al., in prep.) is also given for completeness.

\begin{table*}
  \begin{minipage}{160mm}
    \centering
    \caption{Kolmogorov-Smirnov test results for the Fr({\it R}) distributions, where the sample sizes are shown in brackets.}
    \begin{tabular}{lccccccc}
      \hline 
      & \multicolumn{2}{c}{SNIIn} & \multicolumn{2}{c}{Impostor} & \multicolumn{2}{c}{Interacting Transients}\\
      \hline
      & D$_{R}$ & P$_{R}$ & D$_{R}$ & P$_{R}$ & D$_{R}$ & P$_{R}$ \\
      \hline 
      SNIa (106) & 0.248 & 0.140 & 0.222 & 0.607 & 0.123 & 0.775 \\
      SNIIP (59) & 0.313 & 0.050 & 0.188 & 0.833 & 0.194 & 0.323 \\
      SNIIL (10) & 0.420 & 0.118 & 0.317 & 0.556 & 0.295 & 0.430 \\
      SNIIb (16) & 0.238 & 0.579 & 0.313 & 0.437 & 0.211 & 0.647 \\
      SNIb (41) & 0.215 & 0.424 & 0.311 & 0.274 & 0.193 & 0.426 \\
      SNIc (51) & 0.409 & 0.005 & 0.505 & 0.009 & 0.426 & 0.001 \\
      \hline
    \end{tabular}
  \end{minipage}
\end{table*}                                                                  

\begin{table*}              
  \begin{minipage}{160mm}         
    \centering                         
    \caption{Kolmogorov-Smirnov test results for the Fr({\it H$\alpha$}) distributions, where the sample sizes are shown in brackets.}
    \begin{tabular}{lccccccc}
      \hline 
      & \multicolumn{2}{c}{SNIIn} & \multicolumn{2}{c}{Impostor} & \multicolumn{2}{c}{Interacting Transients}\\
      \hline
      & D$_{H\alpha}$ & P$_{H\alpha}$ & D$_{H\alpha}$ & P$_{H\alpha}$ & D$_{H\alpha}$ & P$_{H\alpha}$ \\
      \hline                                        
      SNIa (106) & 0.256 & 0.131 & 0.285 & 0.294 & 0.131 & 0.720 \\   
      SNIIP (59) & 0.302 & 0.072 & 0.315 & 0.226  & 0.204 & 0.275 \\            
      SNIIL (10) & 0.308 & 0.437 & 0.400 & 0.269 & 0.244 & 0.673 \\  
      SNIIb (16) & 0.167 & 0.930 & 0.354 & 0.287 & 0.222 & 0.588 \\
      SNIb (41) & 0.172 & 0.722 & 0.295 & 0.335 & 0.128 & 0.891 \\     
      SNIc (51) & 0.328 & 0.046 & 0.490 & 0.012 & 0.328  & 0.016 \\               
      \hline                                                               
    \end{tabular}                                                        
  \end{minipage}                                                        
\end{table*}

The results of the KS test on the distributions of SNe with respect to the {\it R}-band light show that the SNIIn population are statistically more similar to the SNIIP population than to the SNIc. However, there is only $\sim$8~per cent probability of SNIIn and SNIIP being drawn from the same parent distribution. The difference between the SNIIn and SNIc populations is the most significant, with the probability of them being drawn from the same parent population being only $\sim$0.5~per cent. This result is driven by the lack of SNIIn in the central regions of their host galaxies, whereas the SNIc distribution peaks towards the nucleus, as seen in various papers \citep[e.g.][]{bart92,ande09,habe10,habe12}. Given the evidence for high-mass progenitors of SNIc \citep{ande12,kunc13}, this stark contrast in galactic host environments between SNIc and SNIIn is hard to reconcile if both classes have similarly high mass progenitors.

As with the {\it R}-band light distribution, SNIIn are also statistically more similar to SNIIP than SNIc with respect to the radial distribution of H$\alpha$ emission, but only marginally. The probability of SNIIn and SNIc being drawn from the same parent distribution when analysing the H$\alpha$ emission is $\sim$5~per cent, and so we consider these to be formally inconsistent distributions.  The comparison between IIn and IIP is marginal, with a 7 per cent chance of their being drawn from the same distribution, an apparent difference but not one that is confirmed at the 95 per cent level.  In this test, the IIn are completely consistent with the radial H$\alpha$ distributions for all other types (SNIa, IIL, IIb and Ib).

In terms of SN Impostors, the distribution with respect to the {\it R}-band light is consistent with all SN subtypes, with the exception of SNIc, where the probability of them being drawn from the same parent population is $\sim$1 per cent. In contrast to the SNIIn distribution, Impostors closely trace the SNIIP population, with a probability of $\sim$83 per cent of being drawn from the same parent sample. The H$\alpha$ distributions are very similar, with only the comparison to the SNIc population showing any significant probability of being drawn from a different overall distribution, with a P value of $\sim$1 per cent. 

Interestingly, a KS test shows that the probability of the SNIIn and SN Impostor fractional {\it R}-band distributions being drawn from the same parent population is only $\sim$4 per cent (D=0.470 and P=0.037). In terms of the fractional H$\alpha$ distributions the values are more consistent, with a P value of $\sim$16 per cent (D=0.375 and P=0.162). However, due to the uncertainty when classifying these events, and previous instances of events being moved from one class to the other, we can combine the classes into a general `interacting transient' class. This allows us to improve the statistics, the results of which are shown in the last two columns of Tables 9 and 10. This reflects the difficulty in distinguishing between the two classes, events of both types often lacking follow-up observations, leading to few Impostors having confirmation of progenitor survival. The distribution of interacting transients in Fr({\it R}) is consistent with being drawn from the SNIIP population with a P value of 0.323 (D=0.194), and has only a $\sim$0.1 per cent probability (D=0.426) of being drawn from the same parent population as SNIc. These results are reflected in the Fr(H$\alpha$) distribution, with a $\sim$2 per cent probability of being drawn from the same parent population as SNIc, and consistent with the distribution of all other SN subtypes.

The radial distributions of SNIIn, Impostors and the combined class, are statistically very different from the distribution of SNIc, both in terms of {\it R}-band and H$\alpha$ light. This is driven by a lack of central `interacting transients' (SNIIn and Impostors) in the sample, which is where the SNIc population peaks. The possible impact of selection effects on this result will be explored in Section 4.

\section{Pixel Statistics}
This section will use H$\alpha$ emission as a tracer of ongoing star formation, and near UV emission as a tracer of recent star formation, to present constraints of the Impostor and SNIIn populations analysed in this paper through the locations of each event, and their association with star forming regions. We will study 37 of the events here as two of the explosions are still in outburst in our observations excluding them from the pixel statistic analysis presented.
 
\subsection{Local H$\alpha$ emission}
If the progenitors of SNIIn are LBV-like as suggested in the literature, one would expect to find them near star formation regions. Higher mass stars have shorter lifetimes and hence are less likely to move away from their place of birth, and the star formation region is also more likely to still be active. Therefore numerous studies have analysed the distribution of massive stars \citep[e.g.][]{bibb12} and CCSNe (e.g. \citealt{ande08}; \citealt{ande12, crow13}) with respect to H$\alpha$ emission. \citet{ande12} obtained large enough samples of the various sub-types of CCSNe to analyse the separate distributions of each with respect to the underlying star formation in the host galaxy. The results seem contrary to the assumption of LBV progenitors for SNIIn. The type IIn SNe are less associated with the H$\alpha$ emission than most other SNII types and have a closer association to recent star formation indicated by archival {\it GALEX} near-UV emission data \citep{ande12}. This result suggests that the progenitors of SNIIn are not the most massive stars predicted elsewhere in the literature. This analysis will be expanded upon and discussed here.


In previous papers \citep{jame06, ande08, ande12} we have developed a statistic, termed the NCR value (for Normalised Cumulative Rank), that quantifies the strength of association between samples of SNe and ongoing star formation as traced by the strength of H$\alpha$ emission at the location of each SN within its host galaxy.  Unlike other studies \citep[e.g.][]{vand92, vand96, bart94, crow13} which have used the distance of each SN from the centre of what was defined to be the nearest HII region, the NCR method is based on fluxes in individual pixels, and specifically on where the SN-containing pixel is placed within a cumulative distribution of all pixels from the H$\alpha$ image.  This effectively incorporates both positional and intensity information, and is normalised to a value between 0 and 1, which represents the fraction of the total SF in the galaxy resulting from regions of lower H$\alpha$ surface brightness than the pixel containing the SN.  Given the likely correlation between HII region age and H$\alpha$ emission strength, shown for example  by \citet{kunc13}, based on the Starburst99 \citep{leit99} and GALEV \citep{ande03} population synthesis models, high NCR values indicate recent SF at the SN location, and hence give evidence for short-lived, high-mass SN progenitors. Thus an NCR value of 1.000 requires not only that a SN lies directly on top of an HII region, but that that region be the most intense \citep[e.g.][]{ande03} within the host galaxy. 

Of the 26 SNIIn contained within our sample, we can calculate the NCR value for 24. SN2010jl does not have a valid NCR value as the event is still bright in H$\alpha$ showing signs of long-lived emission more than two years post explosion. This is also the case with SN2005ip which has well observed long-lived H$\alpha$ emission \citep{stri12} still observable more than 7 years post explosion. These long-lived events are obvious in the H$\alpha$ observations as they still appear as point sources. NCR calculations are therefore only applied to host galaxy images which were taken pre-explosion, or when the H$\alpha$ profile at the site of the transient no longer appears as point source. We are therefore confident that the NCR values presented in this paper reflect the true environment of the events rather than the event itself.

Of all of the CCSN subtypes, SNIIn are least associated with the H$\alpha$ emission and SNIc the most associated according to \citet{ande12}. These results have been updated by using the SNIIn and Impostor samples discussed here, and can be seen in Figure 5. Figure 5 clearly shows the differences in the distributions of each subtype within the host galaxies found in this analysis. The SNIc population almost directly traces the H$\alpha$ emission, represented by the dashed black line, whilst the SNIIn population most closely resembles that of the SNIIPs. The Impostor population is even less associated with H$\alpha$ emission than type Ia SNe. The SNIIn population are the least associated with on-going star formation of all CCSN-subtypes, whereas SNIc are the most associated. The close association of SNIc with H$\alpha$ emission has been found independently by \citet{kang13}. If one follows the general consensus that both SNIc and SNIIn have very high mass stellar progenitors then these results are puzzling. \citet{ande12} do however find that the SNIIn population traces recent star formation in the form of near-UV emission. 

\begin{figure}
 \centering
  \includegraphics[width=0.9\columnwidth,angle=0]{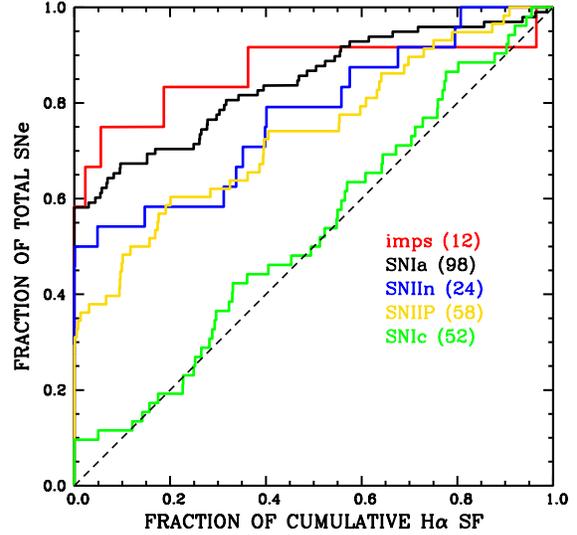}
  \caption{NCR cumulative distributions in terms of H$\alpha$ emission tracing on-going star formation, of interacting transients compared to SNIIP and SNIc, as likely extremes of the CCSN progenitor mass range, and SNIa for reference.}
\end{figure}

We are able to place constraints on the mass of the stars tracing both H$\alpha$ and near-UV using the data given in \citet{goga09}, who find that the age of a HII region is generally less than 16~Myr, compared to star formation giving rise to near-UV emission which is between 16 and 100 Myr. \citet{goga09} use these ages, and apply the isochrones of \citet{mari08} to find the turnoff mass related to these ages. They find that an age of 16 Myr corresponds to a turn off mass of $\sim$12M$_{\odot}$ and therefore H$\alpha$ emission traces stars more massive than 12M$_{\odot}$, and near-UV stars below this limit. This implies that the progenitors of SNIc, which accurately trace the H$\alpha$ emission are more massive than SNIIP and SNIIn which trace the near-UV. The comparison to the Impostors, SNIIP and SNIc can be seen in Figure 6.

\begin{figure}
 \centering
  \includegraphics[width=0.9\columnwidth,angle=0]{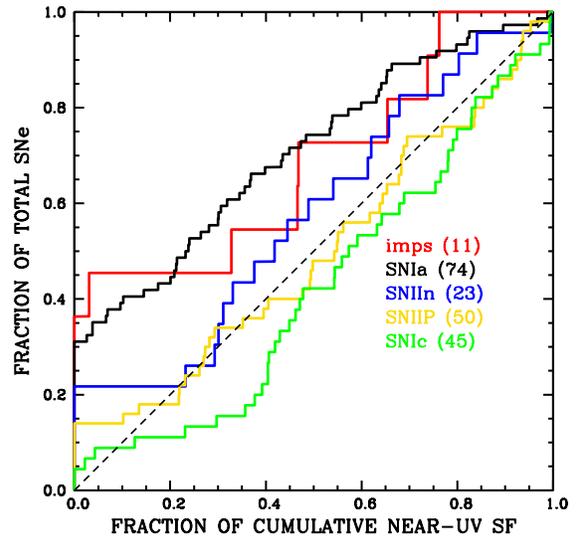}
  \caption{NCR cumulative distributions based on archival {\it GALEX} near-UV images tracing recent star formation, of interacting transients compared to SNIIP and SNIc, as likely extremes of the CCSN progenitor mass range, and SNIa for reference.}
\end{figure}

The NCR values for the SNIIn and SN Impostor samples are given in Table 8, where the following SNIIn are being presented for the first time: SN1996cr, SN2001fa, SN2003G, SN2005db, SN2005gl and SN2008J. The average NCR values for the classes are: NCR$_{SNIIn}$ = 0.225 $\pm$ 0.058 and NCR$_{Impostor}$ = 0.133 $\pm$ 0.082, with the interacting transient class combined to give an average NCR value of 0.194 $\pm$ 0.047. These are consistent with the samples presented in \citet{ande12} and compare to their average NCR value for type IIP SNe of 0.264 $\pm$ 0.039 and 0.469 $\pm$ 0.040 for type Ic SNe. The sample of \citet{kang13} contains seven SNIIn, of which six are within the sample presented here, however, two of these events have been placed within our Impostor group (SN1997bs and SN2001ac). The H$\alpha$ pixel statistic technique used in \citet{kang13} is identical to that used here. For the 6 events we have in common our average NCR value is 0.144 $\pm$ 0.130, compared to 0.086 $\pm$0.054 in \citet{kang13}, which are in close agreement.  

Table 11 presents the KS test results on the distributions of the various NCR values for the subtypes presented in Figures 5 and 6. This shows that the distribution of SNIIn, Impostors, and the classes combined, have a probability of less than 0.1 per cent of tracing the H$\alpha$ (or flat) distribution (represented by the black dashed line in Figure 5). Unsurprisingly this means that the Impostor, and interacting transient, progenitor populations have a probability of less than 0.1 per cent of being drawn from the same population as SNIc progenitors, which follow the H$\alpha$ emission. The probability of the SNIIn population and the SNIc population having the same parent distribution is $\sim$0.2 per cent.  

\begin{table*}              
  \begin{minipage}{160mm}         
    \centering                         
    \caption{Kolmogorov-Smirnov test results for the distributions of H$\alpha$ and near-UV NCR values.}
    \begin{tabular}{lccccc}
      \hline 
      & \multicolumn{2}{c}{H$\alpha$} & \multicolumn{2}{c}{near-UV} \\
      \hline
      & D$_{H\alpha}$ & P$_{H\alpha}$ & D$_{near-UV}$ & P$_{near-UV}$ \\
      \hline                                        
      Impostors-SNIIP & 0.373 & 0.100 & 0.327 & 0.235 \\            
      SNIIn-SNIIP & 0.179 & 0.619 & 0.189 & 0.582 \\
      Interacting-SNIIP & 0.234 & 0.166 & 0.227 & 0.216 \\
      Impostors-SNIc & 0.633 & 0.000 & 0.390 & 0.102 \\
      SNIIn-SNIc & 0.442 & 0.002 & 0.301 & 0.103 \\           
      Interacting-SNIc & 0.491 & 0.000 & 0.322 & 0.027\\           
      Impostors-flat & 0.695 & 0.000 & 0.423 & 0.028 \\      
      SNIIn-flat & 0.498 & 0.000 & 0.216 & 0.214 \\ 
      Interacting-flat & 0.556 & 0.000 & 0.264 & 0.016\\
      SNIIP-flat & 0.392 & 0.000 & 0.139 & 0.294 \\
      SNIc-flat & 0.107 & 0.622 & 0.202 & 0.051 \\
      \hline                                                               
    \end{tabular}                                                        
  \end{minipage}                                                        
\end{table*}                                                         

The association of SNIIn and SN Impostors with host galaxy star formation (in terms of the NCR statistic) is the least of all the CCSNe subtypes, and more consistent with tracing UV emission, suggesting lower mass progenitors. 

\subsection{Dust Embedded Star Formation}
It is possible for star formation regions to be deeply dust embedded, and therefore any H$\alpha$ emission to be absorbed, which could lead to underestimated NCR values, but such HII regions will have strong IR emission. The {\it Spitzer} 24 $\mu$m images are ideal for detecting this emission and as such we have conducted a search on the {\it Spitzer} archive for any host galaxies contained within our interacting transient sample. Of the 36 transient sites for which we have a measured NCR value, 17 are contained in the {\it Spitzer} archive. Of these events we find that there are no observable star formation regions present which are not also detected in the H$\alpha$ observations.

\section{Selection Effects}
The results presented in Sections 2 and 3 must be analysed in terms of any selection effects present in the samples. The samples were taken from the Asiago\footnote{http://graspa.oapd.inaf.it/cgi-bin/sncat.php} and IAU\footnote{http://www.cbat.eps.harvard.edu/lists/Supernovae.html} SN catalogues, and hence contain SN detected in both targeted and untargeted surveys, each with their own inherent biases. The selection effects involved in detecting the Impostor sample are very different to those for the SNIIn subtype as the Impostor explosions are intrinsically much less luminous. As a result the two classes will be analysed individually in this section.

The absolute magnitudes of transient events have been calculated using the discovery apparent magnitudes (taken from the Asiago catalogue and \citealt{lenn12}), and converted using the applicable distance given in Table 2.

\subsection{Host Galaxy Analysis}

Although SNIIn events can span a wide range of absolute magnitudes, many of the most luminous SN events known are of this class. The average V-band absolute magnitude (at discovery) of the SNIIn within this sample is --16.96 (standard deviation = 2.11), compared to the average (discovery) absolute magnitudes for SNIIP of --16.17 (standard deviation = 1.38) and SNIc of --16.46 (standard deviation = 1.07). The SNIIn average discovery magnitude is therefore comparable to other SN types, and not sufficiently `faint' to suggest loss due to the increased luminosity towards the galaxy nuclei, and indeed it is slightly brighter than the SNIIP and SNIc samples. The SN Impostor sample does however have a significantly fainter discovery magnitude with an average of --13.82 mag (standard deviation = 1.51) indicating that a significant fraction of these events could be missed within dusty regions, high surface brightness regions, and at increased distance. This is reflected by the average distance to the host galaxies in this sample being only 14.7 Mpc, compared to the comparison sample values: SNIc - 44.7 Mpc and SNIIP - 25.4 Mpc. The SNIIn population occupies an almost identical redshift range to the SNIc sample, with the average host galaxy lying at a distance of 44.5 Mpc. 

In Table 12 the proportion of each CCSN subtype within the faint class is presented. The probability of a SNIIn event being classed as faint (M$>$--16) is not significantly higher than for other subtypes. The fraction of faint SNIIn ($\sim$37~per cent) is very similar to that of SNIc ($\sim$39~per cent). The class is about half as common as the SNIc subtype, comprising $\sim$7~per cent of all CCSNe, compared to SNIc which make up $\sim$13.5~per cent \citep{li11}. However, with the same fraction of the subtype falling within the faint classification one would not expect to lose SNIIn detections in the central regions where the fraction of detected SNIc remains high in this sample.

\begin{table*}
  \small
  \begin{minipage}{160mm}
  \centering
  \caption{Probability of a `faint' (M$>$--16) CCSN detection being of a particular subtype, based on LOSS SN luminosity functions and rates.}
  \begin{tabular}{ccccc}
    \hline 
    CCSN subtype & \% of all CCSNe & \% of subtype with M$>$--16 & \% CCSNe with M$>$--16 & \% M$>$--16 CCSNe \\
    \hline 
    IIP & 52.0 & $\sim$52 & $\sim$27.0 & 64.1 \\
    IIL & 7.5 & 0 & 0 & 0 \\
    IIn & $\sim$7.0 & $\sim$37 & $\sim$2.6 & 6.1 \\
    IIb & 9.0 & $\sim$24 & $\sim$2.2 & 5.1 \\
    Ib & $\sim$5.0 & 0 & 0 & 0 \\
    Ic & $\sim$13.5 & $\sim$39 & $\sim$5.3 & 12.5 \\
    pec Ibc & $\sim$6.0 & $\sim$86 & $\sim$5.2 & 12.5 \\
    \hline
  \end{tabular}
\end{minipage}
\end{table*}

The diversity in light curve evolution within the SNIIn class means that any single event may have a short decay time equivalent to that of an SNIIb (\citealt{li11} find that SNIbc have even shorter decay times, though \citealt{arca12} find the IIb and Ibc populations to have similarly fast decays). However on average the SNIIn display longer decay times than other CCSNe \citep{kiew12}. Therefore, this allows for detection over a longer time frame than other CCSNe explosions, and so these events should be easier to detect.

Figure 7 displays the distribution of absolute discovery magnitudes of each SN within this sample against their Fr({\it R}) value. In terms of both apparent and absolute magnitude, SNe that are fainter (at discovery) than both the average SNIIn magnitudes, and the SNe corresponding to the faintest SNIIn at discovery within this sample, are detected within the central regions of their hosts. Therefore there is no reason to suspect that faint SNIIn at these magnitudes would not be detected when all other CCSNe events are. Impostors are generally fainter than all other SN types (both at peak and at discovery), although on the whole equivalently faint type IIP and IIn are also detected. It is likely that the lack of Impostor detections within the central regions of the host galaxies is due to the class's fainter average absolute magnitude. The increased extinction towards the central regions of hosts may be sufficient to lose the ability to detect these intrinsically faint events.

It has been suggested that central type IIn SNe could be mistaken for active galactic nuclei (AGN) of type Seyfert 1 \citep[e.g.][]{terl88,terl92}, and this could lead to many being rejected as AGN in SN surveys. The similarity of some AGN spectra and SNIIn, for example QSO 3C 48 and SN1987F \citep{fili89}, are remarkable, and there have been several examples of suspected AGN emission actually being driven by SNIIn explosions (e.g. NGC4151; \citealt{aret94} and NGC7582; \citealt{aret99}). However, the consensus in the literature seems to suggest that this could only apply to low-luminosity AGN with low resolution spectra \citep[e.g][]{fili89,ulri97}. SNIIn do not replicate many other AGN features such as the UV excess and x-ray variability, however, during the course of a SN survey this detail may not be studied before rejection. There is therefore a chance that some central SNIIn have been detected but incorrectly categorised as AGN. The number of events for which this occurs should be small as most SNIIn do not mimic the characteristics of AGN sufficiently (only long-lived, SN1988Z-like events have been found to reflect AGN emission), and most AGN do not look like SNIIn explosions.

\begin{figure}
 \centering
  \includegraphics[width=0.8\columnwidth,angle=270]{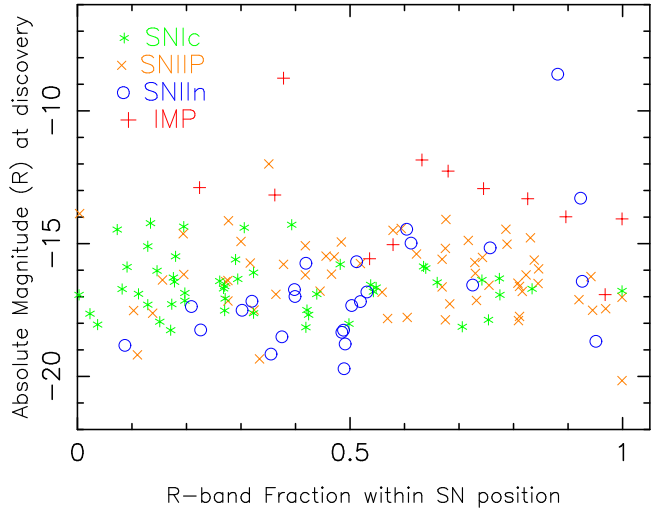}
   \caption{Fractional {\it R}-band light distribution of each SNIIP, SNIIn, SNIc and Impostor in this analysis, against its absolute magnitude at discovery.}
\end{figure}

Figure 8 shows the Fr($R$) distribution of the SN Impostors only, along with a fit to the data shown by the dashed line, and the average absolute discovery magnitude shown by the solid line. The fit to this data has a gradient of --3.8 $\pm$ 2.3, consistent with random scatter, however it is interesting to note that the correlation is counter-intuitive, with the fainter events lying closer to the host galaxy nucleus, given by smaller fractional light values. There are however, no events seen within the central 20 per cent of host galaxy light.

\begin{figure}
 \centering
 \includegraphics[width=0.8\columnwidth,angle=270]{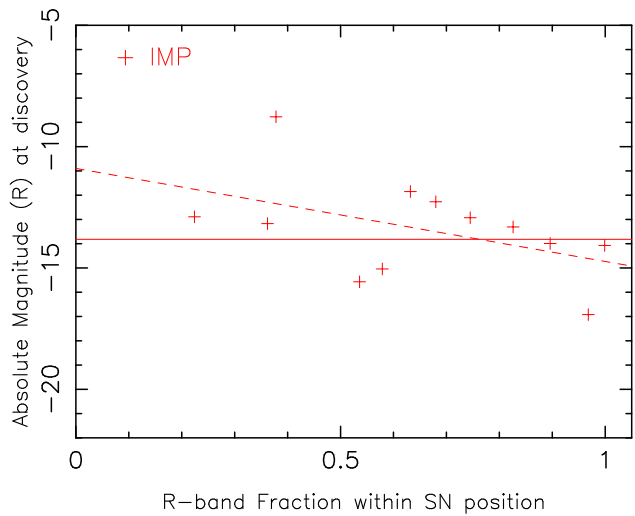}
 \caption{Fr({\it R}) distribution of Impostors with respect to their absolute magnitude at discovery. The solid line represents the average of the group, and the dashed line the fit to the data.}
\end{figure}

The distribution of discovery magnitudes as a function of Fr({\it R}) values for the SNIIn sample is presented in Figure 9, where again the average value is represented by the solid line, and the dashed line shows the fit to the data. The gradient of this fit is statistically significant at 4.9$\pm$1.5, indicating that intrinsically fainter SNIIn are not seen within the central regions of their hosts. Though this could be interpreted as evidence for a selection bias against finding faint SNIIn in the central regions it is interesting to note that the distribution shows that only one bright SNIIn is found outside Fr({\it R})$\simeq$0.5, i.e. outside the effective radius of the host galaxy, defined as the radius within which half of the galaxy's light is emitted \citep{deVau48}). The outer regions of galaxies appear not to host any luminous SNIIn, a result which cannot be driven by any selection bias. 

Faint SNIIn should not be lost within the central regions of their hosts when equally faint SNIIP and SNIc events are detected \citep{habe12}. Supporting this conclusion, the average recession velocity of the SNIIn host galaxies is $\sim$2600 kms$^{-1}$ (range 434-5260 kms$^{-1}$) and at this redshift one would not expect to lose a substantial fraction of SNe. The findings of \citet{mann07} suggest that for this redshift range we would expect to lose $\lesssim$10~per cent of all CCSNe, while \citet{matt12} suggest a slightly higher value of $\sim$19 per cent. 

Therefore it does not seem likely that the absence of SNIIn within the central regions of galaxies, and the lack of association with H$\alpha$ emission can be driven by selection effects. The small sample size, however, must instill caution when interpreting these data.

\begin{figure}
  \centering
  \includegraphics[width=0.8\columnwidth,angle=270]{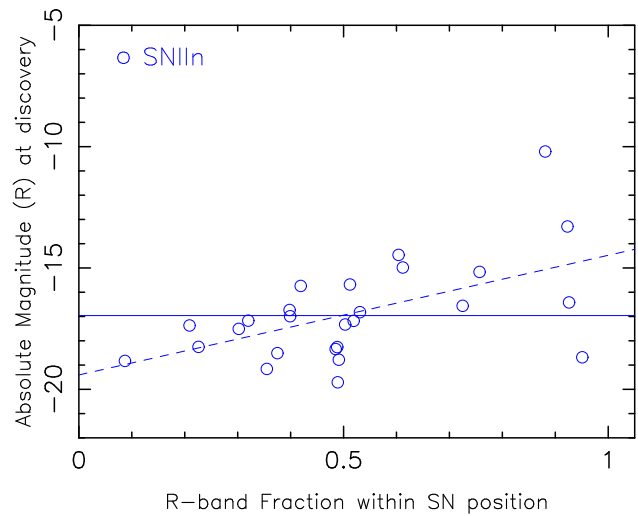}
  \caption{Fr({\it R}) distribution of type IIn SNe with respect to their absolute magnitude at discovery. The solid line represents the average of the group, and the dashed line the fit to the data.}
\end{figure}

\subsection{NCR Analysis}
Selection effects may affect the NCR analysis presented in the last section for two main reasons; the particular subtype of the explosion could be harder to detect against a bright HII region, and at increasing distances it could become harder to detect all explosions against bright HII regions, leading to lower NCR values. A comprehensive analysis of these selection biases is presented in \citet{ande12}. In this paper we concentrate on comparing the interacting transient population with SNIIP and SNIc which should represent the most extreme ends of the CCSNe progenitor mass distribution. The luminosity distributions of the SNIIP, SNIIn and SNIc classes have been discussed and indicate that incompleteness levels and selection effects should be similar for all three types. To specifically test whether any selection bias is present in the NCR analysis presented here with respect to redshift, the sample was split into two bins, a low- and high- redshift bin. For the SNIIn sample the cut-off between these bins was taken as 40 Mpc, dividing the sample into equal halves. The mean NCR value for the low-redshift half is 0.146 $\pm$ 0.079, and for the high-redshift half it is 0.304 $\pm$ 0.081, and therefore no statistically significant difference is present as function of recession velocity. However, it is interesting to note that the higher redshift bin shows slightly more association with H$\alpha$ emission than the low redshift bin.

The selection effects are much more complicated for the Impostor sample, where the peak luminosities are much fainter and hence the transients are less likely to be detected against a bright HII region. For this reason they also represent a much more local sample of events, with the cut off between the equally populated high- and low- redshift bins much lower at $\sim$9 Mpc. Here the low-redshift sample have a mean NCR of 0.101 $\pm$ 0.060, the high-redshift mean is 0.165 $\pm$ 0.160. Again there is no significant redshift dependence here but this is limited by the very small sample in terms of numbers and redshift range. Hence it is much more difficult to draw conclusions for this particular class of object. 

\section{Discussion}

\subsection{Type IIn}
The sample of SNIIn studied here appear to have very different host environments to SNIc, which literature would suggest have the most massive progenitors \citep[e.g.][]{ande12}. If the progenitors of the SNIIn class are all LBV-like, such as that of SN2005gl \citep{galy09}, we would expect to see a similar distribution within the host galaxies, and a higher degree of association with H$\alpha$ emission.
In fact, \citet{stri12} describe at least three main groups within the SNIIn class defined by their peak absolute magnitudes and light curve evolution. These are an ultra-luminous group with M$_{V}$ $\sim$ --22; the SN1988Z-like group with slow light curve evolution (several years); and SN1994W-like events which reach peak magnitudes of $\sim$ --18 and whose light curves display a long ($\sim$100 day) plateau followed by a sharp decline. 
It is not possible to classify all of the SNIIn within our sample into these main groups due to a lack of information in the literature, but where possible the events have been grouped and are discussed below. For those individual SN events which have relevant information in the literature, we will discuss this to find any indications of the potential progenitors of these explosions.

\subsubsection{1988Z-like Events}
SN1988Z was one of the initial SNII events to be separated out into a new class of transient, which became known as type IIn, by \citet{schl90}. This object had a slow light curve evolution, remaining bright for over three years \citep{tura93}. These long-lived events can have a range of peak magnitudes, which have been estimated to fall in the region between --17 and --19 \citep{stri12}. Within our SNIIn sample the following events have been grouped into this class:

{\bf SN1987F:} This SN was part of the initial subtype classification by \citet{schl90} and was discussed extensively in \citet{tura93} as being the only event at that time to have a light curve comparable to SN1988Z. \citet{fili89} quotes an {\it R}-band light curve decay of $\sim$0.005 mag day$^{-1}$ for SN1987F. \citet{kiew12} also find the event to be comparable in both spectroscopic and photometric evolution with SN2005cp, another slow declining event ($\sim$0.014 mag day$^{-1}$ in the V-band; \citealt{kiew12}) with a peak V-band magnitude of --18.2. The peak V-band magnitude of SN1987F was --18.3 \citep{fili89}, making these events remarkably similar.

{\bf SN1995N:} This SN was so long-lived that even its initial discovery \citep{poll95} was made approximately 10 months after the explosion, based upon comparison of the initial observed spectrum with that of the well-observed SN1993N \citep{bene95}. The event was followed up spectroscopically by \citet{fran02} who found emission from the explosion almost 2000 days post explosion. The light curve information presented in \citet{li02} shows the slow decline of this object ($\sim$0.0008 mag day$^{-1}$ in the V-band; \citealt{li02}), and the longevity of the emission. This explosion has been well studied in the infrared regime and much analysis conducted into the dust properties of the explosion. A full discussion of this is beyond the scope of this paper but \citet{gera02} and \citet{fox09} give detailed discussion and analysis.

{\bf SN1996cr:} This event was discovered in archival data more than ten years after the initial explosion \citep{baue07} and the subsequent follow up spectroscopy confirmed the SN to be of type IIn \citep{baue08}, the source remaining bright for over ten years. Hydrodynamic modelling of the light curves and multi-epoch spectra for the event, presented in \citet{dwar10}, match the observations, and these authors conclude that it is unlikely, though not impossible, that the progenitor of the explosion was an LBV star, more likely having a blue supergiant, or WR origin. The late epoch of the classification spectrum for this event leaves uncertainty over this classification.

{\bf SN2005ip} is one of the most long-lived SNIIn known, remaining bright more than 6.5 years post explosion \citep{stri12}. It was first classified as a type IIn by \citet{smit09} who described the event as a type IIL, following its initial steep decline over the first 160 days, followed by interaction with a clumpy CSM, providing the IIn-like spectra. \citet{smit09} concluded that the progenitor was an extreme red supergiant with a mass between 20 and 40 M$_{\odot}$. Contrary to this, \citet{fox09,fox10} looked at the IR fluxes up to 900 days post-explosion to place constraints on the dust formation and surroundings of the SN. They concluded that the high IR fluxes were equivalent to those expected from the interaction of the SN with a pre-existing shell formed by an LBV outburst prior to the eventual SN explosion. Most recently \citet{stri12} conducted a comprehensive analysis of the UV, optical and near-IR evolution of SN2005ip, both spectroscopically and photometrically, concluding that the late time emission from the event was most likely caused by interaction with a clumpy CSM as described originally in \citet{smit09}. As for the SNIIn class as a whole, \citet{stri12} also pointed out the diversity within the SN1988Z-like group, depending upon the mass-loss history of the progenitor stars.

{\bf SN2010jl}: This was a luminous SNIIn event that reached a peak magnitude of M$_{V}$$\sim$--20 \citep{smit11}. Initial spectra classified the event as a type IIn \citep{bene10} with a slowly declining light curve ($\sim$0.009 mag day$^{-1}$ in the V-band, over the first 90 days; \citealt{zang12}). \citet{smit11} reported an analysis of archival pre-explosion Hubble Space Telescope (HST) imaging and concluded that the progenitor star must have a mass of at least 30 M$_{\odot}$, and was likely to have been an LBV. SN2010jl showed an infrared excess soon after explosion and the follow up IR observations presented by \citet{andr11} indicated that a large mass loss event must have taken place prior to explosion, also consistent with an LBV progenitor. The event remains bright, and observations more than 500 days post-explosion indicate that the surrounding CSM is massive and possibly made by an enormous mass loss event (30-50 M$_{\odot}$) just decades prior to explosion \citep{zang12}.  

No information is available in the literature, outside of the initial classification circulars, for {\bf SN2000cl} which \citet{stat01} designated as a type IIn, noting that the spectrum resembled SN1988Z at a similar epoch.

The event {\bf SN2003dv} was classified as a type IIn by \citet{kota03} who described the spectrum as similar to that of SN1998S. However, from photometric follow-up, \citet{li11} described the event as a slow-declining type IIn, similar to SN1988Z. It is in fact used as their template slow-declining type IIn.

The SN1988Z-like events described above have an average NCR index of 0.248 $\pm$ 0.111 (excluding SN2005ip and SN2010jl which are still bright in our observations), and an average Fr({\it R}) value of 0.589 $\pm$ 0.112. 

\subsubsection{1994W-like Events}
SN1994W is the prototype of a group of SNIIn which have a decline of months rather than years \citep{soll98} and often exhibit light-curve plateaus, followed by a drop in luminosity \citep[e.g.][]{kank12}. Although \citet{tadd13} claim that only SN2009kn \citep{kank12} and SN2011ht \citep{maue13b} are confirmed to belong to this group, we use information available in the literature to supplement it with our events which most closely resemble SN1994W. These are discussed below.

{\bf SN1987B:} This SN comprised part of the initial SNIIn classification proposed by \citet{schl90} but \citet{schl96}, a later discussion paper, commented that the spectrum was more similar to SN1994W. The peak V-band magnitude of this event was $\sim$--18.4 (converted from \citealt{schl96}) and the authors also found the event decayed much more quickly than SN1988Z. SN1987B also only exhibited a $\sim$30 day plateau.

{\bf SN1994W:} This event is the prototype of the group but remains the subject of much controversy. After its detection in July 1994 \citep{cort94}, the SN was classified as a type IIn when narrow lines became visible in the spectrum. The classification was changed to a type IIP when the plateau in the light curve was observed, followed by a steep drop in both the B- and V-bands \citep{tsve95}. More recently \citet{dess09} have questioned whether this was a core-collapse event at all, citing the lack of broad lines in time sampled spectra, and the low $^{56}$Ni yields, as evidence that the event may actually have been the result of an interaction between two ejected shells of material, without a core-collapse. Interestingly, we find the environment of this event to have a high association to both H$\alpha$ and near-UV emission. For the purpose of discussion we refer to this event within its name-sake group.

{\bf SN1994Y:} The spectrum of this event is described as having Balmer emission lines without P-Cygni profiles, along with Na features, over a blue continuum \citep{cloc94}. Based on this description, \citet{schl96} described the event as behaving similarly to SN1987B. \citet{ho01} presented multicolour light curves for this event which reveal a short plateau phase ($\sim$30 days in V-band) around maximum, indicating a low ejecta mass in the explosion, similar to SN1994W. 

{\bf SN2005db} had little real-time follow-up, with the spectral classification being made by \citet{blan05} who describe the spectrum as blue and featureless save for some narrow Balmer emission lines. \citet{kiew12} describe the photometric evolution of the event as having a rapid rise and a broad peak, followed by a 30-40 day plateau. It is for this reason we tentatively group the event into the SN1994W-like class of events.

The four events described above have an average NCR index of 0.298 $\pm$ 0.190 and an average Fr({\it R}) of 0.582 $\pm$ 0.129.

\subsubsection{Linearly-declining Events}
Several of the events within our sample show more linear light curve decays, with no long-lived emission like that seen in the SN1988Z-like events, and no plateaus, as are seen in the SN1994W events.

{\bf SN1999el} has a relatively fast light curve decay ($\sim$0.05 mag day$^{-1}$ in the V-band), and there is no plateau visible in the data presented in \citet{dica02}. These authors speculate that the explosion is more reminiscent of an SNIIL which has undergone mass loss just prior to explosion. Based on these observations it is not clear which group described above that event would slot into, if any of them. 

{\bf SN2000P} also shows a rapid decline ($\sim$7 mag in 250 days) but with no plateau features, which is described in \citet{li02}. The initial spectra, which define the event as a type IIn, showed the blue continuum and narrow emission lines \citep{jha00a,capp00} although \citep{jha00} attributed the narrow lines in their initial spectrum to the host galaxy and not the SN.

These two events have an average NCR value of 0.024 $\pm$ 0.024 and an average Fr({\it R}) value of 0.348 $\pm$ 0.028.

\subsubsection{Fast-declining Events}
These events are also known as SN1998S-like, with the prototype being one of the most well studied SNIIn (see \citealt{kiew12} for a full discussion). SN1998S had a fast rise time ($<$ 20 days), and one of the fastest decays ever observed ($\sim$0.05 mag day$^{-1}$; \citealt{kiew12}). Analysis of the pre-explosion mass loss rates yields a value around 10$^{-3}$-10$^{-4}$ M$_{\odot}$y$^{-1}$ \citep{lent01,pool02}.

{\bf SN2003G} was classified as a type IIn by \citet{hamu03iau} as it had a blue continuum and strong H$\alpha$ emission. It was described as spectroscopically similar to the event SN1994W, however, from their photometric follow-up, \citet{li11} describe the event as a type IIn fast decliner, similar to SN1998S.

The only information available for {\bf SN2001fa} is contained in one circular \citep{fili01iau} which describes the spectrum as a type IIn, reminiscent of SN1998S at early times \citep{leon00}.

{\bf SN2005gl} is so far the only type IIn supernova to have had a progenitor detection using archival HST imaging \citep{galy07}, confirmed by late time observations showing the disappearance of the candidate progenitor \citep{galy09}. Although this is the only progenitor to have been confirmed in the group, the `class' of SNIIn it falls into is unclear \citep{stri12}. The event peaked at a magnitude of $\sim$--17 \citep{galy07,smit10} and the photometric evolution showed a steep rise followed by a fast decline ($\sim$0.045 mag day$^{-1}$; \citealt{galy07}), giving a possible similarity with SN1998S \citep{smit10}. The spectral evolution for the event remains confusing, with the initial spectrum clearly showing it to be a type IIn \citep{galy09}, but with the defining narrow lines quickly disappearing. \citet{li07} liken the evolution to that of a type IIL, and \citet{smit10} describe the spectrum at 2 months as similar to that of a type IIP. Regardless of this uncertainty, several independent analyses \citep{smit10,dwar11} agree that the inferred mass loss rates prior to the SN explosion indicate an LBV progenitor, as described in \citet{galy09}, having a mass at least as great as 50~M$_{\odot}$. However, only one pre-explosion image exists for this event and if the progenitor is an LBV star, there is no guarantee that it was not in outburst when this observation was taken. Should the star have been in outburst, \citet{groh13b} estimate the progenitor mass to be closer to 20-25~M$_{\odot}$. Although we find an NCR value of 0.000 for this event it should be noted that the H$\alpha$ observation from which this value was obtained has a low signal-to-noise ratio, and therefore the error on this value may be large. Despite this the position of SN2005gl is between, and not upon, bright HII regions. 

These 3 events have an average NCR value of 0.049 $\pm$ 0.049 and an average Fr($R$) value of 0.343 $\pm$ 0.092.   

\subsubsection{Thermonuclear `IIn'}
The first event to be defined as a thermonuclear IIn, or an SNIa-CSM, was SN2002ic \citep{hamu03}, which subsequently led to the re-classification of several other events, notably SN1997cy and SN1999E. These events all have a spectrum which seems to have two components, the underlying spectrum of an SN1991T-like SNIa, though diluted in nature, with a smoothly varying continuum thought to be caused by the interaction of the shock wave with a hydrogen-rich CSM \citep[e.g.][]{tadd12,silv13}. This also produces narrow hydrogen emission lines which are seen in the spectra. These events often also show silicon features, which are not seen strongly in core-collapse events \citep{hamu03}. 

The initial classification spectrum of {\bf SN2008J} indicated that it was reminiscent of an SNIIn with strong hydrogen lines, complete with narrow components. However, it also noted that the event was similar in nature to the SNIa-CSM event SN2005gj according to the supernova identification code used \citep{blon07}. Extensive analysis presented in \citet{tadd12} and \citet{silv13} confirm this SN to be part of the Ia-CSM group, showing the spectroscopic and photometric evolution to resemble SN2005gj and SN2002ic. The V-band magnitude of the event at peak is also much brighter than most SNIIn events, at --20.3 mag \citep{tadd12}. This is the only thermonuclear event in our sample and has an NCR index of 0.807 and an Fr(R) value of 0.087. The NCR value for this event seems high, with only $\sim$4 per cent of the `normal' SNIa population falling above 0.800 \citep{ande12}. It may be that this is an intrinsically normal SNIa which has passed into a HII region prior to exploding, or that this is an intrinsically unusal SNIa event. It should also be noted that the observation used throughout this analysis was taken within five years of the explosion. Given the longevity of some SNIIn interactions \citep[e.g.][]{stri12} this may be too close to the explosion epoch, and hence the H$\alpha$ fluxes and NCR values may be boosted by emission from the SN itself. We are confident that our sample does not contain a significant fraction of these events, which are described as having magnitudes in the range --19.5$>$M$>$--21.6 \citep{silv13,lelo13}, and hence are much brighter than most of the events in the present study.

\subsubsection{Other IIn Events}
For  16 of the SNIIn in this sample, there is no significant information in the literature other than some notes in the circulars. These SNe are: {\bf SN1993N}, which was noted to have a spectrum similar to SN1988Z \citep{fili94} at the stage when the IIn class was still largely unclear, but with no known long-lasting emission. {\bf SN1994ak} has an initial classifying spectrum showing it clearly as a type IIn \citep{garn95}. {\bf SN1996bu} only has initial discovery circulars where the spectrum was clearly that of a SNIIn, with a blue continuum and narrow Balmer emission lines \citep{naka96}. {\bf SN1997eg} had spectroscopic follow up, presented in \citet{sala02}, which clearly showed the features of a type IIn explosion, however, photometric follow up of the event is lacking in the literature and therefore we are unable to define the exact sub-class of SNIIn in which this event lies. The initial circulars of {\bf SN1999gb} give the event a IIn designation after a spectrum revealed its blue continuum and narrow hydrogen features \citep{jha99}. The classification of {\bf SN2002A} was described by \citet{bene02} as having a spectrum dominated by narrow Balmer emission lines, with broader wings. The spectrum of {\bf SN2002fj} was classified as a type IIn, and described in \citet{hamu02iauc} as being dominated by strong H$\alpha$ emission. {\bf SN2003lo} was designated a type IIn by \citet{math04} who describe a spectrum with blue continuum emission and both broad and narrow hydrogen features. Finally, the only information available for the event {\bf SN2006am} is the spectrum, described in \citet{quim06}, which shows both a blue continuum and Balmer lines with narrow components.

\subsubsection{SNIIn Summary}
This sample contains no ultra-luminous events, which are often the basis of papers on the SNIIn class, many of which focus on extreme events \citep[e.g.][]{smit08}. The properties of the sample of SNIIn analysed here are so diverse that it seems unlikely they can all result from a single type of progenitor system. It is likely that the class of objects known as SNIIn is in fact composed of several different progenitor channels. The lack of a fundamental characteristic defining the group, first required by \citet{schl90}, may indicate that the class is not in fact distinct but merely the result of the quality of data and frequency of observations, or another progenitor channel in an environment with high density CSM. The environmental study conducted here sheds doubt on the LBV progenitor channel for the majority of SNIIn events, with the class overall showing little association with star formation as traced by H$\alpha$ emission. A comparison between the average NCR$_{H\alpha}$ and average Fr({\it R}) values for the different SNIIn types can be seen in Table 13 (number of events in brackets), for comparison the average NCR value for SNIIP is 0.250 $\pm$ 0.038, and for SNIc, 0.466 $\pm$ 0.042.\\
 
{\begin{table}
    \small
    \centering
    \caption{Average NCR and Fr({\it R}) values for SNIIn classes}
    \begin{tabular}{l c c}
      \hline
      Type (Number) & NCR$_{H\alpha}$ & Fr({\it R})\\
      \hline
      1988Z-like (7) & 0.248 $\pm$ 0.111 (5) & 0.589 $\pm$ 0.112 (6)\\
      1994W-like (4) & 0.298 $\pm$ 0.190 & 0.582 $\pm$ 0.129\\
      Ia-CSM (1) & 0.807 & 0.087\\
      IIn-L (2) & 0.024 $\pm$ 0.024 & 0.348 $\pm$ 0.028\\
      1998S-like (3) & 0.049 $\pm$ 0.049 & 0.343 $\pm$ 0.092\\
      \hline
    \end{tabular}
  \end{table}
}

\subsection{Impostors}
SN Impostors can only truly be defined as such if the progenitor star is shown to survive the explosion, however, many of the events are not followed up in sufficient detail to confirm whether the progenitor star has disappeared. Many events are designated as Impostors when the peak magnitude is fainter than expected from a true SN explosion. However, the diversity within the Impostor and SNIIn groups mean that this somewhat arbitrary cutoff is not reliable. SN Impostors clearly require more detailed study; a current literature search of the events contained within our sample is given below.

\subsubsection{2008S-like Events}
SN2008S may define a class with much lower mass progenitors ($\sim$10 M$_{\odot}$) than are expected from LBV eruptions \citep{prie08}. These events are often enshrouded in high extinction regions, leading to an IR excess during the eruption. Aside from the prototype of this group, {\bf SN2008S}, which was discussed in the introduction, here we will discuss other Impostor events which are likely to fall into this group.

Although {\bf SN1999bw} has a possible detection in the infrared taken 5 years post-explosion \citep{suga04}, little information is known about the progenitor star. The literature shows a consensus that the event falls into the class of SN2008S-like \citep{thom09, koch12}. \citet{thom09} argue that the low optical peak luminosity, the narrow Balmer line dominated spectrum, and the IR detection, are all consistent with a SN2008S-like, highly extinguished event.

Little information is available in the literature for {\bf SN2001ac} but there seems to be a consensus that the event is most likely to be a SN2008S-like transient, with a spectrum which resembles that of SN1999bw, and a detection in the mid-IR indicative of the presence of dust \citep{thom09,smith11c,koch12}.

{\bf SN2002bu} is also argued to be a SN2008S-like transient with a mid-IR detection \citep{thom09}. The event is, however, unusual, with the early spectral evolution resembling an LBV-like eruption, but the late time evolution mimicking the cooler SN2008S-like events, and photometrically the decline rate of the light curve is also similar to that of the LBV-like Impostor SN1997bs, but with a colour evolution matching that of SN2008S \citep{smith11c}. The low mass range generally adopted for the progenitors for these events (e.g. $\sim$10~M$_{\odot}$; \citealt{prie08}) is extended for this object with \citet{szcz12} claiming that the local environment of SN2002bu favours stars closer to $\sim$5~M$_{\odot}$ than 10~M$_{\odot}$.

\citet{smith11c} suggest that {\bf SN2010dn} closely resembles SN2008S, with identical spectroscopic features including narrow Ca II emission along with the usual Balmer emission lines. The extensive analysis carried out in \citet{smith11c}, and the high resolution spectra presented in that paper, strongly suggest that the event is heavily dust enshrouded and one of the strongest candidates for inclusion in the SN2008S-like group. 

\subsubsection{$\eta$ Carina-type Impostors}
There are several SN Impostors contained within our sample which do not fall under the umbrella of SN2008S-like events, and are usually assumed (or confirmed) to be associated with LBV progenitors with initial masses of at least $\sim$20~M$_{\odot}$. These events will be discussed in this section.

{\bf SN1954J}, also known as NGC2403-V12, is one of the few Impostor events to have had confirmation that the star still survives \citep{smit01}. Optical and IR images were taken of the event almost 50 years post explosion and the remaining star is faint and red, probably due to the extinction by the dust formed in the initial eruption \citep{smit01}. The star has spectral characteristics similar to $\eta$ Carina, and high-resolution images reveal the star to be very massive, with an estimate of $>$25~M$_{\odot}$ \citep{vand05}. A detailed discussion of this SN Impostor, and the mass constraints placed on it by \citet{smit01} and \citet{vand05}, is presented in \citet{koch12}.

{\bf SN1961V} was a controversial event which has divided the SN community as to whether it was a SN Impostor \citep{vand12}, or a true SN event \citep[e.g.][]{smith11c,koch11}. The event was luminous, reaching a peak magnitude of almost --18 mag and appears to be an outlier amongst the SN Impostor group \citep{smith11c}. \citet{vand12} found a source coincident with the SN Impostor which may be the surviving progenitor of the event. If so, this appears to be an LBV with a high mass ($>$60~M$_{\odot}$). For the purpose of this study we have kept SN1961V in the Impostor group, however, even if the event is a true SN it is thought to have a very high mass progenitor ($>$80~M$_{\odot}$; \citealt{koch11}).

{\bf V1-NGC2366} is located within a HII region in the host NGC2366. It first erupted in 1994 and has remained near maximum light ever since \citep{dris01,smith11c}. Detailed modelling of the UV spectrum \citep{peti06} implies that the event is being powered by a supergiant wind rather than an explosion, consistent with a progenitor star of $\sim$20~M$_{\odot}$ \citep{smith11c}.

{\bf SN1997bs} was first classified as a true SNIIn by LOSS \citep{tref97}, but the data presented in \citet{vand00} question this, instead favouring an Impostor event due to the faint peak magnitude of the event (M$_{V}$$\le$--13.8 mag) and a tentative detection on pre- and post-explosion HST images, which the authors claim is consistent with a luminous supergiant star \citep{vand99}. The post-explosion observations of \citet{li02} question the Impostor hypothesis for this event, finding only one marginal detection in three post-explosion images taken approximately 3 years post explosion. Despite this, the consensus in the literature seems to be that SN1997bs was a SN Impostor, with a very luminous LBV progenitor \citep{vand99,smith11c}.

{\bf SN2002kg} is one of the few SN Impostors with a bona fide connection to an LBV through its presence in archival data \citep{weis05}. The event remained near maximum brightness for $\sim$2 years and originated in a region where the most recent epoch of star formation is consistent with the expectations for a progenitor mass $>$20~M$_{\odot}$ \citep{maun06}. The eruption seems to have very little dust associated with it, and the changes in luminosity of the progenitor LBV over time suggest that the outburst was a normal LBV eruption with little mass loss \citep{koch12}.

\subsubsection{Unclear Impostor Events}
The SN Impostor {\bf SN2003gm} is not easily placed in the $\eta$ Carina-like or SN2008S-like groups. \citet{maun06} discovered a source at the location of SN2003gm using pre-explosion HST imaging, and post-eruption follow-up images, consistent with a yellow supergiant. The age of the region in which the source sits was similar to that of the `LBV' Impostor SN1997bs, and corresponded to a mass for the progenitor of $\sim$25~M$_{\odot}$ \citep{maun06}. These authors also point out that the light curve decay is similar to that of `LBV' Impostor SN2002kg. However, \citet{smith11c} present HST imaging taken more than 5 years post-explosion and discover the source to be reddened by dust formation. Claims that the source is unobscured \citep{thom09}, thus excluding the event from the SN2008S-like class, are also questioned by \citet{koch12} who find mid-IR emission at the site, indicative of dust. We find an NCR$_{H\alpha}$ index for this event of 0.000, showing no association with ongoing star formation, but a high association with recent star formation traced by near UV emission, with an NCR$_{UV}$ of 0.468 \citep{ande12}.

{\bf SN2006bv} was designated SN2008S-like by \citet{thom09} due to its apparent similar spectral characteristics and the lack of a strong LBV progenitor detection, however, information in the literature for this event is sparse and the evidence for classing the event as SN2008S-like is speculative. \citet{smith11c} estimate a peak {\it R}-band magnitude of the event of --15.2, making it one of the most luminous SN Impostors. We find the event to have no association to ongoing star formation with an NCR$_{H\alpha}$ index of 0.000; there are no available near-UV images to define the NCR$_{UV}$ index.

Again the SN Impostor {\bf SN2006fp} was designated as SN2008S-like by \citet{thom09} due to the similar spectral characteristics and the lack of a convincing LBV detection. However, the peak magnitude presented by \citet{smith11c} of --15.47 implies this event is a luminous, giant LBV eruption. We find the NCR$_{H\alpha}$ index to be 0.965, but find that the event is not associated with the recent star formation traced by near-UV, with an NCR$_{UV}$ of 0.000 \citep{ande12}. This may imply that the event is still bright as a point source in our H$\alpha$ imaging, taken almost 2 years after the initial detection of SN2006fp.

\subsubsection{Impostor Summary}
Every one of the SN2008S-like transients listed in Section 5.2.1 has an NCR$_{H\alpha}$ index of 0.000 which, despite the low number of events, suggests strongly that the group is not associated with ongoing star formation. The average NCR$_{UV}$ index for the group is 0.252 $\pm$ 0.155, therefore this group are more associated with recent star formation. For comparison our calculated average NCR$_{UV}$ values for the SNIc and SNIIP samples are 0.576 $\pm$ 0.042 and 0.504 $\pm$ 0.045, respectively. In comparison each of the $\eta$ Carina-like Impostors has a positive NCR$_{H\alpha}$ index, with the average of the group being 0.157 $\pm$ 0.077 (still lower than that for SNIIP, and much lower than that for SNIc). The group also show a stronger degree of association with recent star formation with an average NCR$_{UV}$ index of 0.430 $\pm$ 0.168. Although these differences are derived from low number statistics, the implication that the NCR index may provide a distinguishing feature between different types of SN Impostors is encouraging and when sufficiently well studied samples are available such environmental diagnostics may provide crucial information in constraining the progenitor systems. A comparison between the average NCR$_{H\alpha}$, NCR$_{UV}$ and Fr({\it R}) values for the different SN Impostor types can be seen directly in Table 14 (number of events in brackets). For comparison the average NCR$_{H\alpha}$ value for SNIIP is 0.250 $\pm$ 0.038, and for SNIc, 0.466 $\pm$ 0.042.\\
 
{\begin{table}
    \scriptsize
    \centering
    \caption{Average NCR and Fr({\it R}) values for Impostor classes}
    \begin{tabular}{l c c c}
      \hline
      Type (No.) & NCR$_{H\alpha}$ & NCR$_{UV}$ & Fr({\it R}) \\
      \hline
      2008S-like (5) & 0.000 $\pm$ 0.000 & 0.252 $\pm$ 0.155 & 0.665 $\pm$ 0.119\\
      $\eta$ Car-like (5) & 0.157 $\pm$ 0.077 (4) & 0.430 $\pm$ 0.168 (4) & 0.597 $\pm$ 0.144\\
      \hline
    \end{tabular}
  \end{table}
}

\subsection{Overall Results}
The SNIIn class is diverse in nature, and probably does not encompass a single progenitor channel, but rather a combination of systems each showing one common feature. The identification of an event as a true SN or an Impostor is still unclear, and only with confirmation of the disappearance of a progenitor will a thorough investigation of the two groups as separate classes be possible.

The various sub-types of the interacting transient class, discussed in detail within this section, are summarised in Table 15. We indicate potential characteristics of the classes as defined by the analysis and literature searches carried out during the course of this paper, and described throughout. Any characteristics drawn for a class should therefore be taken with some degree of caution as this is still a very active area of research.

{\begin{table*}
    \small
    \begin{minipage}{160mm}
      \centering
      \caption{Description of interacting transient classes}
      \begin{tabular}{l l l}
        \hline
        Class & Characteristics & Examples from this study\\
        \hline
        SNIIn: Ultraluminous & \parbox[t]{5.5cm}{Peak V-band magnitude $<$ --22\\} & ---\\
        SNIIn: 1988Z-like & \parbox[t]{5.5cm}{Slow LC evolution (several years);\\ --17 $<$ Peak mag $<$ --19\\} & 1987F, 1995N, 1996cr, 2000cl, 20036dv, 2005ip, 2010jl\\
        SNIIn: 1994W-like & \parbox[t]{5.5cm}{LC decay period of months, with plateau\\ followed by drop in luminosity\\} & 1987B, 1994W, 1994Y, 2005db\\
        SNIIn: Ia-CSM & \parbox[t]{5.5cm}{Underlying thermonuclear spectrum \\similar to SN1991T\\} & 2008J \\
        SNIIn: Linear & \parbox[t]{5.5cm}{No long-lived emission, linear LC \\decay with no plateau\\} & 1999el, 2000P\\
        SNIIn: 1998S-like & \parbox[t]{5.5cm}{Fast LC rise ($<$20 days) and decay\\ ($\sim$ 0.05 mag day$^{-1}$)\\} & 2003G, 2001fa, 2005gl\\
        IMP: 2008S-like & \parbox[t]{5.5cm}{Probable progenitor mass $\sim$10M$_{\odot}$, in high\\ extinction region giving an IR-excess\\} & 2008S, 1999bw, 2001ac, 2002bu, 2010dn \\
        IMP: $\eta$-Car-like & \parbox[t]{5.5cm}{LBV progenitor with probable initial\\ mass $>$ 20 M$_{\odot}$\\} & 1954J, 1961V, NGC2366-V1, 1997bs, 2002kg \\
        \hline
      \end{tabular}
    \end{minipage}
  \end{table*}
}

It is clear from this analysis that the locations of interacting transients as a whole are different from the locations of SNIc, both in terms of their radial positions within the hosts, and in terms of their association with H$\alpha$ emission. Taken together with the results of \cite{ande12}, this indicates that most of the interacting events do not have high mass progenitors. 

The NCR value analysis has produced a clear distinction between the SN2008S-like transients and $\eta$ Carina-type Impostors, with all SN2008S-like Impostors lying on regions of zero star formation as traced by H$\alpha$ emission, and $\eta$ Carina-like events, thought to arise from LBV eruptions, showing positive NCR values.

\section{Conclusions}

This paper has presented the most comprehensive host galaxy environment study of interacting transients (SNIIn and SN Impostors) to date. Following a discussion of the selection effects involved in this study, which are more likely to affect the SN Impostor sample than that of the SNIIn sample, we draw the following conclusions:
\begin{itemize}
\item The host galaxies of interacting transients trace the normal star formation in the local Universe, whereas SNIc appear to have more massive hosts.
\item The host galaxies of SN Impostors appear slightly less luminous than CCSN-hosts, and correspondingly have lower metallicities.
\item Inferred local metallicities at the sites of SN Impostors are lower than SNIc, SNIIP and SNIIn sites. SNIIn appear to have slightly more metal-rich sites than SNIIP. 
\item There is a lack of interacting transients in the central regions of host galaxies. The sole event which falls in the central 20 per cent of host galaxy {\it R}-band light is known to be an SNIa-CSM event, therefore no massive progenitors of interacting transients are found in the central regions of their host galaxies.
\item The radial distributions of interacting transients and SNIc (with the highest mass progenitors) are very different (KS; P$<$0.1 per cent), with the former tending to lie in the outer regions of their hosts, while the latter are strongly centrally concentrated.
\item A pronounced difference between interacting transients and SNIc is also seen in the association with ongoing star formation. Unlike the SNIc, the interacting transients do not trace H$\alpha$ emission, and therefore ongoing star formation (KS; P$<$0.1 per cent).
\item All SN Impostors designated SN2008S-like within this sample fall on regions with zero H$\alpha$ emission, whereas those classed as $\eta$ Carina-like fall on positive values, with an average NCR$_{H\alpha}$ of 0.157. Impostors also show higher association with lower mass progenitors, traced by near-UV emission. 
\end{itemize}


\section*{Acknowledgments}
We wish to ackonwledge the referee for their detailed discussion on this paper which has undoubtedly led to improvements.
This research has made use of the NASA/IPAC Extragalactic Database (NED) which is operated by the Jet Propulsion Laboratory, California Institute of Technology, under contract with the National Aeronautics and Space Administration. The Liverpool Telescope is operated on the island of La Palma by Liverpool John Moores University in the Spanish Observatorio del Roque de los Muchachos of the Instituto de Astrof\'isica de Canarias with financial support from the UK Science and Technology Facilities Council. The Isaac Newton Telescope (and Jacobus Kapteyn Telescope) is (was) operated on the island of La Palma by the Isaac Newton Group in the Spanish Observatorio del Roque de los Muchachos of the Instituto de Astrof\'isica de Canarias. Also based on observations made by the ESO 2.2m telescope at the La Silla Observatory (programme ID 084.D.0195). SMH and PAJ acknowledge the UK Science and Technology Facilities Council for PDRA support and research grant support respectively, and JDL for research studentship support. JPA acknowledges fellowship funding from FONDECYT, project number 3110142, and partial support from Iniciativa Cientifica Milenio through the Millennium Centre for Supernova Science (P10-064-F).


\label{lastpage}

\end{document}